\documentclass[11pt]{article}
\usepackage[utf8]{inputenc}
\usepackage[novbox]{pdfsync}   

\usepackage[a4paper,margin=2cm]{geometry}

\usepackage{amsmath,amssymb,amsthm,mathtools,bm,mathrsfs,bbm}
\usepackage{centernot}

\usepackage{graphicx,float,subcaption}
\usepackage{booktabs,longtable,threeparttable,array,multirow}
\usepackage{enumitem}

\usepackage{xcolor,listings,tcolorbox}
\usepackage{algorithm,algorithmicx,algpseudocode}

\usepackage{tikz,tikz-cd,pgfplots}
\usetikzlibrary{graphs,decorations.pathmorphing,decorations.markings,
arrows.meta,automata,positioning}
\tikzcdset{every label/.append style = {font = \small}}
\pgfplotsset{compat=1.18}

\usepackage[version=4]{mhchem}

\usepackage{microtype}
\usepackage{lineno}
\usepackage{nomencl}
\makenomenclature

\usepackage{hyperref}
\usepackage{bookmark}
\usepackage[toc,sort=use]{glossaries}
\makeglossaries

\newif\ifbookbuild
\bookbuildfalse

\newif\ifmadec     \madectrue
\newif\ifslowfast  \slowfasttrue

\newif\ifhnetpart  \hnetparttrue

\RequirePackage{etoolbox}
\usepackage{makecell}
\usepackage[novbox]{pdfsync}
\usepackage{amssymb, amsfonts,mathtools,bbm,eurosym,graphicx,latexsym, amsmath,amsthm,times,nccmath,url}
\usepackage{listings,verbatim,cases,tabularx}
\makeatletter
\@ifpackageloaded{hyperref}{}{%
  \@ifundefined{hyperrefpackage}{%
    \@ifundefined{AAdefNnohyperref}{\usepackage{hyperref}}{}%
  }{}%
}
\makeatother
\makeatletter
\AtBeginDocument{\@ifundefined{hypersetup}{}{\hypersetup{linktoc=all}}}
\makeatother
\usepackage{graphicx,float,fancybox}
\usepackage{epsfig,epstopdf,tikz}
\usepackage[sans]{dsfont}
\usetikzlibrary{positioning,arrows,arrows.meta,calc}
\lstdefinestyle{mathematica}{
 language=Mathematica,
 basicstyle=\small\ttfamily,
 backgroundcolor=\color{gray!10},
 frame=single,
 breaklines=true,
 commentstyle=\color{green!60!black},
 keywordstyle=\color{blue},
 stringstyle=\color{red},
 showstringspaces=false,
 tabsize=2
}
\usepackage{tcolorbox}
\newtcolorbox{keyresult}[1][]{
 colback=blue!5!white,
 colframe=blue!75!black,
 title=#1,
 fonttitle=\bfseries
}
\newtcolorbox{examplebox}[1][]{
 colback=green!5!white,
 colframe=green!75!black,
 title=#1,
 fonttitle=\bfseries
}

\makeatletter
\@ifundefined{theorem}   {\newtheorem{theorem}   {Theorem}}   {}
\@ifundefined{definition}{\newtheorem{definition}{Definition}} {}
\@ifundefined{open}      {}   {}
\@ifundefined{example}   {\newtheorem{example}   {Example}}   {}
\makeatother

\newcounter{tabcounter}
\newcounter{defcounter}

\newcounter{thmcounter}
\newcounter{excounter}
\newcounter{ascounter}
\newcounter{cjcounter}

\newcounter{figcounter}

\newglossarystyle{unifiedstyle}{%
  \setglossarystyle{list}%
  \renewcommand{\glossentry}[2]{%
    \edef\glscursec{\glsentryuseriii{##1}}%
    \ifx\glscursec\glslastsec\else
      \item[]\textbf{\S\,\glscursec\enspace\glsentryuseriv{##1}}%
      \global\let\glslastsec\glscursec
    \fi
    \item[\glsentryuserii{##1} \glsentryuseri{##1}:] ``\glossentryname{##1}'' \glsentrydesc{##1}%
  }%
}

\makeatletter
\def\@firstwordof#1 #2\relax{#1}
\def\@takethree#1#2#3#4\@nil{#1#2#3}

\NewDocumentCommand{\beXa}{s O{} O{} o}{%
  \def\@temparg{#2}%
  \ifx\@temparg\@empty
    \begin{example}%
  \else
    \let\@savedSecTitle\@currentsectionname%
    \refstepcounter{excounter}%
    \IfValueTF{#4}{%
      \edef\@exlblkey{#4}%
    }{%
      \edef\@exfirstword{\@firstwordof#2 \relax}%
      \edef\@exlblkey{%
        \expandafter\@takethree\@exfirstword\@empty\@empty\@empty\@nil}%
    }%
    \def\@tempshort{#3}%
    \ifx\@tempshort\@empty
      \expandafter\xdef\csname @xshort@\@exlblkey\endcsname{\@exlblkey}%
    \else
      \expandafter\def\csname @xshort@\@exlblkey\endcsname{#3}%
    \fi
    \IfBooleanTF{#1}{\let\@extag\@empty}{%
      \edef\@extag{\noexpand\ \noexpand\textup{[\noexpand\texttt{e:\@exlblkey}]}}}%
    \begin{example}[#2\ifshowkeys\@extag\fi]%
    \label{e:\@exlblkey}%
    \protected@write\@auxout{}{%
      \string\@thmtitle{e:\@exlblkey}{\unexpanded{#2}}}%
    \expanded{%
      \noexpand\newglossaryentry{e\theexcounter}{%
        name={\unexpanded{#2}%
          \IfBooleanF{#1}{ [\noexpand\texttt{e:\@exlblkey}]}},%
        sort={\theexample},%
        description={p.~\noexpand\pageref{e:\@exlblkey}},%
        user1={\theexample},%
        user2={Example},%
        user3={\thesection},%
        user4={\expandafter\unexpanded\expandafter{\@savedSecTitle}}%
      }%
    }%
    \protected@write\@auxout{}{\string\glsliveentry{e\theexcounter}}%
  \fi
}
\newcommand{\eeXa}{\end{example}}

\NewDocumentCommand{\beD}{s O{} O{} o}{%
  \def\@temparg{#2}%
  \ifx\@temparg\@empty
    \begin{definition}%
  \else
    \let\@savedSecTitle\@currentsectionname%
    \refstepcounter{defcounter}%
    \IfValueTF{#4}{%
      \edef\@deflblkey{#4}%
    }{%
      \edef\@deffirstword{\@firstwordof#2 \relax}%
      \edef\@deflblkey{%
        \expandafter\@takethree\@deffirstword\@empty\@empty\@empty\@nil}%
    }%
    \def\@tempshort{#3}%
    \ifx\@tempshort\@empty
      \expandafter\xdef\csname @dshort@\@deflblkey\endcsname{\@deflblkey}%
    \else
      \expandafter\def\csname @dshort@\@deflblkey\endcsname{#3}%
    \fi
    \IfBooleanTF{#1}{\let\@deftag\@empty}{%
      \edef\@deftag{\noexpand\ \noexpand\textup{[\noexpand\texttt{d:\@deflblkey}]}}}%
    \begin{definition}[#2\ifshowkeys\@deftag\fi]%
    \label{d:\@deflblkey}%
    \protected@write\@auxout{}{%
      \string\@thmtitle{d:\@deflblkey}{\unexpanded{#2}}}%
    \expanded{%
      \noexpand\newglossaryentry{d\thedefcounter}{%
        name={\unexpanded{#2}%
          \IfBooleanF{#1}{ [\noexpand\texttt{d:\@deflblkey}]}},%
        sort={\thedefinition},%
        description={p.~\noexpand\pageref{d:\@deflblkey}},%
        user1={\thedefinition},%
        user2={Definition},%
        user3={\thesection},%
        user4={\expandafter\unexpanded\expandafter{\@savedSecTitle}}%
      }%
    }%
    \protected@write\@auxout{}{\string\glsliveentry{d\thedefcounter}}%
  \fi
}
\newcommand{\eeD}{\end{definition}}

\newif\ifshowkeys \showkeystrue
\newif\ifcommonred \commonredfalse
\NewDocumentCommand{\beT}{s O{} O{} o}{%
  \def\@temparg{#2}%
  \ifx\@temparg\@empty
    \begin{theorem}%
  \else
    \let\@savedSecTitle\@currentsectionname%
    \refstepcounter{thmcounter}%
    \IfValueTF{#4}{%
      \edef\@thmlblkey{#4}%
    }{%
      \edef\@thmfirstword{\@firstwordof#2 \relax}%
      \edef\@thmlblkey{%
        \expandafter\@takethree\@thmfirstword\@empty\@empty\@empty\@nil}%
    }%
    \IfBooleanTF{#1}{\let\@thmtag\@empty}{%
      \edef\@thmtag{\noexpand\ \noexpand\textup{[\noexpand\texttt{t:\@thmlblkey}]}}}%
    \def\@tempshort{#3}%
    \ifx\@tempshort\@empty
      \expandafter\xdef\csname @tshort@\@thmlblkey\endcsname{\@thmlblkey}%
    \else
      \expandafter\def\csname @tshort@\@thmlblkey\endcsname{#3}%
    \fi
    \begin{theorem}[#2\ifshowkeys\@thmtag\fi]%
    \label{t:\@thmlblkey}%
    \protected@write\@auxout{}{%
      \string\@thmtitle{t:\@thmlblkey}{\unexpanded{#2}}}%
    \expanded{%
      \noexpand\newglossaryentry{t\thethmcounter}{%
        name={\unexpanded{#2}%
          \IfBooleanF{#1}{ [\noexpand\texttt{t:\@thmlblkey}]}},%
        sort={\thetheorem},%
        description={p.~\noexpand\pageref{t:\@thmlblkey}},%
        user1={\thetheorem},%
        user2={Theorem},%
        user3={\thesection},%
        user4={\expandafter\unexpanded\expandafter{\@savedSecTitle}}%
      }%
    }%
    \protected@write\@auxout{}{\string\glsliveentry{t\thethmcounter}}%
  \fi
}
\newcommand{\eeT}{\end{theorem}}

\newcommand{\xpart}[1]{\@ifundefined{r@part:glv}{}{\paragraph{Relation with Parts~I--II.}#1}}

\newcommand{\bookref}[3]{\@ifundefined{r@#1}{#3}{#2}}

\newcommand{\beAs}[1][]{%
  \def\@temparg{#1}%
  \ifx\@temparg\@empty
    \begin{assumption}%
  \else
    \let\@savedSecTitle\@currentsectionname%
    \refstepcounter{ascounter}%
    \begin{assumption}[#1]%
    \label{as:\theascounter}%
    \expanded{%
      \noexpand\newglossaryentry{as\theascounter}{%
        name={\unexpanded{#1}},%
        sort={\theascounter},%
        description={p.~\noexpand\pageref{as:\theascounter}},%
        user1={\theascounter},%
        user2={Assumption},%
        user3={\thesection},%
        user4={\expandafter\unexpanded\expandafter{\@savedSecTitle}}%
      }%
    }%
    \protected@write\@auxout{}{\string\glsliveentry{as\theascounter}}%
  \fi
}
\newcommand{\eeAs}{\end{assumption}}
\newcommand{\beCj}[1][]{%
  \def\@temparg{#1}%
  \ifx\@temparg\@empty
    \begin{conjecture}%
  \else
    \let\@savedSecTitle\@currentsectionname%
    \refstepcounter{cjcounter}%
    \begin{conjecture}[#1]%
    \label{cj:\thecjcounter}%
    \expanded{%
      \noexpand\newglossaryentry{cj\thecjcounter}{%
        name={\unexpanded{#1}},%
        sort={\thecjcounter},%
        description={p.~\noexpand\pageref{cj:\thecjcounter}},%
        user1={\thecjcounter},%
        user2={Conjecture},%
        user3={\thesection},%
        user4={\expandafter\unexpanded\expandafter{\@savedSecTitle}}%
      }%
    }%
    \protected@write\@auxout{}{\string\glsliveentry{cj\thecjcounter}}%
  \fi
}
\newcommand{\eeCj}{\end{conjecture}}
\newcommand{\@mkresult}[5]{%
  \expandafter\NewDocumentCommand\csname be#1\endcsname{s O{} O{} o}{%
    \def\@temparg{##2}%
    \ifx\@temparg\@empty
      \begin{#2}%
    \else
      \let\@savedSecTitle\@currentsectionname%
      \refstepcounter{#4}%
      \IfValueTF{##4}{%
        \edef\@relbl{##4}%
        \IfBooleanTF{##1}{\let\@retag\@empty}{%
          \edef\@retag{\noexpand\ \noexpand\textup{[\noexpand\texttt{#3:##4}]}}}%
      }{%
        \edef\@relbl{\arabic{#4}}%
        \let\@retag\@empty
      }%
      \def\@tempshort{##3}%
      \ifx\@tempshort\@empty
        \expandafter\xdef\csname @sshort@#3\@relbl\endcsname{\@relbl}%
      \else
        \expandafter\def\csname @sshort@#3\@relbl\endcsname{##3}%
      \fi
      \begin{#2}[##2\ifshowkeys\@retag\fi]%
      \edef\@tmplbl{#3:\@relbl}\expandafter\label\expandafter{\@tmplbl}%
      \expanded{%
        \noexpand\newglossaryentry{#3\csname the#4\endcsname}{%
          name={\unexpanded{##2}%
            \IfBooleanF{##1}{\IfValueT{##4}{ [\noexpand\texttt{#3:##4}]}}},%
          sort={\csname the#2\endcsname},%
          description={p.~\noexpand\pageref{#3:\@relbl}},%
          user1={\csname the#2\endcsname},%
          user2={#5},%
          user3={\thesection},%
          user4={\expandafter\unexpanded\expandafter{\@savedSecTitle}}%
        }%
      }%
      \protected@write\@auxout{}{\string\glsliveentry{#3\csname the#4\endcsname}}%
      \protected@write\@auxout{}{\string\@thmtitle{\@tmplbl}{\unexpanded{##2}}}%
    \fi
  }%
  \expandafter\newcommand\csname ee#1\endcsname{\end{#2}}%
}
\@mkresult{L}{lemma}{l}{lemcounter}{Lemma}
\@mkresult{R}{remark}{r}{remcounter}{Remark}
\@mkresult{C}{corollary}{c}{corcounter}{Corollary}
\@mkresult{P}{proposition}{p}{propcounter}{Proposition}
\@mkresult{O}{open}{o}{opcounter}{Problem}

\newcommand{\@currentsectionname}{}
\AtBeginDocument{%
  \let\@epidcrn@orig@refstepcounter\refstepcounter
  \renewcommand{\refstepcounter}[1]{%
    \@epidcrn@orig@refstepcounter{#1}%
    \ifstrequal{#1}{section}{\global\let\@currentsectionname\@currentlabelname}{}%
  }%
}

\newcommand{\beTa}[1][]{%
  \def\@temparg{#1}%
  \ifx\@temparg\@empty
    \begin{table}[htbp]%
  \else
    \let\@savedSecTitle\@currentsectionname%
    \refstepcounter{tabcounter}%
    \begin{table}[htbp]%
    \label{A:\thetabcounter}%
    \expanded{%
      \noexpand\newglossaryentry{A\thetabcounter}{%
        name={\unexpanded{#1}},%
        sort={\thetabcounter},%
        description={p.~\noexpand\pageref{A:\thetabcounter}},%
        user1={\thetabcounter},%
        user2={Table},%
        user3={\thesection},%
        user4={\expandafter\unexpanded\expandafter{\@savedSecTitle}}%
      }%
    }%
    \protected@write\@auxout{}{\string\glsliveentry{A\thetabcounter}}%
  \fi
}
\newcommand{\eeTa}{\end{table}}

\newcommand{\beF}[1][]{%
  \def\@temparg{#1}%
  \ifx\@temparg\@empty
    \begin{figure}[H]%
  \else
    \begin{figure}[H]%
    \let\@savedSecTitle\@currentsectionname%
    \refstepcounter{figcounter}%
    \label{fig:\thefigcounter}%
    \expanded{%
      \noexpand\newglossaryentry{Fig\thefigcounter}{%
        name={\unexpanded{#1}},%
        sort={\thefigcounter},%
        description={p.~\noexpand\pageref{fig:\thefigcounter}},%
        user1={\thefigcounter},%
        user2={Figure},%
        user3={\thesection},%
        user4={\expandafter\unexpanded\expandafter{\@savedSecTitle}}%
      }%
    }%
    \protected@write\@auxout{}{\string\glsliveentry{Fig\thefigcounter}}%
  \fi
}
\newcommand{\eeF}{\end{figure}}
\makeatother
\makeatletter
\newcommand{\@glslive}{}
\newcommand{\@glsaddlive}[1]{\ifglsentryexists{#1}{\glsadd{#1}}{}}
\newcommand{\glsliveentry}[1]{\g@addto@macro\@glslive{\@glsaddlive{#1}}}
\makeatother
\makeatletter
\newcommand{\printboth}{%
  \@glslive   
  \printglossary[style=unifiedstyle, title={0 \hspace{.22em} Definitions, Theorems, Examples and Open Problems}]%
}
\makeatother
\newcommand{\tref}[1]{%
  Thm~\ref{t:#1}(\ifcsname @tshort@#1\endcsname\csname @tshort@#1\endcsname\else#1\fi)%
}

\makeatletter
\newcommand{\@thmtitle}[2]{\global\@namedef{@tlong@#1}{#2}}
\newcommand{\nref}[1]{%
  \ref{#1}%
  \ifcsname @tlong@#1\endcsname
    \ ``\csname @tlong@#1\endcsname''%
  \fi
}
\newcommand{\nm}[1]{%
  \ifcsname @tlong@#1\endcsname
    \csname @tlong@#1\endcsname%
  \else
    \ref{#1}%
  \fi
}
\makeatother

\makeatletter
\@ifundefined{lemma}      {}      {}
\@ifundefined{proposition}{} {}
\@ifundefined{corollary}  {}  {}
\@ifundefined{conjecture} {\newtheorem{conjecture} {Conjecture}}  {}
\@ifundefined{remark}     {}     {}
\@ifundefined{question}   {\newtheorem{question}   {Question}}   {}
\@ifundefined{assumption} {\newtheorem{assumption} {Assumption}} {}
\@ifundefined{hypothesis} {} {}
\@ifundefined{fact}       {}       {}
\makeatother

\makeatletter
\providecommand{\citer}{%
  \@ifundefined{r@t:relayRF}%
    {\cite{AAHrel}}%
    {Theorem~\ref{t:relayRF} of \cite{AAHrel}}%
}
\providecommand{\citeR}{%
  \@ifundefined{r@t:Pstar}%
    {\cite{AAHRah}}%
    {Theorem~\ref{t:Pstar} of \cite{AAHRah}}%
}
\makeatother

\def\beQ{\begin{question}}\def\eeQ{\end{question}}
\def\beA{\begin{assumption}}\def\eeA{\end{assumption}}

\definecolor{funccolor}{RGB}{25,25,112}
\definecolor{desccolor}{RGB}{64,64,64}

\def\bep{\begin{pmatrix}}\def\eep{\end{pmatrix}}
\def\bev{\begin{vmatrix}}\def\eev{\end{vmatrix}}
\def\bea{\begin{eqnarray*}}\def\eea{\end{eqnarray*}}
\def\bc{\begin{cases}}\def\ec{\end{cases}}
\def\BEN{\begin{enumerate}}\def\EEN{\end{enumerate}}
\def\BI{\begin{itemize}}\def\EI{\end{itemize}}

\newcommand{\be}[1]{\begin{equation}\label{#1}}
\newcommand{\ee}{\end{equation}}
\newcommand{\beq}{\begin{eqnarray}}
\def\eeq{\end{eqnarray}}
\def\lbl{\label}
\def\Lra{\Longrightarrow}
\def\Eq{\Leftrightarrow}\def\b{\beta}

\newcommand{\rank}{\operatorname{rank}}

\def\Out{\mathrm{Out}}

\newcommand{\R}{\mathbb{R}}  

\providecommand{\Q}{\mathbb{Q}}

\newcommand{\bff}[1]{{\mbox{\boldmath$#1$}}}

\def\v1{\vec {\bff 1}}

\providecommand{\one}{\mathbf 1}             

\long\def\symbolfootnote[#1]#2{%
\begingroup
\def\thefootnote{\fnsymbol{footnote}}\footnote[#1]{#2}%
\endgroup}

\makeatletter
\newcommand{\ideaitem}[2]{%
  \expandafter\gdef\csname @idea@#1\endcsname{#2}%
  \item\hyperlink{idea:#1}{#2}}
\newcommand{\ideahead}[2][]{%
  \item\hypertarget{idea:#2}{}%
  \def\@ideatmp{#1}%
  \ifx\@ideatmp\@empty
    \@ifundefined{@idea@#2}{%
      \PackageWarning{preamble}{Idea #2 is used in Main Results but was never
        defined by \string\ideaitem\space in the abstract}%
      \textbf{[idea #2 undefined]}%
    }{\textbf{\begingroup\def\ideaterm##1##2{##2}\def\idearest##1{}\@nameuse{@idea@#2}\endgroup}}%
  \else
    \textbf{#1}%
  \fi}
\newcommand{\ideaterm}[2]{\hyperlink{#1}{#2}}
\newcommand{\idearest}[1]{#1}
\makeatother

\newif\ifshort \shortfalse

\providecommand{\cA}{\mathcal A}\providecommand{\cB}{\mathcal B}
\providecommand{\cC}{\mathcal C}

\providecommand{\cS}{\mathcal S}

\providecommand{\Buf}{\operatorname{Buf}}
\providecommand{\Jinv}{J_{\rm inv}}

\newenvironment{frame}[2][]{\subsection{#2}}{}
\newenvironment{columns}[1][]{\par}{\par}
\newenvironment{column}[1]{\par\medskip}{\par\medskip}
\newenvironment{block}[1]{\par\medskip\noindent\textbf{#1.}\ }{\par\medskip}
\providecommand{\slidefont}[1]{}
\providecommand{\framebreak}{\par\medskip}

\usepackage{adjustbox}
\usetikzlibrary{fit,backgrounds}
\showkeysfalse
\hypersetup{hidelinks}

\title{Okada--Mochizuki buffering lattices and strain invasion\\
in epidemic reaction networks}
\author{Florin Avram}
\date{November 2026}

\begin{document}
\maketitle

\begin{abstract}
An Okada--Mochizuki buffering structure is a set of species whose own rates,
when perturbed, leave every equilibrium concentration outside the set
unchanged; these sets form a distributive lattice. We ask when such a lattice
protects a quantity of epidemiological interest, namely the invasion
threshold of a competing strain.

Our main example is a fourteen-species two-strain model with two host groups,
in which three maturation modules sit \emph{inside} the infection loop, so
that no subsystem carrying the thresholds is autonomous. Its resident-face
lattice has six elements and is not a chain, and its two incomparable
maximal buffers separate the rate perturbations that cannot move the
strain-$2$ invasion number from those that can. The largest buffer disjoint
from the invasion-relevant coordinates certifies ten rates, varied
independently and including nonlinear catalytic conversions; the one branch
the lattice does not protect produces an explicit threshold crossing.
Uniform permanence of both strains survives nine of those rates at once.

For the same model we ask what unequal maturation release rates cost. The
positive resident equilibrium stays unique for every catalytic coefficient
and every pair of release rates, so the obstruction is dynamical, not
algebraic. Global attraction on the resident face, hence permanence, is
proved on two explicit regions, neither containing the other: one makes the
resident system cooperative, the other rests on a small-gain estimate that
needs no sign condition and allows arbitrarily large catalysis.
For comparison, a twenty-compartment two-patch model with selective
vaccination and staged immunity has a buffering lattice of $83$ elements, of
which $81$ are proper and nonempty. Adding reinfection reactions one at a
time shrinks it to $47$, then $11$, then $3$. There the buffers lie
downstream of an autonomous infection core, which is why a large lattice
carries little epidemiological information.
\end{abstract}

\tableofcontents

\section{Introduction}

Buffering structures, introduced by \cite{OM16,OM17}, identify biologically
significant subnetworks whose structure implies useful simplifications in
their analysis. Our purpose below is to show that they can also be useful in
more complicated networks arising in eco-epidemiology, microbiology, and
immuno-virology, together with the siphon geometry \cite{AdLS}, already applied in these fields \cite{AAN,AABH25,AAH26}.

After introducing these two geometries, we provide, for the convenience of
researchers new to reaction networks,  some simple examples in which buffering
structures are present but their use is not essential, in Sections \ref{5}, \ref{SEAIR}.
Section~\ref{sec:theory} then reviews the Okada--Mochizuki theory used here:
the lattice of buffered species sets, its canonical blocks, the invasion
certificates it yields on siphon faces, and the permanence criterion applied
to the models that follow.

Possibly the most revealing interaction between siphon geometry and Okada--Mochizuki buffering
 illustrated here is in Section~\ref{sec:twostrain}, by a fourteen-species
two-strain model in which the buffering modules lie inside the infection
loop. Siphon geometry identifies the
invariant resident face and the transverse matrix governing invasion,
while Okada--Mochizuki geometry identifies resident-rate perturbations
that leave this matrix, and hence the invasion threshold, unchanged.
The latter conclusion does not follow from siphon-face triangularity alone.

Section~\ref{sec:newmath} asks what unequal maturation release rates cost in
that model, and answers it twice: the positive resident equilibrium is unique
for every catalytic coefficient and every pair of release rates, and global
attraction on the resident face, hence permanence, holds on two explicit
regions neither of which contains the other --- one making the resident
system cooperative, one resting on a small-gain estimate that imposes no sign
condition.

\paragraph{Accompanying files.} Every computation reported below is carried
out in a Mathematica file of the accompanying folder, each built on the
\texttt{EpidCRN} package: \texttt{TwoStrain.wl} for the structural results of
Section~\ref{sec:twostrain}, \texttt{unequal.wl} for
Section~\ref{sec:unequal}, \texttt{newmath.wl} for
Sections~\ref{sec:newmath} and~\ref{sec:unequalGain}, \texttt{ccf.wl} for the
canonical blocks of Section~\ref{sec:ccf}, \texttt{ChenWu.wl} for
Section~\ref{sec:twomodels}, \texttt{Murotadyn.wl} for
Section~\ref{sec:dynamic} and \texttt{CWG.wl} for Section~\ref{sec:cwg}.
Each file is named again where its computation is reported, and the sections
of the main text say there what their file verifies; the five- and
six-species examples of Sections~\ref{5} and~\ref{SEAIR} are small enough to
be computed in the text.


\section{Reaction networks: definition, siphons, invasion thresholds, buffering index}
\label{sec:defs}
%
%
%

The following definition specializes the influence-specification framework of Shinar and Feinberg (\emph{Concordant chemical reaction networks}, Mathematical Biosciences 240 (2012), Section 9) to differentiable rates satisfying reactant-vanishing boundary conditions; fixed-sign influences form a subclass.

\beD[(Positive) ODEs with a boundary-compatible stoichiometric representation][Reaction network][RN]
We consider autonomous $C^1$ ODEs on the nonnegative orthant
$\mathbb R^{n_S}_{\ge0}$, with species
$x=(x_i)_{i\in\cS}$, $n_S=|\cS|$, admitting a
\emph{stoichiometric representation}
\[
\dot x=f(x)=\Gamma v(x),\qquad
\Gamma=\beta-\alpha,
\]
where $\alpha,\beta\in\mathbb R_{\ge0}^{n_S\times n_R}$
are constant matrices and
$v:\mathbb R_{\ge0}^{n_S}\to\mathbb R_{\ge0}^{n_R}$
is $C^1$ and time-independent. The columns of $\alpha$
and $\beta$ are the input and output vectors of the formal
reactions
\[
\sum_i\alpha_{ij}X_i\longrightarrow
\sum_i\beta_{ij}X_i,\qquad j\in\mathcal R.
\]
Their stoichiometric columns are
$\Gamma_j=\beta_j-\alpha_j$.

We impose the following \emph{boundary compatibility condition}:
\begin{equation}\label{eq:bdcompat}
\alpha_{ij}>0,\quad x_i=0
\quad\Longrightarrow\quad v_j(x)=0,
\qquad x\in\mathbb R_{\ge0}^{n_S}.
\end{equation}
Consequently, for every $i$ and every $x\ge0$ with $x_i=0$,
\[
f_i(x)
=\sum_{j:\alpha_{ij}=0}\beta_{ij}v_j(x)\ge0.
\]
Thus the nonnegative orthant is forward invariant, so the
ODE is positive.

The \emph{reactant set} of reaction $j$ is
$L_j=\{i:\alpha_{ij}>0\}$. Its
\emph{derivative-support set} is
\[
D_j=\{i:\partial v_j/\partial x_i\not\equiv0\},
\qquad
\Out(M)=\{j:D_j\cap M\ne\varnothing\}.
\]
For the subclass with signed influences, each
$\partial v_j/\partial x_i$ is assumed nonnegative,
nonpositive, or identically zero on the positive orthant.
This sign restriction is not required for the buffering
results below.

An \emph{equilibrium} is a zero of $f$ and is
\emph{regular} when $J=Df=\Gamma D_xv$ is nonsingular.
When conservation laws are present, the buffering results
requiring regularity must instead be applied in an
appropriate reduced representation.

We write $\Gamma_R$ for the submatrix of $\Gamma$
with columns indexed by $R\subseteq\mathcal R$.
\eeD

Some comments:

\textbf{KTV and Murota.}
Kuhrs, Telek and Vassena \cite{KTV} assume $D_j=L_j$
in their definition of reaction-rate functions, and use this
equality throughout their reaction-network results, notably to
construct the symbolic reactivity matrix and define
output-complete subnetworks.  Their underlying LM-matrix theorems, based on
Murota's theory, require only the symbolic derivative-support
pattern and do not require $D_j=L_j$.
Our Section~\ref{sec:ccf} demonstrates the applicability of their
results to $D_j\ne L_j$.

\textbf{Reactants, catalysts and inhibitors.}
A species in $D_j\setminus L_j$ with positive derivative is a kinetic
\emph{activator}, and one with negative derivative is an
\emph{inhibitor}; a stoichiometric catalyst instead occurs on both
sides of a reaction. For example, the inhibited transmission of
Section~\ref{sec:twostrain},
\[
v=\frac{\beta_{2g}S_gP_2}{1+\chi C_2},
\qquad L_j=\{S_g,P_2\},\qquad D_j=\{S_g,P_2,C_2\},
\]
satisfies \eqref{eq:bdcompat}, although $C_2$ inhibits the reaction
without being a reactant. Such dependencies are also considered in
\cite{fages2015inferring}.

\textbf{Fixed signs and derivative support.}
The fixed-sign assumption is not needed for Theorems~\ref{t:OMloc},
\ref{t:distlat}, \ref{t:faceRank} and \ref{t:bufInv}, which use only
$D_j$. It is relevant to the signed-influence formulation;
Theorem~\ref{t:sensitivity}
requires additional sign conditions on the derivatives of $F$ and
$V$. The buffering results remain valid for sign-changing derivatives.

\textbf{Reactants versus dependencies.}
Output completeness, buffering indices and equilibrium localization
use $D_j$, whereas siphon invariance uses $L_j$ and
\eqref{eq:bdcompat}. A species influencing a rate need not make that
rate vanish when its concentration is zero.

\textbf{Real stoichiometric columns.}
The models of Section~\ref{sec:twostrain} and
Section~\ref{sec:dynamic} have integer columns, whereas lossy
migration $X_p\to\nu^X_{qp}X_q$ in Sections~\ref{sec:twomodels}
and \ref{sec:cwg} contributes
$-e_{X_p}+\nu^X_{qp}e_{X_q}$, $0<\nu^X_{qp}<1$.
Only Lemma~\ref{l:graphRank} requires the special unit-transfer
column structure.

\subsection{Siphons: faces, resident dynamics and the invasion threshold}

\beD[Siphons, resident dynamics and transverse linearization, regular face][Siphon][siphon]
A species set $\Sigma\subseteq\cS$ is a \emph{siphon} if every reaction that
produces a species of $\Sigma$ has a reactant in $\Sigma$. Its \emph{face}
\[
F_\Sigma=\{x\ge0:x_\Sigma=0\}
\]
is then forward invariant by \eqref{eq:bdcompat}. Conversely, $F_\Sigma$
is invariant if and only if every reaction $j$ with
$L_j\cap\Sigma=\varnothing$ that produces a species of $\Sigma$
satisfies $v_j(y,0)\equiv0$. In particular, invariance implies the siphon
property if, in addition, $v_j(x)>0$ whenever $x_i>0$ for every $i\in L_j$,
regardless of the remaining coordinates; this additional condition is not
assumed here. The species of $\Sigma$ vanish on $F_\Sigma$; the
species of $T=\cS\setminus\Sigma$ are the \emph{resident species} of
$F_\Sigma$. A siphon is \emph{minimal} if it contains no smaller nonempty
siphon.

Let
$T=\cS\setminus\Sigma$. Write $x=(y,z)$ with $y=x_T$ the \emph{resident} and
$z=x_\Sigma$ the \emph{transverse} coordinates.
\begin{enumerate}
\item The \emph{resident network} of $\Sigma$ is obtained by retaining the
species $T$, setting $z=0$, and discarding the reactions whose restricted
rates vanish identically. Its reaction set, rates and stoichiometric matrix
are
\[
\mathcal R_\Sigma=\{j:v_j(y,0)\not\equiv0\},\qquad
v^\Sigma_j(y)=v_j(y,0),\qquad
\Gamma^\Sigma=P_T\Gamma_{\mathcal R_\Sigma}.
\]
\item On $F_\Sigma$ the \emph{resident Jacobian} and the \emph{invasion}, or
\emph{transversal}, \emph{Jacobian} are
\[
J_{\rm tan}(y)=D_yf_y(y,0),\qquad
\Jinv(y)=D_zf_z(y,0),
\]
and $J_\Sigma=J_{\rm tan}=\Gamma^\Sigma D_yv^\Sigma$ is the Jacobian of the
resident network. Invariance of the face gives $f_z(y,0)\equiv0$, hence
$D_yf_z(y,0)=0$ and the exact block triangularity
\begin{equation}\label{eq:siphontri}
J(y,0)=\begin{pmatrix}J_{\rm tan}(y)&J_{\rm mix}(y)\\0&\Jinv(y)\end{pmatrix},
\qquad
\det J(y,0)=\det J_{\rm tan}(y)\,\det \Jinv(y),
\end{equation}
at every point of the face, not only at its equilibria.
Moreover, for $i\ne k$ in $\Sigma$, positivity of the ODE gives
$\partial f_i/\partial z_k(y,0)\ge0$; hence $\Jinv(y)$ is Metzler
without any monotonicity assumption on the rates.
\item The face of $\Sigma$ is \emph{regular} if it carries a
relative-interior equilibrium, positive in every coordinate of $T$, at which
$J_\Sigma$ is nonsingular. When the resident network carries conservation
laws, $J_\Sigma$ is singular for trivial reasons, and nonsingularity is to be
read, as in Definition~\ref{d:RN}, for the resident Jacobian of a reduced
representation in which the conservation laws have been eliminated.
\end{enumerate}
\eeD

\beD[Invasion matrix, invasion exponent, reproduction number and invasion invariance][Invasion exponent and invariance][invExp]
At an equilibrium $e=(y^*,0)$ on the face of a siphon $\Sigma$, the
\emph{invasion matrix} is the transverse diagonal block
$\Jinv=D_zf_z(y^*,0)$ of \eqref{eq:siphontri}
(Definition~\ref{d:siphon}). Its spectral abscissa $\alpha_0=s(\Jinv)$, the
largest real eigenvalue, since $\Jinv$ is Metzler, is the \emph{invasion
exponent}; invasion is possible if $\alpha_0>0$.
For any decomposition $\Jinv=F-V$ with $F\ge0$ and $V$ a nonsingular
$M$-matrix, the \emph{reproduction number} is $R_0=\rho(FV^{-1})$, and
$R_0-1$ and $\alpha_0$ have the same sign. When $\Sigma$ is the set of
species of one strain and the residents are another community $K$, we write
$R_{j\mid K}$ and call it an \emph{invasion number}.

If $e=e(\theta)$ is a smooth equilibrium branch in a reaction-rate parameter
$\theta$, the invasion matrix is \emph{$\theta$-invariant} when
$d\Jinv/d\theta=0$, and the reproduction number is \emph{$\theta$-invariant}
when $dR_0/d\theta=0$. Invariance of $\Jinv$ alone does not imply invariance
of $R_0$ for a chosen splitting (Remark~\ref{r:why}).
\eeD

\subsection{Okada--Mochizuki buffering: output completions, index and localization}

\beD[Output completion, index][Output completion][outComp]
Let $\dot x=\Gamma v(x)$ be a reaction network as in
Definition~\ref{d:RN}.
\begin{enumerate}
\item For $M\subseteq\cS$, the \emph{output completion} is the set of
reactions whose rates involve at least one species of $M$,
\[
\Out(M)=\{j\in\mathcal R:D_{x_M}v_j\not\equiv0\}
=\{j:D_j\cap M\ne\varnothing\},
\]
with $D_j$ the dependence set of Definition~\ref{d:RN} --- not the reactant
set $L_j$.
\item A pair $(M,R)$ of a species set and a reaction set, also called a
subnetwork, is \emph{output-complete} if $\Out(M)\subseteq R$; thus
$(M,\Out(M))$ is the smallest output-complete pair with species set $M$.
\item An output-complete pair has the \emph{index}
$\lambda(M,R)=\rank\Gamma_R-|M|$, and we abbreviate
$\lambda(M)=\lambda(M,\Out(M))$.
\item \emph{The same notions for a resident network.} For a siphon $\Sigma$,
let $T=\cS\setminus\Sigma$ and let $\mathcal R_\Sigma$, $v^\Sigma$,
$\Gamma^\Sigma$ be as in Definition~\ref{d:siphon}. For $M\subseteq T$,
define $\Out_\Sigma(M)$ from the restricted rates $v^\Sigma$, and for
$R\subseteq\mathcal R$ put $R_\Sigma=R\cap\mathcal R_\Sigma$. The
\emph{resident buffering index} is
\[
\lambda_\Sigma(M,R_\Sigma)=\rank\Gamma^\Sigma_{R_\Sigma}-|M|,
\qquad
\lambda_\Sigma(M)=\lambda_\Sigma(M,\Out_\Sigma(M)).
\]
\end{enumerate}

\eeD

\beD[Structural zero and law of localization][Structural zero][strZero]
At a regular equilibrium $x^*$, a parameter $\theta$ of the rate $v_j$
changes the equilibrium by
$\partial_\theta x^*=-J^{-1}\Gamma_j\,\partial_\theta v_j$. The entry
$\partial_\theta x^*_i$ is a \emph{structural zero} if it vanishes for all
admissible rate functions with the given dependence sets $D_j$, and not
merely for particular parameter values. The \emph{law of localization} of \cite{OM16} states that
for a buffering structure $(M,R)$ --- an output-complete pair of index zero,
Definition~\ref{d:bufStruct} --- and every $j\in R$, the entry
$\partial_\theta x^*_i$ is a structural zero for every species $i\notin M$.
Theorem~\ref{t:OMloc} proves this law, with its converse.
\eeD

\emph{Some} of the networks of this paper have only unit-transfer columns
$e_j-e_i$, inflows $e_i$ and outflows $-e_i$, however nonlinear their rates.
The fourteen-species model of Section~\ref{sec:twostrain} and the
forty-five-species model of Section~\ref{sec:dynamic} do, and for those the
index is an incidence-graph count, which is what makes their exhaustive
lattice enumerations possible. The next lemma is stated for exactly that
class, and is used for nothing else.

It does \emph{not} cover lossy migration, and no claim is made that it does.
A column $-e_i+\nu e_j$ with $0<\nu<1$ is not an incidence column, so the
two-patch model of Section~\ref{sec:twomodels} and the twenty-compartment
model of Section~\ref{sec:cwg} fall outside the hypothesis of
Lemma~\ref{l:graphRank} and their lattices are not computed with it. They are
computed with Lemma~\ref{l:fwdclosed}, whose hypothesis is instead an
independent mortality reaction for every species, and which handles a lossy
column directly: provided $X_i\in M$ --- which is automatic for the
source-dependent migration reactions at issue, a reaction of that kind
belonging to $\Out(M)$ only through its source $X_i$ --- the column
$-e_i+\nu e_j$ lies in $\operatorname{span}\{e_X:X\in M\}$ if and only if
$X_j\in M$, the same forward-closure test as for a lossless transfer,
whatever the travel loss. The $83$-element lattice therefore rests
on that argument and not on this lemma.

\beL[The buffering index is a graph count for unit-transfer networks][Graphical buffering index][graphRank]
Suppose that every nonzero stoichiometric column of a network is $\pm e_i$ or
$e_j-e_i$, with arbitrary compatible reaction rates. For a species set $M$,
form the undirected graph $G(M)$ with vertices $\cS\cup\{0\}$ and, for each
reaction in $\Out(M)$, an edge $ij$ for a column $e_j-e_i$ and an edge $0i$
for a column $\pm e_i$; zero columns may be omitted. Let $c(G)$ count the
connected components, isolated vertices included. Then
\begin{equation}\label{eq:OMGraphicRank}
\rank\Gamma_{\Out(M)}=|\cS|+1-c(G(M)),
\qquad
\lambda(M)=|\cS\setminus M|+1-c(G(M)).
\end{equation}
\eeL
\textbf{Proof.} Append a row indexed by $0$, so that each augmented column
sums to zero. The resulting matrix is a signed vertex--edge incidence matrix
of $G(M)$, of rank $|\cS|+1-c(G(M))$; deleting the row $0$ preserves the
rank, because that row is the negative sum of the others. Parallel edges and
discarded zero columns do not change the rank. Subtracting $|M|$ gives the
second identity.

\medskip
The same count applies verbatim to a resident network, with $\cS$ replaced by
$T$ and $\Out$ by $\Out_\Sigma$.

Siphons and buffering structures express different properties, and neither
lattice can be identified with the other without further hypotheses:
\begin{center}\slidefont{\small}
\begin{tabular}{p{.22\linewidth}p{.33\linewidth}p{.33\linewidth}}\toprule
 & Buffering structure & Siphon\\\midrule
Data & species and reaction subset & species subset\\
Main property & steady-state sensitivity localization & invariance of a
boundary face\\
Lattice & distributive, with a closure operator & closed under unions\\
Jacobian consequence & generic inverse zeros, determinant factors & exact
tangential--invasion triangularity at the face\\\bottomrule
\end{tabular}
\end{center}
\section{Okada--Mochizuki buffer(ing structure) in a five-species
eco-epidemiologic example}\lbl{5}
\begin{frame}{Example 1: a five-species network with an internal two-state block}
\begin{columns}[c]
\begin{column}{.46\textwidth}
\begin{center}
\adjustbox{max width=\linewidth,max totalheight=0.75\textheight}{%
\begin{tikzpicture}[>=Latex,
  b/.style={draw,rounded corners,align=center,minimum height=8mm,minimum width=10mm,font=\small},
  lab/.style={font=\scriptsize}]
\node[b,fill=orange!12] (S) at (0,0) {$S$};
\node[b,fill=orange!12] (I) at (3.1,0) {$I$};
\node[b,fill=blue!8] (E1) at (3.1,-2.2) {$E_1$};
\node[b,fill=blue!8] (E2) at (5.7,-2.2) {$E_2$};
\node[b,fill=blue!8] (P) at (5.7,0) {$P$};
\draw[->,thick] (-1.7,0)--node[above,lab]{$\Lambda$}(S);
\draw[->,thick] (S)--node[above,lab]{$\beta SP$}(I);
\draw[->,thick] (I)--node[left,lab]{$qI$}(E1);
\draw[->,thick] (E1.15)--node[above,lab]{$k_{12}$}(E2.165);
\draw[->,thick] (E2.195)--node[below,lab]{$k_{21}$}(E1.-15);
\draw[->,thick] (E1)--node[pos=.35,above,lab,sloped]{$c_1$}(P);
\draw[->,thick] (E2)--node[right,lab]{$c_2$}(P);
\node[draw,dashed,rounded corners,blue!60!black,fit=(E1)(E2),inner sep=6pt,
  label={[lab]below left:storage}] {};
\end{tikzpicture}}
\end{center}

{\slidefont{\tiny} Seven edges, one per non-decay flux. The four decays
$\mu_S,\mu_I,\mu_{E_2},\mu_P$ are not drawn; $P$ catalyses $S\to I$ and
$I$ catalyses the production of $E_1$, which is why those labels carry the
full monomial.}
\end{column}
\begin{column}{.52\textwidth}\slidefont{\tiny}
\textbf{Stoichiometric matrix and rates}, $\dot x=\Gamma v$,
$x=(S,I,E_1,E_2,P)^\top$; inflow column first, decay columns last. Each column
is headed by the label of its arrow in the figure, the decays by their rate:
\[
\adjustbox{max width=\linewidth}{$
\Gamma=\begin{array}{r|r|rrrrrr|rrrr}
&\Lambda&\beta SP&qI&k_{12}&k_{21}&c_1&c_2&\mu_S&\mu_I&\mu_{E_2}&\mu_P\\\hline
S  & 1&-1& 0& 0& 0& 0& 0&-1& 0& 0& 0\\
I  & 0& 1& 0& 0& 0& 0& 0& 0&-1& 0& 0\\
E_1& 0& 0& 1&-1& 1&-1& 0& 0& 0& 0& 0\\
E_2& 0& 0& 0& 1&-1& 0&-1& 0& 0&-1& 0\\
P  & 0& 0& 0& 0& 0& 1& 1& 0& 0& 0&-1
\end{array}$}
\]
\[
v^\top=\big(\Lambda,\ \beta SP,\ qI,\ k_{12}E_1,\ k_{21}E_2,\ c_1E_1,\ c_2E_2,\
\mu_SS,\ \mu_II,\ \mu_{E_2}E_2,\ \mu_PP\big)
\]
with $\mu_S=\mu$, $\mu_I=\gamma+\mu$, $\mu_P=\delta$, and $\mu_{E_2}=d$ the
loss rate of the second internal state, the one that is $0$ in the base model.

\medskip
The column $qI$ produces $E_1$ without consuming $I$, and the column $\beta SP$
moves $S$ to $I$ with $P$ as catalyst: a single $+1$ and a $\pm1$ pair.

\medskip
\textbf{Minimal siphon:} $\{I,E_1,E_2,P\}$, the infection face.
\textbf{Nonnegative fixed points:} at most two, the disease-free one and one
endemic.
\end{column}
\end{columns}
\end{frame}

\begin{frame}{Observation: certain equilibrium coordinates are insensitive to some rates (equilibrium localization)}

For $d=0$, the equilibrium coordinates $S^*,I^*,P^*$ do not involve
$k_{12},k_{21},c_1,c_2$; indeed, adding the two intermediate equations gives
\[\frac d{dt}(E_1+E_2)=qI-c_1E_1-c_2E_2 \Lra c_1E_1^*+c_2E_2^*=qI^*\]
 and hence
\[P^*=\frac{qI^*}{\delta},\qquad
 S^*=\frac{(\gamma+\mu)\delta}{\beta q}.\]

\medskip
Also, the reproduction number
\[R_0=\frac{\beta S_0 q}{a\delta}, S_0=\Lambda/\mu, a=\gamma+\mu\]
 is independent of the same rates.
\end{frame}

\begin{frame}[allowframebreaks]{Which rates cannot affect which equilibrium coordinates?}

\beD[Okada--Mochizuki buffer(ing structure)][Buffering structure][bufStruct]
For a species set $M$, $\Out(M)$, the output completion of $M$, consists of all reactions whose
rates depend on species of $M$. Write $z=x_M$. A pair $(M,R)$ is
\emph{output-complete} iff
\[
\bc R\supseteq\Out(M)\ \Eq\
D_z v_{R^c}\equiv0\ \Eq\
\forall j\in R^c:\ D_z v_j\equiv0
,\\
\lambda(M,R)=\rank\Gamma_R-|M|\quad\text{(index)}.\ec
\]
An output-complete pair with $\lambda=0$ will be called a
\emph{OM buffer(ing structure)}. Equivalently, writing
$\kappa(R)=\dim\{v\in\ker\Gamma:\operatorname{supp}v\subseteq R\}$ for the
dimension of the flux kernel carried by $R$, one has
$\rank\Gamma_R=|R|-\kappa(R)$ and therefore
$\lambda(M,R)=|R|-|M|-\kappa(R)$, which is the form in which the index
appears in \cite{OM16,OM17}.
\eeD
$
M=\{E_1,E_2\},\quad
R=\Out(M)=
\left\{
\begin{array}{c}
E_1\to E_2,\quad E_2\to E_1,\\
E_1\to P,\quad E_2\to P
\end{array}
\right\} \Lra \Gamma_R=\begin{pmatrix}-1&1&-1&0\\1&-1&0&-1\\0&0&1&1\end{pmatrix},
$
so $(M,R)$ is an OM buffer, since the four stoichiometric columns have rank $2$,
$
\Lra \boxed{\lambda(M,R)=2-2=0.}
$

The first condition above $D_z v_{R^c}\equiv0$ is already a localization; with the second, it  implies also equilibrium localization: changes in the
four buffer reaction rates cannot affect the equilibrium coordinates
outside $M$, namely $S^*,I^*,P^*$.
\end{frame}

\begin{frame}{Zero buffering index, triangularization and equilibrium localization}
\beT[Okada--Mochizuki localization][][OMloc]
Let $(M,R)$ be output-complete and $x^*$ a regular equilibrium.
Order $x=(y,z)$, where $z=x_M$, $m=|M|$, and $J=[J_y\ J_z]$.
The following are equivalent:
\begin{enumerate}
\item \textbf{Zero index:} $\lambda(M,R)=\rank\Gamma_R-m=0$.
\item \textbf{Block triangularization:} There exists an invertible
$T=\binom{L}{W}$ such that
\[
TJ=\begin{pmatrix}A_c&0\\B&A_b\end{pmatrix},
\qquad T\Gamma_R=\binom{0}{K},
\qquad \det A_c\det A_b\ne0,
\]
where $A_c$ is $(n-m)\times(n-m)$ and $A_b$ is $m\times m$.
\item \textbf{Equilibrium localization:}
\[
\boxed{P_{M^c}J^{-1}\Gamma_R=0.}
\]
\end{enumerate}
\eeT

In particular, for parameters $\theta$ affecting only rates in $R$,
\[
\partial_\theta y^*
=-P_{M^c}J^{-1}\Gamma_R\,\partial_\theta v_R=0.
\]
\end{frame}

\begin{frame}{Proof of the equivalence}
By the chain rule and output completeness ($D_zv_{R^c}=0$),
\[
\begin{aligned}
J_z&=J\binom{0}{I_m}=\Gamma(D_xv)\binom{0}{I_m}\\
&=\begin{pmatrix}\Gamma_R&\Gamma_{R^c}\end{pmatrix}
\begin{pmatrix}D_yv_R&D_zv_R\\D_yv_{R^c}&0\end{pmatrix}
\binom{0}{I_m}=\Gamma_RD_zv_R.
\end{aligned}
\]
At a regular equilibrium, $\det J\ne0$, hence
$\rank J_z=m\le\rank\Gamma_R$.

\textbf{(1) $\Rightarrow$ (2).}
$\rank\Gamma_R=m\Rightarrow\dim\ker \Gamma_R^\top=n-m$.
Choose $L$ with independent rows spanning this kernel,
and $W$ completing $T=\binom{L}{W}$ to an invertible matrix:
\[
L\Gamma_R=0\Rightarrow LJ_z=0,\qquad
TJ=\begin{pmatrix}LJ_y&0\\WJ_y&WJ_z\end{pmatrix},
\quad T\Gamma_R=\binom{0}{W\Gamma_R}.
\]
$\det A_c\det A_b=\det T\det J\ne0$.
\end{frame}

\begin{frame}{Proof of the equivalence (continued)}
\textbf{(2) $\Rightarrow$ (3).}
\[
\begin{aligned}
J^{-1}\Gamma_R&=(TJ)^{-1}T\Gamma_R\\
&=\begin{pmatrix}A_c^{-1}&0\\-A_b^{-1}BA_c^{-1}&A_b^{-1}\end{pmatrix}
\binom{0}{K}=\binom{0}{A_b^{-1}K}
\Rightarrow P_{M^c}J^{-1}\Gamma_R=0.
\end{aligned}
\]

\textbf{(3) $\Rightarrow$ (1).}
\[
P_{M^c}J^{-1}\Gamma_R=0
\Rightarrow J^{-1}\Gamma_R=\binom{0}{H}
\Rightarrow \Gamma_R=J_zH
\Rightarrow \rank\Gamma_R\le m\le\rank\Gamma_R
\Rightarrow\lambda(M,R)=0.
\]
\end{frame}


\begin{frame}[allowframebreaks]{The same five-species network: lossless storage localizes, inactivation of the stored state does not}
\label{sec:fivloss}
\slidefont{\small}

Example 1 left a parameter unused. The decay column $\mu_{E_2}$ carries the
loss rate $d$ of the second internal state, set to $0$ in the base model, and
the whole localization statement was made at $d=0$. The next proposition says
what that costs: at $d=0$ the pair $M=\{E_1,E_2\}$ is a buffer and the
reproduction number is independent of all four internal rates, while an
arbitrarily small $d>0$ raises the index to $1$, destroys the localization,
and moves the threshold. Write
\[
a=\gamma+\mu,\quad b=\beta S_0,\quad \delta=\mu_P,\quad
E=(E_1,E_2)^\top,\quad c=(c_1,c_2)^\top,
\]
and order the invasion coordinates as $(I,P\mid E_1,E_2)$, so that the
invasion matrix of the siphon $\{I,E_1,E_2,P\}$ at the disease-free
equilibrium is
\begin{equation}\label{eq:fiveBlockMatrix}
\Jinv=\begin{pmatrix}G&uc^\top\\Q&A\end{pmatrix},\quad
G=\begin{pmatrix}-a&b\\0&-\delta\end{pmatrix},\quad
u=\binom01,\quad
Q=\begin{pmatrix}q&0\\0&0\end{pmatrix},\quad
A=K-D,
\end{equation}
with
\[
K=\begin{pmatrix}-k_{12}&k_{21}\\k_{12}&-k_{21}\end{pmatrix},\qquad
D=\operatorname{diag}(c_1,c_2),\qquad
\mathbf1^\top K=0,\qquad c^\top=\mathbf1^\top D.
\]

\beP[Lossless storage localizes the threshold, and inactivation of the stored state destroys both the buffer and the threshold][Storage buffering and inactivation][fiveloss]
Let $D_0=\det(-A)=k_{12}c_2+c_1k_{21}+c_1c_2>0$. Items~1--4 concern lossless
storage, $d=0$; item~5 concerns inactivation, $d>0$.
\begin{enumerate}
\item \textbf{Transfer balance and complete transfer.} Applying
$\mathbf1^\top$ to $\dot E=AE+qe_1I$ gives
\begin{equation}\label{eq:fiveCoreBalance}
\frac{d}{dt}(\mathbf1^\top E)=qI-c^\top E,
\qquad\text{hence}\qquad
c^\top E^*=qI^*,\quad P^*=qI^*/\delta,
\end{equation}
and, equivalently,
\begin{equation}\label{eq:fiveComplete}
\boxed{\ c^\top(-A)^{-1}e_1=1:\ }
\end{equation}
everything entering $E_1$ leaves the storage pair through $c$. This is the
exact analogue, for a two-state store, of the complete transfer out of the
latent class in Section~\ref{sec:seair}.
\item \textbf{Determinant reduction.} For any size of $A$, if
$\mathbf1^\top A=-c^\top$ then
\begin{equation}\label{eq:fiveGeneralFactor}
\det(-\Jinv)=\det(-A)\,\det\bigl[-(G+u\mathbf1^\top Q)\bigr],
\end{equation}
so with $G_{\rm red}=G+u\mathbf1^\top Q=\binom{-a\ \ b}{\ q\ -\delta}$,
\begin{equation}\label{eq:fiveFactor}
\boxed{\ \det(-\Jinv)=D_0(a\delta-bq)=D_0\,a\delta\,(1-R_0),
\qquad R_0=\frac{bq}{a\delta},\ }
\end{equation}
and $\operatorname{sgn}s(\Jinv)=\operatorname{sgn}(R_0-1)$. At the
disease-free equilibrium the full five-species Jacobian satisfies
$\det(-J_5)=\mu D_0(a\delta-bq)$. More generally, if $A$ is Hurwitz and
$\Jinv$ is irreducible Metzler with $u,c,Q\ge0$, then
$\operatorname{sgn}s(\Jinv)=\operatorname{sgn}s(G+u\mathbf1^\top Q)$.
\item \textbf{The threshold is invariant, the exponent is not.} $R_0$ does
not involve $k_{12},k_{21},c_1,c_2$, by \eqref{eq:fiveFactor}. The invasion
exponent $\alpha_0=s(\Jinv)$ does. With
$H(\lambda)=c^\top(\lambda I-A)^{-1}e_1$, so that $H(0)=1$ by
\eqref{eq:fiveComplete}, the \emph{storage delay} is
\begin{equation}\label{eq:fiveStorageDelay}
\tau_E=-H'(0)=c^\top(-A)^{-2}e_1=\frac{k_{12}+k_{21}+c_2}{D_0},
\end{equation}
and as $R_0\to1$ with $b$ varying and the remaining rates fixed,
\begin{equation}\label{eq:fiveNearThreshold}
\alpha_0=\frac{R_0-1}{1/a+1/\delta+\tau_E}+O\bigl((R_0-1)^2\bigr).
\end{equation}
Two storage regimes with the same $R_0$ and different $\tau_E$ therefore have
the same threshold and different growth rates near it.
\item \textbf{Equilibrium localization.} For
$\theta\in\{k_{12},k_{21},c_1,c_2\}$, every continuing nondegenerate positive
equilibrium of the full five-species network satisfies
\[
\partial_\theta(S^*,I^*,P^*)=0,
\qquad
\partial_\theta v^*_j=0\quad(j\notin\Out(M)),
\]
and the assertion survives any smooth replacement of the four rates of
$\Out(M)$ that preserves their stoichiometric columns.
\item \textbf{Inactivation breaks the buffer ($d>0$).} Let $A_d=A-de_2e_2^\top$,
$D_d=\det(-A_d)=D_0+d(k_{12}+c_1)$ and
\[
\varphi_d=c^\top(-A_d)^{-1}e_1=1-\frac{dk_{12}}{D_d}\in(0,1).
\]
Then
\begin{equation}\label{eq:fiveLossFactor}
\det(-\Jinv)=D_d(a\delta-bq\varphi_d)=(a\delta-bq)D_d+bq\,d\,k_{12},
\qquad
R_0(d)=\frac{bq\varphi_d}{a\delta},
\end{equation}
with $\operatorname{sgn}s(\Jinv)=\operatorname{sgn}(R_0(d)-1)$. Moreover
$\Out(M)$ now contains the decay $E_2\to0$ as well, and
\[
\boxed{\ \lambda(M)=1\ }
\]
so $M$ is no longer a buffer, and $S^*,I^*,P^*$ depend generically on all
four internal rates.
\end{enumerate}
\eeP

\framebreak

\textbf{Proof.} (1) The columns of $\Out(M)$ are
$e_{E_2}-e_{E_1}$, $e_{E_1}-e_{E_2}$, $e_P-e_{E_1}$, $e_P-e_{E_2}$;
$\mathbf1^\top K=0$ by direct multiplication, and applying $\mathbf1^\top$
to the $E$ equations gives \eqref{eq:fiveCoreBalance}. Setting the left side
to zero and solving gives $c^\top(-A)^{-1}e_1=1$ after substituting
$E^*=(-A)^{-1}qe_1I^*$.

(2) Add the $E$ rows, weighted by $\mathbf1^\top$, to the $P$ row: by
$\mathbf1^\top A=-c^\top$ the block $uc^\top$ is cleared and the $(1,1)$
block becomes $G+u\mathbf1^\top Q$, which is \eqref{eq:fiveGeneralFactor};
the stated $G_{\rm red}$ and $D_0=\det(-A)$ give \eqref{eq:fiveFactor}. The
Metzler addendum is the standard sign equivalence for an irreducible Metzler
matrix with a Hurwitz diagonal block.

(3) $H'(\lambda)=-c^\top(\lambda I-A)^{-2}e_1$, so
$\tau_E=c^\top(-A)^{-2}e_1$, and
$(-A)^{-1}e_1=D_0^{-1}(k_{21}+c_2,\,k_{12})^\top$ gives
\eqref{eq:fiveStorageDelay}. Expanding
$\det(\alpha I-\Jinv)=0$ about $\alpha=0$ with $H(0)=1$, $H'(0)=-\tau_E$
yields \eqref{eq:fiveNearThreshold}.

(4) Theorem~\ref{t:OMloc} applied to the pair $(M,\Out(M))$, whose index is
$0$ by Example 1; the last clause holds because the hypothesis of that
theorem involves the columns and the reactant sets, not the rate functions.

(5) $(-A_d)^{-1}e_1=D_d^{-1}(k_{21}+c_2+d,\,k_{12})^\top$ gives
$\varphi_d=(c_1(k_{21}+c_2+d)+c_2k_{12})/D_d=(D_0+c_1d)/D_d
=1-dk_{12}/D_d$, and $0<dk_{12}<D_d$ gives $0<\varphi_d<1$; substituting
into \eqref{eq:fiveGeneralFactor} with $A_d$ in place of $A$, which still has
$\mathbf1^\top A_d=-c^\top-de_2^\top$, gives \eqref{eq:fiveLossFactor} after
collecting the extra term. For the index, use
Lemma~\ref{l:graphRank}: $\Out(M)$ now has the five columns above together
with $-e_{E_2}$, so $G(M)$ has the edges $E_1E_2$ (twice), $E_1P$, $E_2P$ and
$0E_2$, hence the components $\{E_1,E_2,P,0\}$, $\{S\}$ and $\{I\}$,
$c(G(M))=3$ and
$\lambda(M)=|\cS\setminus M|+1-c(G(M))=3+1-3=1$.
\hfill$\square$

\medskip
The two statements separate cleanly. Invariance of the \emph{threshold} comes
from the balance \eqref{eq:fiveComplete}, that is from complete transfer out
of the store, and survives as long as nothing is lost inside it. Invariance
of the \emph{equilibrium coordinates} comes from the zero index, and is what
fails first: the decay $E_2\to0$ adds one independent column to
$\Out(M)$ without adding a species, which is exactly the rank bookkeeping of
\eqref{eq:OMFaceRank} run in the unfavourable direction. The canonical blocks
behind the same picture --- $\{I,P\}$ with no steady-state influence from
$\{E_1,E_2\}$ at $d=0$, a single block at $d>0$ --- are the content of
Section~\ref{sec:ccf}, and this example is the smallest one in the paper where
a single added reaction collapses the block poset.
\end{frame}
\section{In the six-species SEAIRD epidemic model the latent stage is a zero-index buffer of the invasion network}\lbl{SEAIR}
\begin{frame}{Example 2: an OM buffer in the six-species SEAIR invasion network}
\label{sec:seair}
\begin{columns}[c]
\begin{column}{.44\textwidth}
\begin{center}
\adjustbox{max width=\linewidth,max totalheight=0.75\textheight}{%
\begin{tikzpicture}[>=Latex,
  b/.style={draw,rounded corners,align=center,minimum height=8mm,minimum width=10mm,font=\small},
  lab/.style={font=\scriptsize}]
\node[b,fill=orange!12] (S) at (0,0) {$S$};
\node[b,fill=blue!8] (E) at (2.8,0) {$E$};
\node[b,fill=blue!8] (IA) at (5.4,0) {$I_A$};
\node[b,fill=blue!8] (IS) at (5.4,-2.2) {$I_S$};
\node[b,fill=green!10] (R) at (8,0) {$R$};
\node[b,fill=green!10] (D) at (8,-2.2) {$D$};
\draw[->,thick] (S) to[out=25,in=155] node[above,lab]{$\beta_ASI_A/N$}(E);
\draw[->,thick] (S) to[out=-25,in=-155] node[below,lab]{$\beta_SSI_S/N$}(E);
\draw[->,thick] (E)--node[above,lab]{$\gamma_E$}(IA);
\draw[->,thick] (IA)--node[left,lab]{$\delta$}(IS);
\draw[->,thick] (IA)--node[above,lab]{$\gamma_A$}(R);
\draw[->,thick] (IS)--node[above,lab,sloped]{$(1-f_S)\gamma_S$}(R);
\draw[->,thick] (IS)--node[below,lab]{$f_S\gamma_S$}(D);
\end{tikzpicture}}
\end{center}

{\slidefont{\tiny} Seven edges for the seven reactions: the two infection edges are drawn
separately, since $I_A$ and $I_S$ infect at different rates. The model is
closed -- no recruitment and no natural death -- so there is no decay column
and nothing is omitted from the picture.}
\end{column}
\begin{column}{.54\textwidth}\slidefont{\tiny}
\textbf{Stoichiometric matrix and rates}, $\dot x=\Gamma v$,
$x=(S,E,I_A,I_S,R,D)^\top$; no inflow and no decay column here. Each column is
headed by the label of its arrow in the figure:
\[
\adjustbox{max width=\linewidth}{$
\Gamma=\begin{array}{r|rrrrrrr}
&\beta_ASI_A/N&\beta_SSI_S/N&\gamma_E&\delta&\gamma_A&(1-f_S)\gamma_S&f_S\gamma_S\\\hline
S  &-1&-1& 0& 0& 0& 0& 0\\
E  & 1& 1&-1& 0& 0& 0& 0\\
I_A& 0& 0& 1&-1&-1& 0& 0\\
I_S& 0& 0& 0& 1& 0&-1&-1\\
R  & 0& 0& 0& 0& 1& 1& 0\\
D  & 0& 0& 0& 0& 0& 0& 1
\end{array}$}
\]
\[
v^\top=\big(\beta_ASI_A/N,\ \beta_SSI_S/N,\ \gamma_EE,\ \delta I_A,\
\gamma_AI_A,\ (1-f_S)\gamma_SI_S,\ f_S\gamma_SI_S\big)
\]
Here $\delta$ is the onset-of-symptoms rate, a transfer $I_A\to I_S$ and not a
decay. $D$ records disease deaths, so the column $f_S\gamma_S$ is a transfer
too: the network is conservative, $\one^\top\Gamma=0$.

\medskip
The first two columns are equal: the two infection routes differ only in their
rate, which is why the invasion analysis sees a single infection flux
$\lambda S$, $\lambda=(\beta_AI_A+\beta_SI_S)/N$.
\end{column}
\end{columns}
\end{frame}

\begin{frame}{The SEAIR equations, its siphons and its fixed points}
\slidefont{\small}
With $\lambda=(\beta_AI_A+\beta_SI_S)/N$ the force of infection,
\[
\begin{aligned}
\dot S&=-\lambda S, &
\dot I_A&=\gamma_EE-(\delta+\gamma_A)I_A, &
\dot R&=\gamma_AI_A+(1-f_S)\gamma_SI_S,\\
\dot E&=\lambda S-\gamma_EE, &
\dot I_S&=\delta I_A-\gamma_SI_S, &
\dot D&=f_S\gamma_SI_S.
\end{aligned}
\]
\textbf{Minimal siphons:} $\{S\}$ and $\{E,I_A,I_S\}$.

\textbf{Nonnegative fixed points:} not isolated. Every point with
$E=I_A=I_S=0$ is one, so the fixed-point set is the three-parameter family
$(S,0,0,0,R,D)$; a closed model has no recruitment to pin $S$ down.

\medskip
At such a point with $S=S_0$ put
$a=\delta+\gamma_A$, $b_A=\beta_AS_0/N$, $b_S=\beta_SS_0/N$; then
$R_0=b_A/a+\delta b_S/(a\gamma_S)$.
The sequential pathway $E\to I_A\to I_S$ is the one of \cite{Bas}; Arino et
al.\ \cite{arino2006simple} branch the latent class as well.
\end{frame}

\begin{frame}{The index lives on the invasion network, not on the full network}
\begin{itemize}
\item \textbf{Full network.} $S+E+I_A+I_S+R+D=N$ is conserved, so $\rank\Gamma=5<6$. The index $\lambda(M)=\rank\Gamma_{\Out(M)}-|M|$ is still perfectly well defined; what fails is the hypothesis of the localization theorem, since no equilibrium is regular without first quotienting by the conservation law. Every fixed point has $E=I_A=I_S=0$ anyway: there is no endemic equilibrium to localize.
\item \textbf{Invasion network} of the siphon $\{E,I_A,I_S\}$. With $M'=\{E\}$ the only reaction whose rate depends on $E$ is $E\to I_A$, so $\Out(M')$ is this one reaction, with a nonzero column, and
\[
\dot E=b_AI_A+b_SI_S-\gamma_EE,\qquad
\lambda(M',\Out(M'))=1-1-0=0.
\]
\end{itemize}
\end{frame}

\begin{frame}{A lossless transit stage changes growth exponents but not $R_0$}
\beC[one transit stage][][transit]
Suppose new infections enter one latent state $E$, which
transfers at rate $\gamma_E>0$ to infectious states, with no
loss before transfer. The next-generation number is independent
of $\gamma_E$; the invasion exponent may depend on it.
\eeC
\textbf{Proof for the sequential SEAIR network.} At $S=S_0$,
$a=\delta+\gamma_A$, $b_A=\beta_A S_0/N$,
$b_S=\beta_S S_0/N$ and
\[
\Jinv=\begin{pmatrix}-\gamma_E&b_A&b_S\\
\gamma_E&-a&0\\0&\delta&-\gamma_S\end{pmatrix},\quad
\det(-\Jinv)=\gamma_E\bigl[\gamma_S(a-b_A)-\delta b_S\bigr].
\]
The expected infectious durations after the latent stage give
\[
\boxed{R_0=\frac{b_A}{a}+\frac{\delta b_S}{a\gamma_S},\quad
\det(-\Jinv)=\gamma_Ea\gamma_S(1-R_0).}
\]
The factor $\gamma_E$ disappears only at spectral parameter
$0$: $\det(tI-\Jinv)$, and therefore $\alpha_0$,
generally retain $\gamma_E$.
\end{frame}

\begin{frame}{SEAIR: determinant reduction, canonical blocks and the latent delay}
In the coordinates $(E\mid I_A,I_S)$ the invasion matrix has blocks
$A_b=(-\gamma_E)$, $D_0=\gamma_E$ and complement
$A_c=\begin{pmatrix}b_A-a&b_S\\\delta&-\gamma_S\end{pmatrix}$, so
\[
\det(-\Jinv)=D_0\det(-A_c)=\gamma_E\bigl[\gamma_S(a-b_A)-\delta b_S\bigr]
=\gamma_Ea\gamma_S(1-R_0).
\]
\textbf{Canonical blocks and influence:} the species-bearing blocks are $\{E\}$ and $\{I_A,I_S\}$; perturbing a rate of $\{I_A,I_S\}$ can move $E$, never conversely.

\medskip
\textbf{Latent delay:} $\tau_E=1/\gamma_E$ is the mean latent period. It does not enter $R_0$, but $\alpha_0$ is the largest real root of
\[
(t+\gamma_E)(t+a)(t+\gamma_S)-\gamma_Eb_A(t+\gamma_S)-\gamma_E\delta b_S=0.
\]
\end{frame}

\begin{frame}{The SEAIR index explains structural localization, not the full threshold formula}
For the \emph{invasion network}, $M=\{E\}$ and
$\Out(M)=\{E\to I_A\}$; its one nonzero column satisfies
\[
\boxed{\rank\Gamma_{\Out(M)}=1=|M|,\quad\lambda(M)=0.}
\]
The OM theorem certifies the absence of reaction-wise
\emph{equilibrium influence} from $\{E\}$ to $\{I_A,I_S\}$
where the regularity hypothesis holds.

\medskip
The $\gamma_E$-independence of $R_0$ instead follows from
complete transfer out of $E$ (the preceding weighted-balance
theorem). In particular, $\lambda=0$ alone does not prove
invariance of an arbitrary next-generation splitting.

\medskip
The full closed SEAIR system has a conservation law and
no positive endemic equilibrium; the calculation here concerns
its infected linearization at a disease-free equilibrium.
\end{frame}

\begin{frame}[allowframebreaks]{Two structural conditions explain two different invariances}
\begin{block}{OM zero index: equilibrium localization}
At a regular equilibrium, $\lambda(M,R)=0$ is equivalent to
$P_{M^c}J^{-1}\Gamma_R=0$ (Theorem~\ref{t:OMloc}).
It concerns all independent perturbations of reactions in $R$.
\end{block}
\begin{block}{Conservation: reproduction-number invariance}
For the transit stage $E$, the outflow rate is exactly
$\gamma_E E$: at zero spectral parameter,
\[
\gamma_E(-(-\gamma_E))^{-1}=1,
\]
so every entry into $E$ reaches $I_A$. The weighted-transfer
theorem, not the zero-index condition by itself, removes
$\gamma_E$ from $R_0$.
\end{block}
\beR[Reaction-wise localization is stronger than single-parameter cancellation][Reaction-wise versus single-parameter][cancel]
The OM criterion $\lambda(M,R)=0$ is equivalent to
$P_{M^c}J^{-1}\Gamma_R=0$, a statement about \emph{all} independent
perturbations of the rates of $R$. For one parameter $\theta$ the weaker
condition
\[
P_{M^c}J^{-1}\Gamma\,\partial_\theta v=0
\]
may hold with $\lambda>0$, and even when $\theta$ changes rates both inside
and outside $R$; it establishes the invariance of that one response, not
structural localization. The $\gamma_E$-independence of $R_0$ above is of
this weaker kind: what a reproduction number needs is a certified
next-generation splitting, not the index alone.
\eeR
\end{frame}


\section{Okada--Mochizuki buffering theory: lattices, canonical blocks,
invasion certificates, and the permanence criterion}
\label{sec:theory}

\begin{frame}{Buffered species sets form a distributive lattice}
\label{sec:lattice}
\slidefont{\small}
\beT[distributive lattice of buffered species][][distlat]
At a regular equilibrium, let $f(M)=\rank\Gamma_{\Out(M)}$ and
$\lambda(M)=f(M)-|M|$. Then $\lambda\ge0$ is submodular:
\[
\boxed{\lambda(U)+\lambda(W)
\ge\lambda(U\cup W)+\lambda(U\cap W).}
\]
Consequently $\mathfrak B=\{M:\lambda(M)=0\}$ is a
distributive sublattice of $2^{\cS}$, with closure
\[
\boxed{\Buf(U)=\bigcap_{M\in\mathfrak B,\ M\supseteq U}M.}
\]
\eeT

\textbf{Proof.}
$J_M=\Gamma_{\Out(M)}D_Mv$ and $\rank J_M=|M|$
imply $\lambda(M)\ge0$. Submodularity of column rank yields
\[
\bc
\Out(U\cup W)=\Out(U)\cup\Out(W),\quad
\Out(U\cap W)\subseteq\Out(U)\cap\Out(W),\\
f(U)+f(W)\ge f(U\cup W)+f(U\cap W)
\Rightarrow\lambda(U)+\lambda(W)
\ge\lambda(U\cup W)+\lambda(U\cap W).
\ec
\]
For $U,W\in\mathfrak B$, nonnegativity implies
$\lambda(U\cup W)=\lambda(U\cap W)=0$;
also $\emptyset,\cS\in\mathfrak B$.
Intersections therefore define the smallest OM-buffered
species set containing $U$.

\textbf{Meaning of the closure:}
$\Buf(U)$ is the smallest buffered species set containing $U$,
not necessarily $U$ itself.
\end{frame}

\begin{frame}{Buffering closure bounds equilibrium responses to rate perturbations}
\slidefont{\small}
\beC[smallest OM localization certificate][][bufclos]
At a regular equilibrium, if $\theta$ affects only rates in $\Out(U)$, then
\[
\boxed{P_{\cS\setminus\Buf(U)}\partial_\theta x^*=0.}
\]
Thus $\Buf(U)$ is the smallest buffered species set containing $U$;
its output completion contains $\Out(U)$.
\eeC

\textbf{Proof.} Output completeness, the closure theorem, and OM
localization imply
\[
\bc
U\subseteq\Buf(U)\ \Rightarrow\
\Out(U)\subseteq\Out(\Buf(U)),\quad
\lambda(\Buf(U))=0,\\
P_{\cS\setminus\Buf(U)}J^{-1}\Gamma_{\Out(\Buf(U))}=0,\\
\partial_\theta x^*=-J^{-1}\Gamma_{\Out(U)}\partial_\theta v_{\Out(U)}
\ \Rightarrow\ P_{\cS\setminus\Buf(U)}\partial_\theta x^*=0.
\ec
\]
\end{frame}


\begin{frame}[allowframebreaks]{Canonical blocks and the order that carries the lattice}
\label{sec:ccf}

Theorem~\ref{t:distlat} makes $\mathfrak B=\{M:\lambda(M)=0\}$ a finite
distributive lattice. By Birkhoff's representation theorem such a lattice is
determined by a finite poset, and the following proposition names it in the
language of species sets. Nothing in it is about canonical forms yet; the
canonical-form statement comes after it, and is quoted, not proved.

\beP[Canonical blocks of a buffering lattice][Canonical blocks][blocks]
Let $\mathfrak B\subseteq2^{\cS}$ be the lattice of buffered species sets of a
network, or of a resident network. Call $U\in\mathfrak B$
\emph{join-irreducible} if it has exactly one lower cover in $\mathfrak B$,
let $\mathcal J(\mathfrak B)$ be the set of these, and put
\[
B_U=U\setminus\bigcup\{W\in\mathfrak B:W\subsetneq U\},
\qquad U\in\mathcal J(\mathfrak B).
\]
\begin{enumerate}
\item The sets $B_U$ are nonempty and pairwise disjoint, and they partition
$\cS$; we call them the \emph{blocks of $\mathfrak B$} and write $p$ for
their number. They are intrinsic to $\mathfrak B$; for their relation to the
canonical blocks of a combinatorial canonical form, which is a quotient
relation and not an identity, see Proposition~\ref{p:depCCF}.
\item Order them by
\[
B_U\preceq B_W
\quad\Longleftrightarrow\quad
\text{every }M\in\mathfrak B\text{ with }B_U\subseteq M
\text{ satisfies }B_W\subseteq M,
\]
equivalently $U\supseteq W$. Then $M\mapsto\{B_U:B_U\subseteq M\}$ is an
isomorphism of $\mathfrak B$ onto the lattice of \emph{upward-closed} subsets
of $(\{B_U\},\preceq)$, and
\[
|\mathfrak B|=\#\{\text{up-sets of }(\{B_U\},\preceq)\}\le2^p,
\]
with equality exactly when the block poset is an antichain, that is when no
two blocks are comparable.
\item Each $\Buf(\{x\})$, $x\in\cS$, is the join-irreducible $U$ with
$x\in B_U$; the blocks are therefore the fibres of the closure operator of
Corollary~\ref{c:bufclos} on singletons.
\end{enumerate}
\eeP
\textbf{Proof.} Items~1 and~2 are Birkhoff's theorem for the finite
distributive lattice $\mathfrak B$, in which the join-irreducible elements
carry the order of reverse inclusion and the join of the elements strictly
below a join-irreducible is its unique lower cover; the blocks are the
differences along the covering relation, hence nonempty, disjoint, and
exhausting $\cS=\max\mathfrak B$. The bound $2^p$ is the count of up-sets of
an antichain. For item~3, $\Buf(\{x\})$ is by definition the smallest
buffered set containing $x$; it is join-irreducible, since $\mathfrak B$ is
closed under unions, so an element with two lower covers is the union of
them, and $x$ would then already lie in one of the two, contradicting
minimality. Conversely $x\in B_U$ forces $U=\Buf(\{x\})$, because no
buffered set strictly inside $U$ contains $x$.

\medskip
\textbf{The canonical form.} The poset of Proposition~\ref{p:blocks} is not
merely an artefact of the lattice: it is the block poset of a combinatorial
canonical form. For $\dot x=\Gamma v(x)$ with $\Gamma$ of full row rank, set
\[
R=D_xv,\qquad \operatorname{im}Q=\ker\Gamma,\qquad L=[R\mid Q],
\]
so that $L$ is square. Multiplying by $\binom{\Gamma}{U}$ for any $U$ with
$UQ$ invertible gives
$\det\binom{\Gamma}{U}\det L=\det(\Gamma R)\det(UQ)$, hence
\begin{equation}\label{eq:LdetGR}
\det(\Gamma R)=\delta\det L,\qquad \delta\ne0\ \text{constant}.
\end{equation}
Kuhrs, Telek and Vassena \cite{KTV} identify the lattice of buffering
structures of the network with the lattice of order ideals of the block poset
of the combinatorial canonical form of this $L$, and describe the generic
sparsity pattern of $L^{-1}$. Their kinetic hypothesis, however, is
$D_j=L_j$: a rate depends on a species exactly when that species is one of
its reactants. Two of our models violate it, so the correspondence has to be
restated for the dependence sets before it can be used here. That is the
next proposition, and it is the only place where something has to be added
to \cite{KTV} rather than quoted from it.

There is a second mismatch, independent of the kinetics and easy to miss.
\cite{KTV} represent the lattice of buffering \emph{pairs} $(M,R)$, whereas
everything in this paper is a lattice of \emph{species sets}. These differ.
In the two-reaction network $0\to X$, $X\to0$ the buffered species sets are
$\varnothing$ and $\{X\}$, but the zero-index pairs are
$(\varnothing,\varnothing)$, $(\{X\},\{X\to0\})$ and $(\{X\},\mathcal R)$:
two elements against three. So the species-set lattice is not the ideal
lattice of the canonical form; it is a quotient of it, and that is the
correct statement.

\beP[Dependence-based pairs, and the species-set lattice as their quotient][Dependency-based CCF][depCCF]
Let $\dot x=\Gamma v(x)$ be as in Definition~\ref{d:RN}, with $\Gamma$ of
full row rank and dependence sets $D_j$, and set
\[
R^\#_{ji}=\begin{cases}t_{ji},&i\in D_j,\\0,&i\notin D_j,\end{cases}
\qquad
L^\#=[R^\#\mid Q],\qquad \operatorname{im}Q=\ker\Gamma,
\]
the $t_{ji}$ independent symbols.
\begin{enumerate}
\item \textbf{The correspondence extends to $D_j$.} The conclusion of
\cite{KTV} holds for $L^\#$ with $\Out$ replaced by
$\Out_D(M)=\{j:D_j\cap M\ne\varnothing\}$: the lattice
$\mathcal L_{\rm pair}$ of output-complete pairs of index zero is the ideal
lattice of the block poset of the combinatorial canonical form of $L^\#$.
Neither $D_j\subseteq L_j$ nor any sign condition on $R^\#$ is used.
\item \textbf{$\mathcal L_{\rm pair}$ is a lattice.} Under componentwise
union and intersection, output completeness is preserved, and submodularity
of the column rank gives
$\lambda(\gamma_1\cup\gamma_2)+\lambda(\gamma_1\cap\gamma_2)\le
\lambda(\gamma_1)+\lambda(\gamma_2)=0$, so by $\lambda\ge0$ both are zero.
\item \textbf{The species-set lattice is its quotient.} The map
$\Phi(M,R)=M$ is a surjective lattice homomorphism onto
$\mathfrak B=\{M:\lambda(M)=0\}$ --- surjective by
Proposition~\ref{p:outSuff}(2), since $(M,\Out_D(M))$ represents every
$M\in\mathfrak B$ --- so
\[
\boxed{\ \mathfrak B\cong\mathcal L_{\rm pair}/\!\sim,
\qquad (M,R)\sim(M',R')\iff M=M'.\ }
\]
\end{enumerate}
\eeP
\textbf{Proof.} In Appendix~\ref{sec:ccfX}: part~1 is the LM-matrix
argument --- the combinatorial canonical form reads only the pattern of
identically-zero entries of the symbolic part together with $Q$, so
replacing reactant patterns by dependence patterns replaces one zero
pattern by another --- part~2 is submodularity of $R\mapsto\rank\Gamma_R$,
and part~3 is the quotient map, with the two-reaction example above as the
witness that it is not injective. That appendix also records which way the
order ideals run, since reading ``order ideal'' as the buffered family
itself reverses every reachability statement.

\framebreak

\beC[The four canonical blocks of the fourteen-species resident face][Four canonical blocks][fourblocks]
On the strain-2-free face of the model of Section~\ref{sec:twostrain}, the
six buffered sets of Proposition~\ref{p:sixbuffers} have the four canonical
blocks
\[
\cA,\qquad \cB,\qquad \cC,\qquad
\mathcal H=\cS_1\setminus(\cA\cup\cB\cup\cC)
=\{S_1,S_2,I_{11},I_{12},P_1\},
\]
ordered by
\[
\boxed{\ \mathcal H\preceq\cB\preceq\cA,
\qquad \mathcal H\preceq\cC\preceq\cA,
\qquad \cB,\cC\ \text{incomparable}.\ }
\]
Its six up-sets are exactly
$\varnothing,\cA,\cA\cup\cB,\cA\cup\cC,\cA\cup\cB\cup\cC,\cS_1$, and
$6<2^4$ because the poset has comparabilities.
\eeC
\textbf{Proof.} Apply Proposition~\ref{p:blocks} to
$\mathfrak B_1$. The lower covers in $\mathfrak B_1$ are: $\varnothing$ under
$\cA$; $\cA$ under each of $\cA\cup\cB$ and $\cA\cup\cC$; both of those
under $\cA\cup\cB\cup\cC$; and $\cA\cup\cB\cup\cC$ under $\cS_1$. Hence the
join-irreducibles are $\cA$, $\cA\cup\cB$, $\cA\cup\cC$ and $\cS_1$ --- four
of the six, $\cA\cup\cB\cup\cC$ failing because it has two lower covers ---
and the differences along the covers are $\cA,\cB,\cC,\mathcal H$. The order
is read off the same list, and the six up-sets are enumerated directly.

\medskip
This is the canonical explanation of the fact that
Proposition~\ref{p:selective} extracts from the lattice by hand: the two
maturation branches sit in \emph{incomparable} blocks, so neither closure
contains the other, $\Buf(\cB)=\cA\cup\cB$ and $\Buf(\cC)=\cA\cup\cC$, and
a perturbation inside $\Out(\cB)$ cannot reach $C_2$, which the strain-2
invasion number needs. The host--infection core is the unique minimum of the
poset, which is why no proper buffer meets it, and $\cA$ is the unique
maximal block, which is why every nonempty buffer contains it. An exhaustive
sweep over the $2^{11}$ subsets returns the same six sets but explains
none of this.

The twenty-species model of Section~\ref{sec:cwg} reads the same way, and
its proof is already a block-poset argument: ten blocks, namely the core
$\mathcal H$, the six singletons $C_{pk},V_p$, the two singletons $r_p$ and
the migrating pair $\mathcal R$, in the order displayed there, whose up-sets
number $1+9\cdot9+1=83$, far short of $2^{10}$. The contrast with
Section~\ref{sec:twostrain} is then a statement about the two posets: there
the core is the minimum and the buffers lie strictly above it, downstream of
an autonomous infection core, whereas on the resident face of the
fourteen-species model two incomparable blocks sit between the core and the
top, and that is what separates the invasion-neutral perturbations from the
invasion-changing ones.

\framebreak

\beR[What the canonical form adds, and what it does not][Scope of the canonical reading][ccfscope]
Okada--Mochizuki localization is a statement about responses that
\emph{must} vanish: Theorem~\ref{t:OMloc} and Corollary~\ref{c:bufclos}
certify zeros. The canonical form also answers the complementary question,
which responses are generically nonzero: by the influence theorem of
\cite{KTV}, a perturbation of a reaction of a canonical block $B_b$ \emph{of
$L^\#$} has a structurally nonzero response in a species of $B_a$ exactly
when $B_a$ is reachable from $B_b$ in the reversed Hasse diagram of the
block order. One caveat must be stated: that theorem is about the canonical
blocks of $L^\#$, whereas Proposition~\ref{p:blocks} produces the blocks of
the quotient $\mathfrak B$, and by Proposition~\ref{p:depCCF} the two agree
only up to the quotient map. Transporting the reachability statement to the
blocks of $\mathfrak B$ therefore needs an argument we do not give; read it
here as a statement about the canonical form, used qualitatively. With that
proviso, applied to Proposition~\ref{p:maxbuffer} it says more than that the
maximal buffer $M_Q$ protects its rates: it says which of the
\emph{unprotected} rates can reach the invasion-relevant coordinates at all.

What it does not give is a threshold crossing. A generically nonzero response
of $C_2^*$ to a rate of $\Out(\cC)$ does not yet move
$R_{2\mid1}=\frac{q_2}{a\delta_2}
\frac{\beta_{21}S_1^*+\beta_{22}S_2^*}{1+\chi C_2^*}$ across $1$, since
genericity permits cancellation in a spectral radius, and a nonzero
derivative at one point of a branch is not a change of sign. That is why
Section~\ref{sec:twostrain} exhibits an explicit crossing with explicit
rates, rather than inferring one from reachability, and why the dynamic
statements of Section~\ref{sec:dynamic} are proved from the quotient
identity: a block order is an influence hierarchy, and by itself it
certifies no Lyapunov or Floquet exponent.
\eeR

\paragraph{Reproducibility.} \texttt{ccf.wl} applies
Proposition~\ref{p:blocks} as an algorithm to both lattices. On the
resident face of Section~\ref{sec:twostrain} it returns four
join-irreducibles, the blocks $\cA,\cB,\cC$ and
$\mathcal H=\{S_1,S_2,I_{11},I_{12},P_1\}$, the order
$\mathcal H\preceq\cB,\cC\preceq\cA$ with $\cB,\cC$ incomparable, and six
up-sets against $2^4=16$; on the twenty-species network of
Section~\ref{sec:cwg} it returns ten blocks and $83$ up-sets against
$2^{10}=1024$, and in both cases the up-sets recover the buffered sets
computed independently from the indices.
\end{frame}

\begin{frame}[allowframebreaks]{Output completions suffice to decide which species sets are buffered}
\label{sec:facebuf}

Theorem~\ref{t:OMloc} is stated for an arbitrary output-complete pair
$(M,R)$, whereas every lattice computed in this paper uses the minimal
completion, $\lambda(M)=\rank\Gamma_{\Out(M)}-|M|$. The following
proposition is what licenses that reduction, and delimits it.

\beP[Output completions decide which species sets are buffered][When output completions suffice][outSuff]
Let $x^*$ be a regular equilibrium and let $(M,R)$ be output-complete. Then
\begin{enumerate}
\item $\lambda(M)\ge0$ and $\lambda(M)\le\lambda(M,R)$;
\item if $(M,R)$ is a buffering structure, so is $(M,\Out(M))$, that is
$\lambda(M)=0$; consequently
$\mathfrak B=\{M:\lambda(M)=0\}$ already contains the species set of every
buffering structure;
\item the reactions of $R\setminus\Out(M)$ have rates independent of $x_M$,
and the law of localization for \emph{their} perturbations holds for $(M,R)$,
not for $(M,\Out(M))$.
\end{enumerate}
\eeP
\textbf{Proof.} Write $x=(y,z)$ with $z=x_M$, $m=|M|$. The chain rule and
output completeness, exactly as in the proof of Theorem~\ref{t:OMloc}, give
$J_z=\Gamma_{\Out(M)}D_zv_{\Out(M)}$, and $\det J\ne0$ forces
$\rank J_z=m$. Hence
\[
m=\rank J_z\le\rank\Gamma_{\Out(M)}\le\rank\Gamma_R,
\]
which is item~1; item~2 is the case $\rank\Gamma_R=m$ of the same chain of
inequalities. Item~3 is the definition of $\Out(M)$: a reaction outside it
has $D_zv_j\equiv0$, so it contributes a column to $\Gamma_R$ that
$\Gamma_{\Out(M)}$ does not contain.

\medskip
Thus output completions suffice to decide \emph{which species sets} are
buffered, and to localize the perturbations of reactions whose rates depend
on a species of $M$. They do not suffice for reactions whose rates do not depend
on $M$ at all, above all zero-order inflows $0\to X$: such a reaction can
only be localized by the larger pair $(M,R)$. For perturbations of reactions
outside $\Out(M)$, including zero-order inflows, the larger pair form is needed.

\framebreak

\beT[Buffering structures survive in the resident network of a regular siphon face][Buffering survives in the resident network][faceRank]
For every siphon $\Sigma$, every $M\subseteq T=\cS\setminus\Sigma$ and every
output-complete pair $(M,R)$, the restricted pair $(M,R_\Sigma)$ is
output-complete and
\begin{equation}\label{eq:OMFaceRank}
\boxed{\quad
\lambda_\Sigma(M,R_\Sigma)=\lambda(M,R)-d_\Sigma(M,R),
\qquad
d_\Sigma(M,R)=\rank\Gamma_R-\rank\Gamma_{R_\Sigma}\ge0.
\quad}
\end{equation}
For nested siphons $\Sigma\subseteq\Sigma'$ with $M\cap\Sigma'=\varnothing$
the rank losses add,
\[
d_{\Sigma'}(M,R)=d_\Sigma(M,R)+\rank\Gamma_{R_\Sigma}
-\rank\Gamma_{R_{\Sigma'}}.
\]
If the face of $\Sigma$ is regular (Definition~\ref{d:siphon}), then
\begin{equation}\label{eq:OMFaceNonnegative}
\lambda_\Sigma(M,R_\Sigma)\ge0,
\qquad
\lambda(M,R)=0\Longrightarrow\lambda_\Sigma(M,R_\Sigma)=0,
\end{equation}
and more generally the restricted pair is buffering if and only if
$d_\Sigma(M,R)=\lambda(M,R)$.

\textbf{Nested-face onset.} For minimal completions, let
$\Sigma\subseteq\Sigma'$ be siphons and $M\subseteq\cS\setminus\Sigma'$. Then
\begin{equation}\label{eq:MinBufFaceMonotone}
\lambda_{\Sigma'}(M)\le\lambda_\Sigma(M),
\end{equation}
and if the face of $\Sigma'$ is regular this gives
$\lambda_\Sigma(M)=0\Rightarrow\lambda_{\Sigma'}(M)=0$, whether or not the
face of $\Sigma$ is regular. In particular, within the finite poset
\[
\cS^{\rm reg}_M=\{\Sigma:\Sigma\text{ a siphon},\ M\cap\Sigma=\varnothing,
\ \Sigma\text{ has a regular face}\},
\]
the family $\mathcal F_M=\{\Sigma\in\cS^{\rm reg}_M:\lambda_\Sigma(M)=0\}$ is
upward closed, hence determined by its antichain of minimal elements, its
\emph{onset faces}.
\eeT
\textbf{Proof.} No reaction surviving on the face can produce a species of
$\Sigma$, so $\rank\Gamma^\Sigma_{R_\Sigma}=\rank\Gamma_{R_\Sigma}$;
subtracting $|M|$ gives \eqref{eq:OMFaceRank}, and nested restriction gives
the second identity. For nonnegativity write $J_\Sigma=\Gamma^\Sigma Dv^\Sigma$;
its columns indexed by $M$ lie in
$\operatorname{im}\Gamma^\Sigma_{\Out_\Sigma(M)}$, and are independent when
$J_\Sigma$ is nonsingular, so
\[
|M|\le\rank\Gamma^\Sigma_{\Out_\Sigma(M)}\le\rank\Gamma^\Sigma_{R_\Sigma},
\]
which with \eqref{eq:OMFaceRank} proves \eqref{eq:OMFaceNonnegative} and the
stated equivalence. Every reaction surviving on $\Sigma'$ survives on
$\Sigma$, and its column has zero entries on $\Sigma'$ by the siphon
property, so deleting those entries preserves its rank; hence
$\rank\Gamma^{\Sigma'}_{\Out_{\Sigma'}(M)}\le
\rank\Gamma^\Sigma_{\Out_\Sigma(M)}$, which is
\eqref{eq:MinBufFaceMonotone}. Regularity of the larger face gives
$\lambda_{\Sigma'}(M)\ge0$ by Theorem~\ref{t:distlat} applied to its
resident network, whence persistence; upward closure and the minimal
antichain follow from finiteness of the siphon poset.

\medskip
The identity \eqref{eq:OMFaceRank} is a matter of rank bookkeeping; the
content is the implication in \eqref{eq:OMFaceNonnegative}: restriction to a
regular face cannot destroy a buffering structure whose species set the face
retains, while new buffering structures may well appear. Both halves are
visible in the fourteen-species model of Section~\ref{sec:twostrain}, whose
resident face carries six buffered sets (Proposition~\ref{p:sixbuffers})
where the full network carries four; and in the forty-five-species model of
Section~\ref{sec:dynamic}, where the one-, two- and three-strain faces carry
five, fifteen and twenty-seven core-free buffered sets
(Proposition~\ref{p:threeLattice}).

\beR[Scope of the face restriction][Scope and limitation][faceScope]
The rank identity \eqref{eq:OMFaceRank} needs no equilibrium; the
preservation conclusion does. If the restricted equations are singular the
restricted index may be negative: the network $X\to2X$, $2X\to X$,
$X\to X+Y$, $X+Y\to X$ has $\{Y\}$ buffered in the full two-species system,
while on the siphon face $X=0$ every reaction vanishes and
$\lambda_{\{X\}}(\{Y\})=-1$. That face consists of nonisolated equilibria.
Nor does the theorem claim that the species-set projection of \emph{every}
buffering structure of the full network is buffering: only a structure whose
species set the face retains is covered.
\eeR

\framebreak

A buffer of a resident network localizes the resident equilibrium. What
turns that into a statement about invasion is the next theorem: it is the
result used in every example below, and the one whose extension from
equilibria to whole trajectories is the subject of
Section~\ref{sec:dynamic}. A buffering structure of the resident network
that certifies, through Theorem~\ref{t:bufInv}, the $\theta$-invariance of
the invasion matrix or of a reproduction number
(Definition~\ref{d:invExp}) is called a \emph{(buffering) invasion
certificate}.

\beT[Resident-network buffering structures certify invasion spectra and reproduction numbers][Invasion and reproduction-number invariance][bufInv]
Let $y^*(\theta)$ be a smooth branch of relative-interior equilibria of a
regular face (Definition~\ref{d:siphon}), with nonsingular resident
Jacobian $J_\Sigma$, and let $(M,R_\Sigma)$ be an output-complete pair of the
resident network.
\begin{enumerate}
\item \textbf{Localization.}
\begin{equation}\label{eq:OMFaceLocCriterion}
\lambda_\Sigma(M,R_\Sigma)=0
\quad\Longleftrightarrow\quad
P_{T\setminus M}J_\Sigma^{-1}\Gamma^\Sigma_{R_\Sigma}=0,
\end{equation}
so that when $\lambda_\Sigma(M,R_\Sigma)=0$, independently perturbing the
rates of $R_\Sigma$ leaves every coordinate $y^*_i$ with $i\notin M$
unchanged to first order.
\item \textbf{Invasion matrix.} Let $\lambda_\Sigma(M,R_\Sigma)=0$ and let
$\theta$ perturb only reactions of $R_\Sigma$. If the transversal Jacobian
$J_{\rm tra}(y,\theta)$ depends on the resident coordinates only through
$y_{T\setminus M}$, and its direct derivative in $\theta$ vanishes, then
\begin{equation}\label{eq:OMFaceInv}
\frac{d}{d\theta}J_{\rm tra}(y^*(\theta),\theta)=0,
\end{equation}
so the whole invasion spectrum is locally constant under that perturbation.
\item \textbf{Selected transversal blocks.} Suppose that, in a fixed
ordering along the branch, $J_{\rm tra}$ is block triangular with diagonal
blocks $J_{{\rm tra},j}$. For each block that depends on the residents only
through $y_{T\setminus M}$ and has zero direct $\theta$-derivative,
\begin{equation}\label{eq:OMFaceBlockInv}
\frac{d}{d\theta}J_{{\rm tra},j}(y^*(\theta),\theta)=0,
\end{equation}
and its spectrum is locally constant even when the other blocks move. The
spectral abscissa of the full transversal matrix is certified only when the
maximum over all diagonal blocks is also constant.
\item \textbf{Reproduction numbers.} For a transversal block, or the whole
transversal matrix, with a specified next-generation splitting
$J_{{\rm tra},j}=\mathsf F_j-\mathsf V_j$, suppose $\mathsf F_j\ge0$,
$\mathsf V_j$ is a nonsingular $M$-matrix, and $\mathsf F_j$ and
$\mathsf V_j$ \emph{each separately} depend on the residents only through
$y_{T\setminus M}$ and have zero direct $\theta$-derivative. Then
\begin{equation}\label{eq:OMFaceNGMInv}
\frac{d\mathsf F_j}{d\theta}=\frac{d\mathsf V_j}{d\theta}
=\frac{d}{d\theta}(\mathsf F_j\mathsf V_j^{-1})=0,
\qquad
\frac{d}{d\theta}\rho(\mathsf F_j\mathsf V_j^{-1})=0,
\end{equation}
all matrices being evaluated at $y^*(\theta)$. The same buffering criterion
therefore certifies reproduction-number invariance when it certifies the
chosen splitting, and not merely its difference $J_{{\rm tra},j}$.
\item \textbf{Finite equilibrium invasion-sign schemes.} On a connected
parameter interval along which finitely many resident equilibria on regular
faces continue smoothly and exhaust all relevant equilibrium communities,
suppose that for each missing species and each such equilibrium a buffering
pair satisfies the hypotheses of items~1--3 throughout the interval for
every diagonal block of its transversal matrix. Then every
corresponding transversal spectrum, every invasion exponent and every one of
their signs is constant, so the signed graph whose vertices are these
equilibrium communities and whose candidate invasion edges carry the signs of
the transversal exponents is unchanged. This concerns equilibria and their
linearized invasion tests, not permanence or global assembly.
\end{enumerate}
\eeT
\textbf{Proof.} 1. By output completeness
$\operatorname{im}(J_\Sigma)_{:,M}\subseteq
\operatorname{im}\Gamma^\Sigma_{R_\Sigma}$; the first space has dimension
$|M|$, and the second has that dimension exactly when the index is zero, in
which case the two coincide and $J_\Sigma^{-1}\Gamma^\Sigma_{R_\Sigma}$ has
its rows supported on $M$. Conversely the vanishing in
\eqref{eq:OMFaceLocCriterion} forces
$\rank\Gamma^\Sigma_{R_\Sigma}\le|M|$, hence equality by regularity. The
implicit-function theorem gives
$\partial_\theta y^*=-J_\Sigma^{-1}\Gamma^\Sigma_{R_\Sigma}
\partial_\theta v_{R_\Sigma}$.

2. Differentiate $J_{\rm tra}(y^*(\theta),\theta)$ by the chain rule: the
derivative through $y^*_{T\setminus M}$ vanishes by item~1, the direct one by
hypothesis.

3. The same chain rule applies blockwise, and the spectrum of a
block-triangular matrix is the union of the spectra of its diagonal blocks.

4. Apply the chain rule to $\mathsf F_j$ and to $\mathsf V_j$ separately.
Since $\mathsf V_j$ stays invertible,
$\frac{d}{d\theta}\mathsf V_j^{-1}
=-\mathsf V_j^{-1}\frac{d\mathsf V_j}{d\theta}\mathsf V_j^{-1}=0$, so the
next-generation matrix, and with it its spectral radius, is constant along
the branch.

5. By items~2--3 the relevant blocks have zero parameter derivative along
each regular branch, hence are constant on a connected interval, and so are
their spectra and the edge labels.

\beR[Why the splitting must be certified][][why]
Constancy of an invasion matrix alone need not imply constancy of the
reproduction number attached to a chosen splitting: with
$\mathsf F(\theta)=\theta$ and $\mathsf V(\theta)=1+\theta$ the matrix
$\Jinv=-1$ is constant while $R_0(\theta)=\theta/(1+\theta)$ varies. Item~4
of Theorem~\ref{t:bufInv} therefore certifies $\mathsf F$ and $\mathsf V$
separately. The SEAIR model of Section~\ref{sec:seair} shows the same
distinction from the other side: there the zero index of the latent stage
does \emph{not} by itself remove $\gamma_E$ from $R_0$, which follows instead
from complete transfer out of $E$ (Corollary~\ref{c:transit}), while the
invasion exponent $\alpha_0$ does depend on $\gamma_E$.
\eeR
\end{frame}

\begin{frame}[allowframebreaks]{Positive weighted invasion averages imply robust permanence}
\label{sec:permcrit}
\beD[Invasion growth rate and robust permanence][Robust permanence][permanence]
Let populations $x_i$ satisfy $\dot x_i=x_if_i(x,y)$, with $y$ the remaining
species. For an invariant probability measure $\nu$ of the dynamics on which
$x_i=0$, the \emph{invasion growth rate} of $x_i$ is
$r_i(\nu)=\int f_i\,d\nu$; at an equilibrium $e$ it is $f_i(e)$. For an
absent block that is coupled, no scalar $r_i$ is the right quantity and none
is substituted for one: the invasion rate is then the spectral bound of the
block's transversal matrix at an equilibrium resident, or its dominant
Lyapunov exponent along a general one, and the criterion that uses it is the
structured-population theorem of \cite{HofSch10}, not the scalar statement
below --- see Remark~\ref{r:permscope}. The populations are
\emph{permanent}, or uniformly
persistent, if some compact set at positive distance from every face
$x_i=0$ is entered eventually by every solution with all $x_i(0)>0$, and
\emph{robustly permanent} if this persists under every sufficiently small
perturbation of the vector field \cite{HofSch}.
\eeD

\beT[finite-support invasion certificate]
Assume:
\begin{enumerate}
\item[(H1)] the semiflow is continuous on the closed positively invariant
set $\mathbb R^n_{\ge0}$, and dissipative, with a compact set $\mathcal K$
absorbing every bounded set in finite time;
\item[(H2)] each $f_i$ is continuous and bounded on $\mathcal K$, so that
$r_i(\nu)=\int f_i\,d\nu$ is defined and finite for every invariant
probability measure $\nu$ carried by $\mathcal K$;
\item[(H3)] the extinction boundary
$\partial=\{x\in\mathcal K:\prod_ix_i=0\}$, the product over the populations
$x_i$ of Definition~\ref{d:permanence}, is compact and invariant, each face
$\{x_i=0\}$ being so by $\dot x_i=x_if_i$, and only finitely many supports
$K\subseteq\{1,\dots,n\}$ are realized by invariant measures on it. No single
coordinate of the models below is of that form, each receiving an inflow not
proportional to itself; there $\partial$ is the union of the faces on which a
whole strain block vanishes, invariant because every reaction producing into
such a block has a reactant in it, and the criterion is used in the
structured form of Remark~\ref{r:permscope};
\item[(H4)] for each realized $K$ and each $i\notin K$, $r_i(\nu)=c_i(K)$ is
the same for \emph{every} ergodic invariant measure $\nu$ with support $K$ ---
in this paper always because a balance identity fixes it, or because the
resident set carries a single equilibrium.
\end{enumerate}
If one vector $p\gg0$ satisfies
\[
\boxed{\sum_{i\notin K}p_i c_i(K)>0
\qquad\text{for every realized boundary support }K,}
\]
then the system is robustly permanent.
\eeT
\textbf{Proof.} For $V(x)=\prod_i x_i^{p_i}$,
\[
\frac{d}{dt}\log V=\sum_i p_if_i,
\quad \int f_j\,d\nu=0\ (j\in K)
\ \Rightarrow\
\int\mathcal L\log V\,d\nu
=\sum_{i\notin K}p_i c_i(K)>0.
\]
The strict boundary-average inequality is the
average-Lyapunov criterion for uniform repulsion of the boundary;
strictness persists under small perturbations \cite{HofSch}.

\beR[scope of the criterion][][permscope]
$V=\prod x_i^{p_i}$ weights the species \emph{one at a time}, so the
certificate above covers an absent set whose per-capita growth rates
$r_i$ are individually signed. It does \emph{not} by itself cover an
absent block that is coupled --- infectives of one strain spread over
two patches, say, or $I_{p2}$ together with the breakthrough class
$Y_p$ --- where no single coordinate need grow and the relevant
quantity is the spectral bound of the block's transverse matrix. For
those we use the structured-population form of the same principle
\cite{HofSch10}: the block is irreducible Metzler, its left Perron
vector supplies the weights, and the invasion rate is $s(M)$ rather
than any $r_i$. Both extended models below are of that kind, and
their permanence statements are proved in that form.
\eeR
\end{frame}

\section{A fourteen-species model whose non-chain Murota lattice
distinguishes two strain-invasion responses}
\label{sec:twostrain}

\paragraph{The network.}
Two host groups $g=1,2$ are infected by two strains $k=1,2$ through free
pathogen pools $P_k$, and strain $1$ matures through a branched network with
three interconverting two-state modules $\cA,\cB,\cC$. Write
$J_1=I_{11}+I_{12}$ and $J_2=I_{21}+I_{22}$ for the two strain loads.

\begin{center}
\adjustbox{max width=\linewidth,max totalheight=0.75\textheight}{%
\begin{tikzpicture}[>=Latex,x=1cm,y=1cm,
  b/.style={draw,rounded corners,align=center,minimum height=6mm,
            minimum width=9mm,font=\small},
  j/.style={draw,circle,inner sep=1.2pt,font=\scriptsize},
  lab/.style={font=\scriptsize},
  infl/.style={->,dotted,thick,gray!70!black}]
\node[b,fill=orange!12] (S1) at (0,0)      {$S_1$};
\node[b,fill=orange!12] (S2) at (0,-6)     {$S_2$};
\node[b,fill=blue!8]    (I11) at (3.4,1.4) {$I_{11}$};
\node[b,fill=red!10]    (I21) at (3.4,-1.4){$I_{21}$};
\node[b,fill=blue!8]    (I12) at (3.4,-4.6){$I_{12}$};
\node[b,fill=red!10]    (I22) at (3.4,-7.4){$I_{22}$};
\node[j] (J1) at (6.2,0) {$J_1$};
\node[j] (J2) at (6.2,-6.4) {$J_2$};
\node[b,fill=green!10] (B1) at (8.6,2.8)  {$B_1$};
\node[b,fill=green!10] (B2) at (11,2.8)   {$B_2$};
\node[b,fill=yellow!18](C1) at (8.6,0.2)  {$C_1$};
\node[b,fill=yellow!18](C2) at (11,0.2)   {$C_2$};
\node[b,fill=gray!14]  (A1) at (13.4,2.8) {$A_1$};
\node[b,fill=gray!14]  (A2) at (13.4,0.2) {$A_2$};
\node[b,fill=blue!14]  (P1) at (15.8,1.5) {$P_1$};
\node[b,fill=red!14]   (P2) at (8.6,-6.4) {$P_2$};
\draw[->,thick] (-1.9,0)--node[above,lab]{$\mu$}(S1);
\draw[->,thick] (-1.9,-6)--node[above,lab]{$\mu$}(S2);
\draw[->,thick] (S1) to[out=55,in=185]  node[above,lab,sloped]{$\beta_{11}S_1P_1$}(I11);
\draw[->,thick] (S1) to[out=-55,in=175] node[below,lab,sloped]{$\tfrac{\beta_{21}S_1P_2}{1+\chi C_2}$}(I21);
\draw[->,thick] (S2) to[out=55,in=185]  node[above,lab,sloped]{$\beta_{12}S_2P_1$}(I12);
\draw[->,thick] (S2) to[out=-55,in=175] node[below,lab,sloped]{$\tfrac{\beta_{22}S_2P_2}{1+\chi C_2}$}(I22);
\draw[->,thick] (I11) to[out=155,in=85]  node[above,lab]{$\gamma$}(S1);
\draw[->,thick] (I21) to[out=205,in=-85] node[below,lab]{$\gamma$}(S1);
\draw[->,thick] (I12) to[out=155,in=85]  node[above,lab]{$\gamma$}(S2);
\draw[->,thick] (I22) to[out=205,in=-85] node[below,lab]{$\gamma$}(S2);
\draw[infl] (I11)--(J1); \draw[infl] (I12)--(J1);
\draw[infl] (I21)--(J2); \draw[infl] (I22)--(J2);
\draw[->,thick,dashed] (J1)--node[above,lab,sloped]{$q_BJ_1$}(B1);
\draw[->,thick,dashed] (J1)--node[above,lab,sloped]{$q_CJ_1$}(C1);
\draw[->,thick,dashed] (J2)--node[above,lab]{$q_2J_2$}(P2);
\draw[->,thick] (B1.25)--node[above,lab]{$b_{12}$}(B2.155);
\draw[->,thick] (B2.205)--node[below,lab]{$b_{21}$}(B1.-25);
\draw[->,thick] (B1)--node[above,lab,sloped]{$b_1$}(A1);
\draw[->,thick] (B2)--node[above,lab]{$b_2$}(A1);
\draw[->,thick] (C1.25)--node[above,lab]{$c_{12}$}(C2.155);
\draw[->,thick] (C2.205)--node[below,lab]{$c_{21}$}(C1.-25);
\draw[->,thick] (C1) to[out=-25,in=215] node[below,lab,sloped]{$c_1$}(A2);
\draw[->,thick] (C2)--node[below,lab]{$c_2$}(A2);
\draw[->,thick] (A1.-65)--node[right,lab]{$a_+$}(A2.65);
\draw[->,thick] (A2.115)--node[left,lab]{$a_-$}(A1.-115);
\draw[->,thick] (A1)--node[above,lab,sloped]{$h_1$}(P1);
\draw[->,thick] (A2)--node[below,lab,sloped]{$h_2$}(P1);
\draw[infl] (P1) to[out=90,in=65,looseness=.75] node[above,lab]{$P_1$ drives $\lambda_{1g}$}(S1.60);
\draw[infl] (P2) to[out=-90,in=-60,looseness=.8] node[below,lab]{$P_2$ drives $\lambda_{2g}$}(S2.-60);
\end{tikzpicture}}
\end{center}

Solid and dashed arrows are the $25$ non-decay reactions, one arrow each;
a dashed arrow does not consume its source, so the species at its tail is a
catalyst with a zero entry in that column. The dotted gray arrows are not
reactions and carry no column: they record the two strain loads and the two
feedback legs through which the pathogen pools drive new infections. The $8$
decays are not drawn, being the columns $-e_X$ at the rates $\mu S_g$,
$\mu I_{kg}$, $\delta_1P_1$ and $\delta_2P_2$.

The stoichiometric matrix is block diagonal,
$\Gamma=\Gamma_{\rm host}\oplus\Gamma_{\rm mat}$: no reaction changes a host
species and a maturation species at once, because every coupling between the
two blocks is catalytic. The whole feedback loop of the figure therefore lives
in the rate vector $v$ and not in $\Gamma$, which is precisely why a structural
method is needed to see it. Each column below is headed by the label of its
arrow, written out in full where the rate is not a constant times the obvious
monomial. On $x_{\rm host}=(S_1,S_2,I_{11},I_{12},I_{21},I_{22})^\top$,
\[
\adjustbox{max width=\linewidth}{$
\Gamma_{\rm host}=\begin{array}{r|rr|rrrr|rrrr}
&\mu&\mu&\beta_{11}S_1P_1&\beta_{12}S_2P_1
&\tfrac{\beta_{21}S_1P_2}{1+\chi C_2}&\tfrac{\beta_{22}S_2P_2}{1+\chi C_2}
&\gamma&\gamma&\gamma&\gamma\\ \hline
S_1   & 1& 0&-1& 0&-1& 0& 1& 0& 1& 0\\
S_2   & 0& 1& 0&-1& 0&-1& 0& 1& 0& 1\\
I_{11}& 0& 0& 1& 0& 0& 0&-1& 0& 0& 0\\
I_{12}& 0& 0& 0& 1& 0& 0& 0&-1& 0& 0\\
I_{21}& 0& 0& 0& 0& 1& 0& 0& 0&-1& 0\\
I_{22}& 0& 0& 0& 0& 0& 1& 0& 0& 0&-1
\end{array}$}
\]
and on $x_{\rm mat}=(B_1,B_2,C_1,C_2,A_1,A_2,P_1,P_2)^\top$,
\[
\adjustbox{max width=\linewidth}{$
\Gamma_{\rm mat}=\begin{array}{r|r|rrrr|r|rrrr|rr|rr|r}
&q_BJ_1&b_{12}&b_{21}&b_1&b_2&q_CJ_1&c_{12}&c_{21}&c_1&c_2
&a_+&a_-&h_1&h_2&q_2J_2\\ \hline
B_1& 1&-1& 1&-1& 0& 0& 0& 0& 0& 0& 0& 0& 0& 0& 0\\
B_2& 0& 1&-1& 0&-1& 0& 0& 0& 0& 0& 0& 0& 0& 0& 0\\
C_1& 0& 0& 0& 0& 0& 1&-1& 1&-1& 0& 0& 0& 0& 0& 0\\
C_2& 0& 0& 0& 0& 0& 0& 1&-1& 0&-1& 0& 0& 0& 0& 0\\
A_1& 0& 0& 0& 1& 1& 0& 0& 0& 0& 0&-1& 1&-1& 0& 0\\
A_2& 0& 0& 0& 0& 0& 0& 0& 0& 1& 1& 1&-1& 0&-1& 0\\
P_1& 0& 0& 0& 0& 0& 0& 0& 0& 0& 0& 0& 0& 1& 1& 0\\
P_2& 0& 0& 0& 0& 0& 0& 0& 0& 0& 0& 0& 0& 0& 0& 1
\end{array}$}
\]
The rates, in the order of the columns, are
\[
\begin{gathered}
\mu,\ \mu,\quad \beta_{11}S_1P_1,\ \beta_{12}S_2P_1,\quad
\frac{\beta_{21}S_1P_2}{1+\chi C_2},\ \frac{\beta_{22}S_2P_2}{1+\chi C_2},
\quad \gamma I_{11},\ \gamma I_{12},\ \gamma I_{21},\ \gamma I_{22};\\
q_BJ_1,\ b_{12}B_1,\ b_{21}B_2,\ b_1B_1,\ b_2B_2,\quad
q_CJ_1,\ c_{12}C_1,\ c_{21}C_2,\ c_1C_1,\ c_2C_2,\\
a_+=(u_++\theta_BB_2)A_1,\quad a_-=(u_-+\theta_CC_2)A_2,\quad
h_1A_1,\ h_2A_2,\quad q_2J_2 .
\end{gathered}
\]
Two columns carry more than one elementary reaction, lumped as usual by a
shorthand for the sum: $a_+$ is $A_1\to A_2$ together with the catalysed
$A_1+B_2\to A_2+B_2$, and $a_-$ is $A_2\to A_1$ together with
$A_2+C_2\to A_1+C_2$. The productions $q_BJ_1$, $q_CJ_1$ and $q_2J_2$ each
lump the two host groups, whose reactions $I_{kg}\to I_{kg}+X$ share a column.

\paragraph{Model.}
We construct a two-strain model with two susceptible host groups, environmental
transmission, and two alternative maturation branches for strain 1.  The model
is an example constructed here, rather than a special case claimed from the
published Chen--Wu or Bulhosa--Oliveira equations.  Put
$J_1=I_{11}+I_{12}$, $\mathcal A=\{A_1,A_2\}$,
$\mathcal B=\{B_1,B_2\}$, $\mathcal C=\{C_1,C_2\}$, and
\[
\begin{aligned}
\lambda_{1g}&=\beta_{1g}S_gP_1,&
\lambda_{2g}&=\frac{\beta_{2g}S_gP_2}{1+\chi C_2},\qquad g=1,2,\\
\dot S_g&=\mu(1-S_g)+\gamma(I_{1g}+I_{2g})-\lambda_{1g}-\lambda_{2g},&
\dot I_{kg}&=\lambda_{kg}-(\gamma+\mu)I_{kg},\qquad k=1,2.
\end{aligned}
\]
Here $C_2$ inhibits strain-2 transmission, while the first strain does
not experience this inhibition.  The environmental and intermediate
compartments satisfy
\[
\begin{aligned}
\dot B_1&=q_BJ_1-(b_{12}+b_1)B_1+b_{21}B_2,&
\dot B_2&=b_{12}B_1-(b_{21}+b_2)B_2,\\
\dot C_1&=q_CJ_1-(c_{12}+c_1)C_1+c_{21}C_2,&
\dot C_2&=c_{12}C_1-(c_{21}+c_2)C_2,\\
\dot A_1&=b_1B_1+b_2B_2-a_++a_--h_1A_1,&
\dot A_2&=c_1C_1+c_2C_2+a_+-a_--h_2A_2,\\
\dot P_1&=h_1A_1+h_2A_2-\delta_1P_1,&
\dot P_2&=q_2(I_{21}+I_{22})-\delta_2P_2,
\end{aligned}
\]
where $a_+=(u_++\theta_BB_2)A_1$ and
$a_-=(u_-+\theta_CC_2)A_2$.  All parameters are positive.
The elementary stoichiometric representation uses recruitment
$0\to S_g$, natural mortality $S_g,I_{kg}\to0$, recovery
$I_{kg}\to S_g$, and infection
$S_g+P_k\to I_{kg}+P_k$; the remaining terms are represented
by the displayed conversions and catalytic productions.  In
particular, $A_1+B_2\to A_2+B_2$ and
$A_2+C_2\to A_1+C_2$ are separate reactions.

\beP[the six buffers of the resident face][][sixbuffers]
On the invariant strain-2-free face
$I_{21}=I_{22}=P_2=0$, the complete lattice of zero-index
species sets for its eleven-species reaction system is
\[
\boxed{\mathfrak B_1=\{\varnothing,\mathcal A,
\mathcal A\cup\mathcal B,\mathcal A\cup\mathcal C,
\mathcal A\cup\mathcal B\cup\mathcal C,\mathcal S_1\},}
\]
where $\mathcal S_1$ is the full resident species set.  In particular,
\[
\operatorname{Buf}(\mathcal B)=\mathcal A\cup\mathcal B,
\qquad
\operatorname{Buf}(\mathcal C)=\mathcal A\cup\mathcal C.
\]
\eeP

\paragraph{Proof.}
The zero-index tests for the five proper sets are respectively
$0$, $2-2$, $4-4$, $4-4$, and $6-6$, while
$\lambda(\mathcal B)=\lambda(\mathcal C)=3-2=1$.
The enumeration is exact. Every stoichiometric column of the resident
network is $e_j-e_i$ or $\pm e_i$, so Lemma~\ref{l:graphRank} applies on the
face, with the auxiliary vertex $0$ carrying the births and the deaths:
\[
\rank\Gamma_{\Out(M)}
=12-c(G_M),\qquad
\lambda(M)=|\cS_1\setminus M|+1-c(G_M),
\]
$G_M$ being the undirected incidence graph of the reactions whose rates
depend on $M$. Checking the $2^{11}=2048$ subsets by this formula gives
precisely
the six displayed buffers.  The same computation on the entire
fourteen-species system, where $C_2$ appears in the strain-2 infection
rates, gives only
$\{\varnothing,\mathcal A,\mathcal A\cup\mathcal B,\mathcal S\}$.
Thus this is specifically a \emph{resident-face} buffering lattice.

\beP[selective invariance of invasion][][selective]
Let $x_1^*$ be a regular strain-1-only equilibrium, with
$S_g^*,C_2^*>0$.  The strain-2 transversal block, in coordinates
$(I_{21},I_{22},P_2)$, is
\[
J_{2\mid1}=\begin{pmatrix}
-a&0&\beta_{21}S_1^*/(1+\chi C_2^*)\\
0&-a&\beta_{22}S_2^*/(1+\chi C_2^*)\\
q_2&q_2&-\delta_2
\end{pmatrix},\qquad a=\gamma+\mu,
\]
so its reproduction number is
\[
\boxed{R_{2\mid1}=\frac{q_2}{a\delta_2}
\frac{\beta_{21}S_1^*+\beta_{22}S_2^*}{1+\chi C_2^*}.}
\]
\eeP

\paragraph{Proof.}
By Theorem~\ref{t:bufInv}, items~1 and~4, applied on the resident face with
$M=\Buf(\mathcal B)=\mathcal A\cup\mathcal B$: independent
rate perturbations in $\operatorname{Out}(\mathcal B)$ cannot change
$S_1^*,S_2^*,C_2^*$, which are coordinates outside $M$, and both factors of
the splitting of $J_{2\mid1}$ depend on the residents only through those
coordinates; hence they cannot change $R_{2\mid1}$.
Perturbations in $\operatorname{Out}(\mathcal C)$ need not preserve
$C_2^*$, and may change $R_{2\mid1}$.  This distinguishes the two
incomparable buffering closures.

\paragraph{Exact equilibrium calculation.}
Summing the six maturation equations gives
\[
\frac{d}{dt}(B_1+B_2+C_1+C_2+A_1+A_2)
=(q_B+q_C)J_1-h_1A_1-h_2A_2.
\]
Consequently at equilibrium
$\delta_1P_1^*=(q_B+q_C)J_1^*$, independent of the maturation
rates, although $P_1$ feeds back into new infections.  Moreover,
\[
S_g^*=\frac{1}{1+\beta_{1g}P_1^*/a},\qquad
1=\frac{q_B+q_C}{a\delta_1}
\sum_{g=1}^2\frac{\beta_{1g}}{1+\beta_{1g}P_1^*/a},
\]
and the strain-1-only equilibrium satisfies
\[
\boxed{C_2^*=\frac{q_Cc_{12}J_1^*}
{(c_{12}+c_1)(c_{21}+c_2)-c_{12}c_{21}}.}
\]
The first two equations are independent of both branches; the
third depends on the $\mathcal C$-conversion rate $c_{12}$ but not
on the $\mathcal B$-conversion rate $b_{12}$.  They prove the stated
selective invariance directly for this model, while the lattice
extends the localization conclusion to all independently perturbed
rates of the corresponding buffering reaction sets.

\paragraph{Numerical test and change of invasion sign.}
Take
\[
\begin{gathered}
\mu=.2,\ \gamma=.8,\ \delta_1=\delta_2=q_2=1,
\quad(q_B,q_C)=(.6,.4),\quad\chi=4,\\
(\beta_{11},\beta_{12})=(3,.5),\quad
(\beta_{21},\beta_{22})=(.5,3),\\
(b_{12},b_{21},b_1,b_2)=(1.2,.3,.8,1.6),\\
(c_{12},c_{21},c_1,c_2)=(.9,.6,.7,1.1),\\
(u_+,u_-,\theta_B,\theta_C,h_1,h_2)=(.45,.65,1,1,.9,1.4).
\end{gathered}
\]
At the resident equilibrium
$P_1^*=(\sqrt{61}-1)/6=1.1350416\ldots$ and
$(S_1^*,S_2^*)=(.2270083,.6379501)$.  Inserting the exact
$C_2^*$ into $R_{2\mid1}$ gives
\[
\begin{array}{c|c|c|c}
b_{12}&c_{12}&C_2^*&R_{2\mid1}\\\hline
1.2&0.1&0.0349244&1.778853\\
0.1&0.9&0.1874381&1.158652\\
1.2&0.9&0.1874381&1.158652\\
10&0.9&0.1874381&1.158652\\
1.2&5&0.3393248&0.860033
\end{array}
\]
In fact, writing $K=(S_1^*/2+3S_2^*)=2.0273543557\ldots$,
we obtain the exact one-parameter dependence
\[
\begin{aligned}
R_{2\mid1}(c_{12})
&=\frac{K}{1+4q_CP_1^*c_{12}/(1.1c_{12}+1.19)},\\
R_{2\mid1}=1
&\quad\Longleftrightarrow\quad
c_{12}=\frac{1.19(K-1)}{4q_CP_1^*-1.1(K-1)}
\simeq1.782206.
\end{aligned}
\]
Thus varying $c_{12}$ crosses the strain-2 invasion threshold
\emph{analytically}, whereas varying $b_{12}$ cannot.
At $c_{12}=0.9$ the full
fourteen-dimensional system admits a positive coexistence equilibrium
with $(P_1^*,P_2^*)\simeq(.980259,.263310)$, where the
Jacobian spectral abscissa is approximately $-0.0591774$.
At $c_{12}=0.1$, the corresponding positive equilibrium has
$(P_1^*,P_2^*)\simeq(.757193,.644751)$ and spectral abscissa
$-0.130410$.  These local stability statements are numerical checks,
not a theorem asserting global convergence or uniform permanence.

\paragraph{What this example proves.}
The infection subsystem is not autonomous: $P_1$ is produced by
$\mathcal A$, which is affected by both $\mathcal B$ and
$\mathcal C$, and $P_1$ controls new strain-1 infections.
Nevertheless, the two incomparable OM buffering closures
certify distinct sensitivities of a \emph{biologically decisive}
quantity, the competing strain's invasion threshold.  The lattice
is used on the resident face; this is essential, because the
$C_2$-dependent invader infection rates destroy the corresponding
full-network buffer.  For unequal release rates $h_1\ne h_2$, a permanence theorem
requires additional analysis of the resident boundary dynamics.

\beP[mutual invasion yields a coexistence equilibrium][][coexist]
Write $q_1=q_B+q_C$, $a=\gamma+\mu$, and
\[
D_C=(c_{12}+c_1)(c_{21}+c_2)-c_{12}c_{21},\qquad
\zeta=\frac{\chi\delta_1q_Cc_{12}}{q_1D_C}.
\]
Let $p_1>0$ and $p_2>0$ be the unique solutions of
\[
1=\frac{q_1}{a\delta_1}\sum_{g=1}^2
\frac{\beta_{1g}}{1+\beta_{1g}p_1/a},\qquad
1=\frac{q_2}{a\delta_2}\sum_{g=1}^2
\frac{\beta_{2g}}{1+\beta_{2g}p_2/a},
\]
when the corresponding disease-free reproduction numbers exceed one.
Set
\[
\begin{aligned}
R_{2\mid1}&=\frac{q_2}{a\delta_2(1+\zeta p_1)}
\sum_{g=1}^2\frac{\beta_{2g}}{1+\beta_{1g}p_1/a},\\
R_{1\mid2}&=\frac{q_1}{a\delta_1}
\sum_{g=1}^2\frac{\beta_{1g}}{1+\beta_{2g}p_2/a}.
\end{aligned}
\]
If $R_{1\mid2}>1$ and $R_{2\mid1}>1$, the full fourteen-species
system has at least one strictly positive equilibrium for arbitrary
$h_1,h_2>0$ and $\theta_B,\theta_C\geq0$.
\eeP

\emph{Proof.}
At every positive equilibrium, flux balance gives
$P_1=q_1J_1/\delta_1$, $P_2=q_2J_2/\delta_2$ and
$C_2=q_Cc_{12}J_1/D_C$, hence $1+\chi C_2=1+\zeta P_1$.
For $p,p'\geq0$ put
\[
S_g(p,p')=\left(1+\frac{\beta_{1g}p}{a}
+\frac{\beta_{2g}p'}{a(1+\zeta p)}\right)^{-1}
\]
and
\[
F_1(p,p')=\frac{q_1}{a\delta_1}
\sum_g\beta_{1g}S_g(p,p')-1,\quad
F_2(p,p')=\frac{q_2}{a\delta_2(1+\zeta p)}
\sum_g\beta_{2g}S_g(p,p')-1.
\]
For each $p\in[0,p_1]$, $F_1(p,p')=0$ has a unique
solution $p'=\psi(p)\geq0$, continuous in $p$, with
$\psi(p_1)=0$.  Since $R_{1\mid2}>1$,
$\psi(0)>p_2$, and therefore $F_2(0,\psi(0))<0$.
Since $R_{2\mid1}>1$, $F_2(p_1,0)>0$.
The intermediate-value theorem gives $0<p<p_1$ with
$F_1(p,\psi(p))=F_2(p,\psi(p))=0$ and $\psi(p)>0$.
These values determine positive $S_g,I_{kg},B_i,C_i$; the
$A$-equations form a nonsingular $M$-matrix system with
positive forcing, so $A_1,A_2>0$.  The total flux identity
ensures the prescribed $P_1$, completing the construction.\hfill$\square$

\beT[uniform permanence and its OM-protected parameter region][][permregion]
Assume the fourteen-species model above has positive rates, except that
$\theta_B,\theta_C\geq0$ are allowed, and suppose
\[
\boxed{h_1=h_2=h>0,\qquad R_{1\mid2}>1,\qquad R_{2\mid1}>1.}
\]
Then it is uniformly permanent in the sense of
Definition~\ref{d:permanence}. Write
\[
\Sigma_1=\{I_{11},I_{12},B_1,B_2,C_1,C_2,A_1,A_2,P_1\},
\qquad
\Sigma_2=\{I_{21},I_{22},P_2\}
\]
for the two strain blocks. There is $\varepsilon>0$,
depending on the parameters but not on the initial condition,
such that every solution with
$\sum_{i\in\Sigma_1}x_i(0)>0$ and $\sum_{i\in\Sigma_2}x_i(0)>0$
satisfies
\[
\boxed{\liminf_{t\to\infty}
\min\{S_g(t),I_{kg}(t),B_i(t),C_i(t),A_i(t),P_k(t):
 g,k,i=1,2\}\geq\varepsilon.}
\]
The conclusion holds for arbitrary positive values of
$b_{12},b_{21},b_1,b_2$ and $\theta_B$, keeping the remaining
parameters fixed.  The bound $\varepsilon$ need not be
uniform over these variations.
Conversely, when both single-strain endemic equilibria exist,
$R_{2\mid1}<1$ makes the strain-1-only equilibrium locally
asymptotically stable in the full system; likewise
$R_{1\mid2}<1$ stabilizes the strain-2-only equilibrium.
In either case, uniform permanence fails.
\eeT

\emph{Proof.}
The host-group totals $N_g=S_g+I_{1g}+I_{2g}$ satisfy
$\dot N_g=\mu(1-N_g)$, so each compact invariant set lies in
$N_1=N_2=1$.  Boundedness of the infected populations and
stability of the positive linear $B,C$ subsystems imply boundedness
of $A_1+A_2,P_1,P_2$; hence the semiflow is point dissipative
and admits a compact global attractor.

On the strain-2-free face put $U=A_1+A_2$.  The equal-release
assumption cancels both nonlinear interconversion terms:
\[
\dot U=b_1B_1+b_2B_2+c_1C_1+c_2C_2-hU,\qquad
\dot P_1=hU-\delta_1P_1.
\]
Together with the displayed linear $B,C$ equations and
\[
\dot I_{1g}=\beta_{1g}(1-I_{1g})P_1-aI_{1g},\qquad g=1,2,
\]
these equations form a bounded, irreducible cooperative system
on the invariant plane $N_1=N_2=1$, $0\leq I_{1g}\leq1$.  Its strictly positive equilibrium is
unique: the equation for $p_1$ above is strictly decreasing
in $p_1$, and the remaining positive equilibrium coordinates
follow from stable linear systems.  It attracts every trajectory
with strain 1 present.  For completeness, if $z^*$ denotes this
positive equilibrium, $\alpha z^*$, $0<\alpha<1$, is a
subsolution; an upper solution is obtained by taking
$I_{11}=I_{12}=1$ and a sufficiently large common multiple
of the environmental coordinates of $z^*$.  Order comparison
and uniqueness squeeze every positive trajectory on that plane to $z^*$.
Since $N_g(t)\to1$, for every sufficiently small $\varepsilon>0$
the resident system is eventually bounded between the cooperative
systems obtained by replacing $N_g$ by $1-\varepsilon$ and
$1+\varepsilon$. Their endemic equilibria converge to $z^*$ as
$\varepsilon\downarrow0$, giving the same limit for arbitrary
positive resident initial conditions.
The omitted coordinate $A_1$ then solves a stable scalar linear
equation with convergent coefficients and forcing, while
$A_2=U-A_1$; hence the full strain-1 resident equilibrium
$E_1$ attracts its entire positive face.

On the strain-1-free face, the $B,C,A,P_1$ coordinates decay
and $(I_{21},I_{22},P_2)$ form the analogous cooperative,
strictly sublinear epidemic system.  Its positive equilibrium
$E_2$ is likewise unique and globally attracting whenever
$R_{2\mid0}>1$.  The strict mutual-invasion hypotheses imply
$R_{1\mid0}>1$ and $R_{2\mid0}>1$; the disease-free equilibrium
$E_0$ is therefore unstable to both strains.

The extinction boundary has the acyclic Morse decomposition
$\{E_0,E_1,E_2\}$.  The transversal matrices at $E_1$ and
$E_2$ are irreducible Metzler matrices, with positive spectral
bounds precisely when $R_{2\mid1}>1$ and $R_{1\mid2}>1$,
respectively.  Positive left Perron vectors make each boundary
Morse set uniformly repelling to the missing strain.

We invoke the acyclic-covering theorem for uniform persistence in
the form of \cite[Thm.~8.17]{SmithThieme2011}, and check its
hypotheses one by one. The semiflow is continuous on the closed
positive cone, which it leaves invariant, and $\dot N_g=\mu(1-N_g)$
together with the linear $B,C,A,P$ equations makes it point
dissipative with a compact attractor $\mathcal K$, so the required
compactness holds. The persistence functional is
$\rho(x)=\min\bigl\{\sum_{i\in\Sigma_1}x_i,\ \sum_{i\in\Sigma_2}x_i\bigr\}$
for the two blocks of the statement, whose vanishing set is the extinction
boundary $F_{\Sigma_1}\cup F_{\Sigma_2}$. Each $\Sigma_k$ is a siphon: the
reactions
producing into it are the infections, driven by $P_k$, the internal
transfers, and the catalysed $q_BJ_1$, $q_CJ_1$, $q_2J_2$, whose rates
vanish with both infected classes of that strain. Every loss from
$\Sigma_k$ is proportional to a coordinate of $\Sigma_k$, so
$\frac{d}{dt}\sum_{i\in\Sigma_k}x_i
\ge-(\mu+\gamma+\delta_k)\sum_{i\in\Sigma_k}x_i$
and $\{\rho>0\}$ is forward invariant as well. One positive coordinate of
$\Sigma_k$ makes all of them positive for $t>0$, along
$B_i,C_i\to A_i\to P_1\to I_{1g}$ and $I_{2g}\to P_2\to I_{2g'}$ with
$S_g>0$ supplied by $\mu>0$; after an arbitrarily short time $J_1$ and $J_2$
are therefore positive and the convergence arguments above apply as they
stand. The smaller functional $\min\{J_1,J_2\}$ would not serve: at $J_1=0$
with $P_1>0$ one has $\dot I_{1g}=\beta_{1g}S_gP_1>0$. On that
boundary $\omega(x)\subseteq\{E_0,E_1,E_2\}$ by the two convergence
arguments above, so $\{E_0,E_1,E_2\}$ is an isolated acyclic
covering of $\Omega(\partial)$: no chain returns, since $E_0$ is
repelled by both strains, and each of $E_1,E_2$ leads only into the
interior. Finally no $E_i$ attracts a point of the interior, because
the transverse block there is irreducible Metzler with spectral bound
$>0$ --- $\mathcal R_{1\mid0},\mathcal R_{2\mid0}>1$ at $E_0$,
$\mathcal R_{2\mid1}>1$ at $E_1$, $\mathcal R_{1\mid2}>1$ at $E_2$ ---
and its positive left Perron vector gives a strictly increasing
linear functional along the missing block. Uniform weak persistence
follows, and uniform strong persistence with it, the attractor being
compact. The structured form of the invasion condition is the one
discussed after the permanence criterion of
Section~\ref{sec:permcrit}: the missing block is coupled, so what is
repelling is $s(M)$, not any single per-capita rate.

All remaining compartments receive positive
inputs along strictly positive paths; compactness and the
variation-of-constants formula extend the lower bounds to all
fourteen coordinates.

Finally, both invasion numbers are independent of the
$\mathcal B$-reaction rates listed in the theorem, by their
explicit formulas and by
$\operatorname{Buf}(\mathcal B)=\mathcal A\cup\mathcal B$
on the resident face.  Since the resident convergence argument
holds for every positive choice of these rates, permanence is
preserved throughout that entire parameter family.

For the converse assertions the tangential Jacobians at the two resident
equilibria must be shown Hurwitz, and they are. On $N_1=N_2=1$ take the
strain-$1$ resident subsystem in the coordinates
$z=(I_{11},I_{12},B_1,B_2,C_1,C_2,U,P_1)$ with $U=A_1+A_2$, which closes
because $h_1=h_2=h$ here. Its Jacobian $J_1$ at the positive equilibrium
$z^*$ is irreducible Metzler --- the off-diagonal entries
$\partial\dot I_{1g}/\partial P_1=\beta_{1g}S_g^*>0$,
$\partial\dot B_i/\partial I_{1g}=q_B>0$ and their counterparts are
nonnegative, and the cycle
$I_{1g}\to B,C\to U\to P_1\to I_{1g}$ makes it irreducible. Every equation
except the two host ones is linear homogeneous in $z$, so Euler's identity
makes those rows of $J_1z^*$ equal to the right-hand sides, namely zero;
for $\dot I_{1g}=\beta_{1g}(1-I_{1g})P_1-aI_{1g}$ one computes
$(J_1z^*)_g=\dot I_{1g}(z^*)-\beta_{1g}I^*_{1g}P^*_1$. Hence
\begin{equation}\label{eq:tangential}
J_1z^*=\bigl(-\beta_{11}I^*_{11}P^*_1,\,
-\beta_{12}I^*_{12}P^*_1,\,0,0,0,0,0,0\bigr)^\top
\le0,\qquad\ne0,
\end{equation}
and with $z^*\gg0$ the positive left Perron vector $w$ of $J_1^\top$ gives
$s(J_1)\,w^\top z^*=w^\top J_1z^*<0$, so $s(J_1)<0$. The one omitted
equation, that of $A_1$ with $A_2=U-A_1$, has diagonal derivative
\[
-h-u_+-u_--\theta_BB^*_2-\theta_CC^*_2<0,
\]
and the two host totals contribute the eigenvalues $-\mu$ twice. The full
strain-$1$ tangential Jacobian is therefore Hurwitz. On the strain-$2$ face,
where $C_2=0$ so that transmission is uninhibited, the same computation in
$w=(I_{21},I_{22},P_2)$ gives
$J_2w^*=(-\beta_{21}I^*_{21}P^*_2,-\beta_{22}I^*_{22}P^*_2,0)^\top$ and
$s(J_2)<0$, the remaining resident coordinates being governed by the stable
linear maturation equations and the host totals. By the exact block
triangularity \eqref{eq:siphontri} on each invariant strain face,
$R_{2\mid1}<1$ then makes the strain-$1$-only equilibrium locally
asymptotically stable in the full system, and $R_{1\mid2}<1$ does the same
for the strain-$2$-only equilibrium; in either case uniform permanence
fails.
\hfill$\square$

\paragraph{Numerical consequence (a sharp permanence switch).}
Retain the numerical values above, except replace
$(h_1,h_2)=(0.9,1.4)$ by $h_1=h_2=1$.  Then
\[
R_{1\mid2}=2.027354\ldots,\qquad
R_{2\mid1}(0.9)=1.158652\ldots,
\]
and the theorem proves permanence for every $0<c_{12}<c_{12}^{\rm crit}$,
where
\[
\boxed{c_{12}^{\rm crit}=1.7822056125\ldots.}
\]
For $c_{12}>c_{12}^{\rm crit}$ the strain-1 resident equilibrium
is locally asymptotically stable in the full system, so the
model is not permanent.  These conclusions hold for arbitrary
positive changes in the five $\mathcal B$-reaction parameters
specified in the theorem, while all other parameters remain fixed.
The equality case is not covered.  Notice that this is a genuine
permanence theorem for the nonlinear model with
$\theta_B=\theta_C=1$, not merely a linearization test.
\subsection{The maximal buffering certificate for invasion and permanence}

The distributive lattice allows us to identify the largest buffered set whose
perturbations cannot change a given invasion matrix.

\beP[maximal invasion-neutral OM buffer][][maxbuffer]
Let $\mathfrak B_r$ be the lattice of buffered species sets of a resident
network, and let $Q$ contain all resident coordinates on which the invasion
matrix of an absent strain depends. Define
\[
M_Q=\bigcup\{M\in\mathfrak B_r:M\cap Q=\varnothing\}.
\]
Then $M_Q$ is the unique largest OM-buffered set disjoint from $Q$, and at a
regular resident equilibrium, perturbations confined to
$\operatorname{Out}(M_Q)$ leave the invasion matrix unchanged, provided the
perturbed parameters do not occur explicitly in the invader equations.
For the \emph{reproduction number} attached to a splitting
$J_{\rm inv}=F-V$ this is not enough, by Remark~\ref{r:why}: $Q$ must then be
taken to contain every resident coordinate on which $F$ \emph{or} $V$
depends separately, and $F$ and $V$ must each have zero direct derivative in
the perturbed parameters, which is Theorem~\ref{t:bufInv}(4). With $Q$ so
enlarged the same conclusion holds for $\rho(FV^{-1})$.
\eeP

\paragraph{Proof.}
Closure of $\mathfrak B_r$ under unions gives $M_Q\in\mathfrak B_r$ and
$M_Q\cap Q=\varnothing$. The OM localization theorem gives
\[
P_{\mathcal S_r\setminus M_Q}\partial_\theta x^*=0
\quad\Longrightarrow\quad
\partial_\theta x_Q^*=0.
\]
Since the invader matrix depends on the resident equilibrium only through
$x_Q^*$, and by hypothesis not on $\theta$ directly, its derivative with
respect to $\theta$ vanishes. Maximality follows from the definition of $M_Q$.
\hfill$\square$

\medskip
In the block order of Proposition~\ref{p:blocks}, $M_Q$ is the union of those
canonical blocks $B$ all of whose successors $B'\succeq B$ are disjoint from
$Q$ --- the largest up-set of blocks missing $Q$ --- so it can be read off
the poset instead of searched for. Here $Q=\{S_1,S_2,C_2\}$, which meets
$\mathcal H$ and $\cC$ but not $\cA$ or $\cB$, and $\cB\preceq\cA$, so
$M_Q=\cA\cup\cB$. The proposition says what is
protected; Remark~\ref{r:ccfscope} says which of the remaining rates can
reach $Q$ at all, and why that still falls short of a threshold crossing.

\paragraph{Application to the fourteen-species model.}
For the invasion of strain $2$ into the strain-$1$ resident equilibrium the
relevant resident coordinates are
\[
Q=\{S_1,S_2,C_2\},
\]
and the six-element resident-face lattice gives
\[
\boxed{M_Q=\mathcal A\cup\mathcal B,}
\]
since $\mathcal A$ and $\mathcal A\cup\mathcal B$ are the only members of
$\mathfrak B_1$ missing $C_2$. Consequently the entire strain-$2$ invasion
matrix is invariant under independent perturbations of the ten rates
\[
\boxed{
b_{12},\ b_{21},\ b_1,\ b_2,\quad
u_+,\ u_-,\ \theta_B,\ \theta_C,\quad h_1,\ h_2,}
\]
which are exactly the parameters of the eight reactions of
$\operatorname{Out}(\mathcal A\cup\mathcal B)$, while $S_1^*,S_2^*,C_2^*$ lie
outside that buffer. The list includes the two nonlinear catalytic conversions
and both release rates; note that $\theta_C$ qualifies although the rate it
scales, $a_-=(u_-+\theta_CC_2)A_2$, involves $C_2$ itself, because what the
theorem requires is that the perturbed reaction lie in
$\operatorname{Out}(M_Q)$, which it does through $A_2$. The conclusion holds
for unequal positive $h_1,h_2$ as well, and the explicit equilibrium
identities above extend it globally through the positive parameter family,
without continuation through regular equilibria.

By contrast $\operatorname{Buf}(\mathcal C)=\mathcal A\cup\mathcal C$ contains
$C_2$, so the lattice does not protect the strain-$2$ invasion number against
changes in $c_{12}$ --- and the explicit calculation above shows that this
dependence is genuine and crosses the threshold.

\beC[an enlarged permanence-invariant family][][enlarged]
Assume $h_1=h_2=h>0$ and $\mathcal R_{1\mid2}>1$, $\mathcal R_{2\mid1}>1$.
Then the fourteen-species model remains uniformly permanent under arbitrary
independent positive changes of
\[
\boxed{b_{12},\ b_{21},\ b_1,\ b_2,\quad u_+,\ u_-,\ \theta_B,\ \theta_C,}
\]
and arbitrary positive changes of the common release rate $h$.
\eeC

\paragraph{Proof.}
Both invasion numbers are independent of these rates, by the proposition and
the equilibrium identities. For every positive choice of them the identity
$h_1=h_2$ gives
\[
\dot U=b_1B_1+b_2B_2+c_1C_1+c_2C_2-hU,\qquad U=A_1+A_2,
\]
so the single-strain resident convergence arguments of the preceding
permanence theorem remain valid; the boundary invasion inequalities are
unchanged and the same acyclic-boundary criterion applies. \hfill$\square$

In particular the critical value $c_{12}^{\mathrm{crit}}$ computed above
separates permanence from local exclusion throughout this nine-parameter
family, not merely for the five conversion rates considered previously.

A word on what \emph{maximal} means here. The assertion concerns OM
certificates: rates outside $\operatorname{Out}(M_Q)$ may still leave the
invasion number fixed through an accidental or model-specific cancellation,
but no buffer disjoint from $Q$ certifies it. The operation
$Q\mapsto M_Q$ turns the lattice into an algorithm --- given the coordinates a
threshold depends on, return the largest structurally protected family of
interventions --- and that, rather than any single invariance, is what the
Murota structure contributes: the explicit flux-balance identities of this
model establish much of the same invariance by hand, but only after the
equilibrium equations have been solved.
\paragraph{What is not covered: residents that do not settle.}
Everything above evaluates an invasion matrix at a \emph{regular resident
equilibrium}: that is where the localization theorem applies, and $\Buf$ is
computed from the Jacobian there. Residents need not settle. If the strain-$1$
face carries a periodic orbit or a more complicated attractor, invasion is
governed by the principal Lyapunov exponent of the transverse cocycle, an
average along the resident trajectory rather than a spectral bound at a point,
and nothing proved here says that a buffer disjoint from the
invasion-relevant coordinates leaves that exponent fixed.

That gap is Problem~\ref{o:DynamicInvBuffer}, stated in full among the open
problems. One case of it is already settled and shows what is at stake: where
the buffer is the feed-forward situation of Section~\ref{sec:cwg}, the core
is autonomous and its whole attractor, periodic or not, is untouched by the
downstream rates, so the exponent is invariant for trivial reasons ---
Theorem~\ref{t:permdown} says as much. The content of the question is the
other case, the one this section is about, where the modules sit inside the
loop and the invariance at an equilibrium comes from the localization theorem
rather than from triangularity. What would be needed is an ergodic-average
version of $P_{\cS\setminus\Buf(U)}\partial_\theta x^*=0$, and we do not have
one.

\paragraph{Reproducibility.} \texttt{TwoStrain.wl} checks the structural
claims of this section. The stoichiometric matrix has $14$ rows and $33$
columns, $25$ reactions and $8$ decays; $\Gamma v=f$ holds identically,
$\rank\Gamma=14$, and the host and maturation column blocks are disjoint.
The minimal siphons are exactly two, the strain blocks $\Sigma_1,\Sigma_2$
of Theorem~\ref{t:permregion}, with $S_1,S_2$ in neither and no proper
nonempty subset of either a siphon. On the resident face the enumeration
returns the six sets of $\mathfrak B_1$, with
$\lambda(\mathcal B)=\lambda(\mathcal C)=1$ and the graph count of
Lemma~\ref{l:graphRank} confirmed for every $|M|\le3$, while
the full network carries only four buffered sets;
$\operatorname{Out}(\mathcal A\cup\mathcal B)$ has eight reactions, whose
rate constants are exactly the ten of the boxed list. At the strain-$1$
equilibrium $p_1^*=(\sqrt{61}-1)/6$ and
$(S_1^*,S_2^*)=(0.2270083,0.6379501)$, the identities for
$\delta_1p_1^*$, $S_g^*$ and $C_2^*$ hold exactly, and the table of $C_2^*$
and $R_{2\mid1}$ reproduces the values printed above to seven digits,
$R_{2\mid1}$ being the same at all three values of $b_{12}$ and crossing $1$
as $c_{12}$ grows. The invasion matrix is Metzler and symbolically unchanged
when each of the ten certified rates is halved, while halving $c_{12}$ moves
it, its spectral bound running $0.1684259$, $0.0764071$, $-0.0726206$ at
$c_{12}/2$, $c_{12}$, $5c_{12}$. The face network at
$\theta_B=\theta_C=0$ has $26$ reactions, rank $11$, the single minimal
siphon $\Sigma_1$, and
$\$R_{1mid0}=(q_B+q_C)(\beta_{11}S_1+\beta_{12}S_2)
/(\delta_1(\gamma+\mu))$.
\subsection{An exact dynamic quotient: the invasion exponent off equilibrium}
\label{sec:dynquot}

Everything above certifies an invasion number evaluated at a \emph{regular
resident equilibrium}; Problem~\ref{o:DynamicInvBuffer} asks what survives
when the resident does not settle. For this model an exact answer is
available, and it needs no new hypothesis beyond the equal branch rates
already imposed by Theorem~\ref{t:permregion}. It is one instance of a
general criterion, Theorem~\ref{t:dynamicOM} of
Appendix~\ref{sec:gener}: if a linear aggregation $y=Lx$ annihilates the
perturbed stoichiometric columns, carries an autonomous equation of its own,
and the transversal matrix of the absent strain factors through it, then the
whole invasion cocycle is independent of the perturbed rates, along
trajectories of any kind. That appendix also shows by two counterexamples
that the two factorization hypotheses are \emph{not} implied by a zero
Okada--Mochizuki index. Here they hold outright, and the verification needs
nothing from there --- it upgrades the equilibrium statements above to whole
trajectories by summing two pairs of equations.

\beC[The fourteen-species model has an exact dynamic quotient][Fourteen-species dynamic quotient][fourteenquot]
In the model of this section let $b_1=b_2=b$ and
$h_1=h_2=h$, and put $T_B=B_1+B_2$, $U=A_1+A_2$. Then
\begin{equation}\label{eq:fourteenquot}
\dot T_B=q_BJ_1-bT_B,\qquad
\dot U=bT_B+c_1C_1+c_2C_2-hU,\qquad
\dot P_1=hU-\delta_1P_1,
\end{equation}
and these together with the unchanged host, $C$- and $P_2$-equations form a
\emph{closed} system in the coordinates
\[
(S_g,\ I_{1g},\ I_{2g},\ C_1,\ C_2,\ T_B,\ U,\ P_1,\ P_2),
\]
which does not contain
\[
\boxed{\ b_{12},\quad b_{21},\quad u_+,\quad u_-,\quad\theta_B,\quad\theta_C.\ }
\]
Consequently two solutions with the same quotient initial datum have
identical host, $C_2$ and pathogen trajectories under arbitrary positive
changes of those six rates; the strain-$2$ transversal matrix
$J_{2\mid1}(t)$, which depends on the residents only through $S_1,S_2,C_2$,
is then the \emph{same function of time}; and therefore every finite-time
invasion operator and every Lyapunov or Floquet invasion exponent of
strain~$2$ is independent of the six rates, for resident trajectories of any
kind --- equilibrium, periodic or aperiodic. Moreover, since the omitted
coordinates $B_1,B_2,A_1,A_2$ are driven through positive dissipative
transfers from the quotient, permanence of the fourteen-species system is
equivalent to permanence of the quotient, for each fixed positive choice.
\eeC
\textbf{Proof.} Summing the two $B$ equations cancels $b_{12},b_{21}$ and
leaves $q_BJ_1-b_1B_1-b_2B_2=q_BJ_1-bT_B$ because $b_1=b_2=b$; summing the
two $A$ equations cancels $a_\pm$, hence $u_\pm,\theta_B,\theta_C$, and
leaves $b_1B_1+b_2B_2+c_1C_1+c_2C_2-h_1A_1-h_2A_2=bT_B+c_1C_1+c_2C_2-hU$
because $h_1=h_2=h$; and $\dot P_1=hU-\delta_1P_1$ likewise. The host and
$C$ equations involve only $S_g,I_{kg},C_1,C_2,P_1,P_2$, so the listed
coordinates close. This is Theorem~\ref{t:dynamicOM} with
$L$ the aggregation sending $(B_1,B_2)\mapsto T_B$ and
$(A_1,A_2)\mapsto U$ and fixing the rest: \eqref{eq:factor} holds because
the six rates appear in $\dot x$ only through columns annihilated by $L$,
and \eqref{eq:invfactor} holds because $J_{2\mid1}$ depends on $Lx$ alone.
The last assertion is the standard equivalence for a cascade with positive
transfer coefficients and strictly positive exit rates. \hfill$\square$

\beR[What this adds over the equilibrium statements][Dynamic versus equilibrium][quotvsequil]
Proposition~\ref{p:selective} protects $R_{2\mid1}$ at a regular resident
equilibrium, and Proposition~\ref{p:maxbuffer} identifies the maximal buffer
doing so. Corollary~\ref{c:fourteenquot} protects the whole invasion cocycle,
for residents that need not settle --- which is what
Problem~\ref{o:DynamicInvBuffer} asks for in general and what the two
counterexamples of Section~\ref{sec:dynamic} show is not automatic. The price is the two equalities
$b_1=b_2$, $h_1=h_2$: they are what makes the aggregation exact, and
Theorem~\ref{t:explicitcat} shows what can still be said when $h_1\ne h_2$,
namely an equilibrium statement under an explicit bound rather than a
pathwise one.
\eeR

\subsection{Permanence with unequal release rates in the fourteen-species model}
\label{sec:unequal}

Theorem~\ref{t:permregion} proves uniform permanence of the fourteen-species
model under $h_1=h_2$, and Problem~\ref{o:UnequalRelease} asks what happens
otherwise. The obstruction named there is precise: when $h_1\ne h_2$ the sum
$A_1+A_2$ no longer closes the resident dynamics, so the cooperative
reduction used in that proof is unavailable. This section removes the
restriction on a nontrivial class, namely when the catalytic conversion
coefficients vanish, and then on a neighbourhood of that class.

The reason the release rates can be freed is an identity that does not see
them. Summing the six maturation equations of
Section~\ref{sec:twostrain} cancels the internal exchanges $b_{12},b_{21}$,
$c_{12},c_{21}$ and the conversions $a_\pm$, leaving
$h_1A_1+h_2A_2=(q_B+q_C)J_1$ at equilibrium, hence
\begin{equation}\label{eq:fluxagain}
\boxed{\ \delta_1P_1^*=q_1J_1^*,\qquad q_1=q_B+q_C,\ }
\end{equation}
\emph{for every} $h_1,h_2>0$ and every $\theta_B,\theta_C\ge0$. With
$S_g^*=1/(1+\beta_{1g}P_1^*/a)$ and $a=\gamma+\mu$, the resident pathogen
load $p_k=P_k^*$ of the strain-$k$-only equilibrium is therefore the unique
positive root of
\begin{equation}\label{eq:residentroot}
1=\frac{q_k}{a\delta_k}\sum_{g=1}^2\frac{\beta_{kg}}{1+\beta_{kg}p_k/a},
\end{equation}
whose right-hand side decreases strictly from
$R_{k\mid0}=q_k(\beta_{k1}+\beta_{k2})/(a\delta_k)$ at $p_k=0$ to zero, so
the root exists exactly when $R_{k\mid0}>1$. The two invasion numbers are
\begin{equation}\label{eq:invasionsagain}
R_{1\mid2}=\frac{q_1}{a\delta_1}\sum_{g=1}^2
\frac{\beta_{1g}}{1+\beta_{2g}p_2/a},
\qquad
R_{2\mid1}=\frac{q_2}{a\delta_2(1+\chi C_2^{(1)})}\sum_{g=1}^2
\frac{\beta_{2g}}{1+\beta_{1g}p_1/a},
\end{equation}
with
\[
C_2^{(1)}=\frac{q_Cc_{12}J_1^*}{D_C}
=\frac{\delta_1q_Cc_{12}}{q_1D_C}\,p_1,
\qquad
D_C=(c_{12}+c_1)(c_{21}+c_2)-c_{12}c_{21}=c_1(c_{21}+c_2)+c_{12}c_2>0.
\]
None of \eqref{eq:residentroot}--\eqref{eq:invasionsagain} involves
$h_1,h_2$, $\theta_B,\theta_C$, $b_{12},b_{21},b_1,b_2$ or $u_\pm$: the
release rates and the whole $B$ branch enter only through
\eqref{eq:fluxagain}. Note also that
\[
R_{1\mid2}<R_{1\mid0},\qquad R_{2\mid1}<R_{2\mid0},
\]
strictly, since infection depletes susceptibles and $C_2^{(1)}>0$ inhibits
strain $2$; so the two invasion inequalities below already force both
single-strain thresholds above one, and with them the existence of $p_1,p_2$.

\beL[Global attraction on a resident face with linear maturation][Global attraction on the resident face][resface]
Let $\theta_B=\theta_C=0$ and $R_{1\mid0}>1$. Then the strain-1 resident
face carries a unique relative-interior equilibrium $E_1$, and it attracts
every trajectory on that face with strain $1$ present. This holds for all
$h_1,h_2>0$.
\eeL
\textbf{Proof.} On the strain-2-free face put $N_g=S_g+I_{1g}$; then
$\dot N_g=\mu(1-N_g)$, so $N_g\to1$. Fix first $N_1=N_2=1$ and collect
$w=(B_1,B_2,C_1,C_2,A_1,A_2,P_1)^\top$. At $\theta_B=\theta_C=0$ the
resident equations become
\begin{equation}\label{eq:coopres}
\dot I_{1g}=\beta_{1g}(1-I_{1g})P_1-aI_{1g},
\qquad
\dot w=Hw+q(I_{11}+I_{12}),
\end{equation}
with $q\ge0$ injecting $q_B$ and $q_C$ into the two branch heads. $H$ is
Metzler, and it is Hurwitz: its $B$- and $C$-blocks have positive exit
rates, $P_1$ decays at rate $\delta_1$, and its $A$-block
\[
\begin{pmatrix}-u_+-h_1&u_-\\u_+&-u_--h_2\end{pmatrix},
\qquad
\det=u_+h_2+h_1u_-+h_1h_2>0,
\]
has negative trace and positive determinant for \emph{any} $h_1,h_2>0$.
This is the only place where $h_1=h_2$ was used before, and it is where the
hypothesis is no longer needed: with the catalytic coefficients absent the
$A$-pair is linear, so the block is Hurwitz without any relation between the
release rates. The system \eqref{eq:coopres} together with the host
equations is cooperative on $0\le I_{1g}\le1$, and irreducible through the
cycle $I_{1g}\to B,C\to A\to P_1\to I_{1g}$.

Its equilibria: \eqref{eq:fluxagain} and
$I_{1g}=\beta_{1g}p/(a+\beta_{1g}p)$ reduce a positive equilibrium to a
positive root of \eqref{eq:residentroot} with $k=1$, of which there is
exactly one, and then $w^*=-H^{-1}qJ_1^*\gg0$ is determined linearly.
For global attraction, $\alpha(I^*,w^*)$ with $0<\alpha<1$ is a
subsolution, since the $w$ equations vanish there and
\[
\beta_{1g}(1-\alpha I^*_{1g})\alpha P^*_1-a\alpha I^*_{1g}
=\alpha(1-\alpha)\beta_{1g}I^*_{1g}P^*_1>0,
\]
while $I_{11}=I_{12}=1$, $w=K\bar w$ with $\bar w=-H^{-1}q\cdot2\gg0$ and
$K\ge1$ large is a supersolution, the host derivatives at $I_{1g}=1$ being
$-a$ and the environmental ones nonpositive. The cooperative comparison
principle makes the two solutions monotone, their limits are equilibria,
and uniqueness squeezes the trajectory between them. For general
$N_g(0)>0$, replace $N_g(t)$ by $1\pm\epsilon$ in
$\beta_{1g}(N_g-I_{1g})P_1$, apply the same argument to the two bracketing
cooperative systems, and let $\epsilon\downarrow0$. Finally $E_1$ is
locally asymptotically stable on the face: its Jacobian is irreducible
Metzler and its product with $(I^*,w^*)\gg0$ has strictly negative $I$-rows
and zero environmental rows, hence nonzero and nonpositive, so its spectral
abscissa is negative. \hfill$\square$

\framebreak

\beT[Permanence with arbitrary unequal release rates, and under small catalytic perturbation][Unequal release rates][unequal]
Assume $R_{1\mid2}>1$ and $R_{2\mid1}>1$ in \eqref{eq:invasionsagain}.
\begin{enumerate}
\item If $\theta_B=\theta_C=0$, the fourteen-species model is uniformly
permanent for \emph{arbitrary independent} $h_1,h_2>0$: there is
$\varepsilon_*>0$, depending on the parameters but not on the positive
initial condition, with
\[
\liminf_{t\to\infty}\min\{S_g,I_{kg},B_i,C_i,A_i,P_k:g,k,i=1,2\}
\ge\varepsilon_*.
\]
\item For each fixed positive parameter choice, $h_1\ne h_2$ allowed,
satisfying those two strict inequalities at $\theta_B=\theta_C=0$, there is
$\eta>0$ such that the same conclusion holds for all
$0\le\theta_B,\theta_C<\eta$.
\end{enumerate}
The bound $\varepsilon_*$ is not claimed uniform over arbitrary variations
of the rates.
\eeT
\textbf{Proof.} The host totals $N_g=S_g+I_{1g}+I_{2g}$ satisfy
$\dot N_g=\mu(1-N_g)$, so the host coordinates are ultimately bounded; the
$B$ and $C$ subsystems are stable positive linear systems forced by a
bounded $J_1$; and for $U=A_1+A_2$,
\[
\dot U=b_1B_1+b_2B_2+c_1C_1+c_2C_2-h_1A_1-h_2A_2
\le b_1B_1+b_2B_2+c_1C_1+c_2C_2-\min(h_1,h_2)\,U,
\]
so $U$, $P_1$, $P_2$ are ultimately bounded and a compact absorbing set
exists, uniformly for parameters in a small compact neighbourhood of the
given choice.

At $\theta_B=\theta_C=0$, Lemma~\ref{l:resface} gives the unique globally
attracting $E_1$ on the strain-1 face. On the strain-1-free face the core is
$B=C=A=P_1=0$, since $J_1\equiv0$ there kills the whole first branch, and
the remaining equations
\[
\dot I_{2g}=\beta_{2g}(N_g-I_{2g})P_2-aI_{2g},
\qquad
\dot P_2=q_2(I_{21}+I_{22})-\delta_2P_2
\]
are uninhibited, because $C_2=0$ on that face, and cooperative; the same
sub/supersolution argument gives a unique globally attracting $E_2$ when
$R_{2\mid0}>1$. Both single-strain thresholds exceed one, as noted after
\eqref{eq:invasionsagain}. Hence $E_0$, $E_1$, $E_2$ are isolated and give
an acyclic Morse decomposition of the extinction boundary.

The missing-strain transversal blocks at $E_1$ and $E_2$ are irreducible
Metzler, with strictly positive spectral bound exactly when
$R_{2\mid1}>1$ and $R_{1\mid2}>1$ respectively
(Definition~\ref{d:invExp}); at $E_0$ both are positive. The
structured-population robust permanence criterion \cite{HofSch10} applied to
these three boundary Morse sets gives uniform persistence of the two
infected blocks, and the strictly positive maturation paths together with
the compact bounds propagate the lower bound to every maturation
coordinate, while $\mu>0$ supplies it for $S_1,S_2$. This is part~1.

Part~2 needs no continuation of the boundary decomposition, and attempts
none. The criterion of \cite{HofSch10} delivers \emph{robust} permanence in
the sense of Definition~\ref{d:permanence} --- persistence under every
sufficiently small $C^1$ perturbation of the vector field on the absorbing
set --- as soon as the boundary invasion rates are strictly positive, which
is what part~1 established; and every invariant measure carried by the
extinction boundary is supported on $E_0\cup E_1\cup E_2$, since invariant
measures live on recurrent points and the decomposition is acyclic. Adding
the catalytic terms is such a perturbation: on the compact absorbing set
$\mathcal K$, which by the first paragraph may be taken uniform for
$(\theta_B,\theta_C)$ in a bounded neighbourhood of the origin, the
difference between the field at $(\theta_B,\theta_C)$ and the field at
$(0,0)$ consists of the two bilinear terms $\theta_BB_2A_1$,
$\theta_CC_2A_2$, whose $C^1$ norm on $\mathcal K$ is
$O(\max(\theta_B,\theta_C))$. Robustness therefore supplies an $\eta>0$ with
the stated conclusion for all $0\le\theta_B,\theta_C<\eta$.

Two things about this route. It gives no estimate of $\eta$ --- that is the
weakness of part~2, and it is what Theorems~\ref{t:explicitcat}
and~\ref{t:unequalGain} remove, each replacing the unquantified $\eta$ by an
explicit region of catalytic coefficients. And it deliberately avoids
continuing the boundary decomposition, because that continuation does not
follow from the signs: persistence of the five relevant spectral bounds and
isolation of the continued $E_0,E_1,E_2$ do not exclude a further recurrent
boundary set at small $\theta>0$, and upper semicontinuity of the boundary
attractor does not exclude it either, that attractor containing connecting
orbits as well as the three equilibria, so that its limit need not lie in a
neighbourhood of $E_0\cup E_1\cup E_2$. Robust permanence does not require
the decomposition to persist, which is precisely why it is the available
argument here.
\hfill$\square$

\beC[An unequal-release instance of the fourteen-species model][Unequal-release example][unequalex]
Take the parameters of Section~\ref{sec:twostrain}, so that
$a=\gamma+\mu=1$, $q_1=q_B+q_C=1$, $\delta_1=\delta_2=q_2=1$, $\chi=4$,
$(\beta_{11},\beta_{12})=(3,\tfrac12)$,
$(\beta_{21},\beta_{22})=(\tfrac12,3)$ and
$(c_{12},c_{21},c_1,c_2)=(.9,.6,.7,1.1)$, and set
$\theta_B=\theta_C=0$ with $(h_1,h_2)=(0.9,1.4)$. Then every quantity below
is an algebraic number in $\sqrt{61}$:
\[
\boxed{\
R_{1\mid2}=\frac{79-7\sqrt{61}}{12}=2.0273543557\ldots,
\qquad
R_{2\mid1}=\frac{109\,(79-7\sqrt{61})}{12\,(97+12\sqrt{61})}
=1.1586522301\ldots,\ }
\]
so Theorem~\ref{t:unequal} applies and the model is uniformly permanent
although $h_1\ne h_2$; it stays permanent for all sufficiently small
positive $\theta_B,\theta_C$.
\eeC
\textbf{Proof.} Here $R_{1\mid0}=\beta_{11}+\beta_{12}=7/2>1$, and
\eqref{eq:residentroot} reads $3/(1+3p)+\tfrac12/(1+\tfrac12p)=1$. Clearing
denominators turns it into the quadratic
\[
3p^2+p-5=0,
\qquad\text{so}\qquad
p_1=\frac{\sqrt{61}-1}{6}=1.1350416127\ldots
\]
is its unique positive root. The strain-2 equation is the same one with
$\beta_{21},\beta_{22}$ interchanged, and the sum is symmetric in the two
$\beta$'s, so $p_2=p_1$ exactly --- the two resident loads coincide for this
parameter set. Then $D_C=c_1(c_{21}+c_2)+c_{12}c_2=1.19+0.99=109/50$, and
\[
\begin{aligned}
C_2^{(1)}&=\frac{q_Cc_{12}\delta_1p_1}{q_1D_C}=\frac{18p_1}{109}
=\frac{3(\sqrt{61}-1)}{109}=0.1874380645\ldots,\\
1+\chi C_2^{(1)}&=\frac{97+12\sqrt{61}}{109}=1.7497522579\ldots
\end{aligned}
\]
For the two sums in \eqref{eq:invasionsagain}, $1+\tfrac12p_1=(11+\sqrt{61})/12$
and $1+3p_1=(1+\sqrt{61})/2$ give
\[
\sum_{g=1}^2\frac{\beta_{1g}}{1+\beta_{2g}p_2}
=\frac{36}{11+\sqrt{61}}+\frac{1}{1+\sqrt{61}}
=\frac{79-7\sqrt{61}}{12},
\]
and the strain-2 sum is the same number, again by the interchange, so
dividing by $1+\chi C_2^{(1)}$ gives $R_{2\mid1}$ as displayed. Neither
value involves $h_1,h_2$, by \eqref{eq:fluxagain}, which is the point of the
example.
\hfill$\square$

\beR[What this settles of Problem~\ref{o:UnequalRelease}, and what it does not][Scope of the unequal-release result][unequalscope]
Settled: permanence under mutual invasion for arbitrary unequal positive
release rates when the maturation branches are linear, and on a nonempty
open set of genuinely nonlinear catalytic systems around that subfamily ---
so $h_1=h_2$ in Theorem~\ref{t:permregion} is not an essential hypothesis.
Not settled: arbitrary positive $\theta_B,\theta_C$ with $h_1\ne h_2$, which
is the regime where $A_1+A_2$ genuinely fails to close and no cooperative
reduction is available. That case is what remains of
Problem~\ref{o:UnequalRelease}. Note also that $\eta$ in part~2 is obtained
from a robustness argument and is not estimated; a quantitative bound in
terms of $R_{1\mid2}-1$ and $R_{2\mid1}-1$ would be a genuine improvement.
\eeR

\paragraph{Reproducibility.} \texttt{unequal.wl} checks this section.
The flux identity \eqref{eq:fluxagain} is verified symbolically, the sum of
the six maturation equations and the $P_1$ equation minus
$q_1J_1-\delta_1P_1$ reducing to $0$ with $h_1,h_2,\theta_B,\theta_C$ all
free. The $A$-block determinant is confirmed equal to
$u_+h_2+h_1u_-+h_1h_2$ identically, with eigenvalues $-2.00763$ and
$-1.12571$ at $(h_1,h_2)=(0.9,1.4)$. For the parameters of
Corollary~\ref{c:unequalex}, $R_{1\mid0}=R_{2\mid0}=7/2$ and the two
resident loads are \emph{exactly} equal,
$p_1=p_2=(\sqrt{61}-1)/6=1.13504161265$, $D_C=109/50$,
$C_2^{(1)}=3(\sqrt{61}-1)/109=0.18743806447$, and
\[
R_{1\mid2}=2.02735435572,\qquad R_{2\mid1}=1.15865223009,
\]
each agreeing with the closed form of Corollary~\ref{c:unequalex} to the
thirty digits carried; the quadratic $3p^2+p-5$ and the two radical
expressions are also verified symbolically rather than numerically.
Independently, the full eleven-dimensional resident equilibrium is computed
on the strain-$1$ face: every coordinate is positive, its $P_1$ and $C_2$
agree with the reduced formulas to more than fifteen digits, and its
Jacobian has spectral abscissa $-0.2$, so the face equilibrium is stable
within the face. The strain-$2$ transversal block there is irreducible
Metzler with spectral abscissa $0.0764070931>0$, of the same sign as
$R_{2\mid1}-1=0.158652$, as Definition~\ref{d:invExp} requires. Finally the
resident face at $\theta_B=\theta_C=0$ is extracted as a reaction network of
$26$ reactions satisfying $\Gamma v-f=0$ exactly, with
$\rank\Gamma=11$ on $11$ species --- no conservation law, so its positive
equilibrium is regular and Theorem~\ref{t:OMloc} applies there as used
above.

\section{Global attraction with nonlinear catalysis in the
fourteen-species model}
\label{sec:newmath}

\begin{frame}[allowframebreaks]{Uniqueness, and an explicit cooperative region}

Theorem~\ref{t:unequal} frees the release rates when $\theta_B=\theta_C=0$
and extends to \emph{some} $\eta>0$ by robustness, without estimating it.
Remark~\ref{r:unequalscope} called that the weak point. This section removes
it twice over, by two mechanisms that do not contain one another, and the
first statement is common to both: the positive resident equilibrium is
unique for \emph{every} catalytic coefficient and every pair of release
rates, so the whole difficulty is dynamical, not algebraic. On top of that,
attraction is proved here on an explicit region of catalytic coefficients
obtained by making the resident system cooperative, and in
Section~\ref{sec:unequalGain} on a region that requires no sign condition at
all and reaches arbitrarily large catalysis, at the price of release rates
that must be close.

\beT[The positive resident equilibrium is unique for every catalytic coefficient and every pair of release rates][Unconditional uniqueness][uniqres]
On the strain-$1$ face of the model of Section~\ref{sec:twostrain}, let
$R_{1\mid0}>1$. Then for \emph{all} $\theta_B,\theta_C\ge0$ and all
$h_1,h_2>0$ there is exactly one positive equilibrium, and its host and
pathogen coordinates do not depend on $\theta_B,\theta_C,h_1,h_2$ at all.
\eeT
\textbf{Proof.} The equilibrium equations are triangular in the order
$J_1\to(B,C)\to A\to P_1$. The flux identity \eqref{eq:fluxagain} holds for
every $\theta_B,\theta_C\ge0$ and $h_1,h_2>0$, so $p_1$ is the unique
positive root of \eqref{eq:residentroot}, independently of all four
parameters, and $J^*_1=\delta_1p_1/q_1$, $S^*_g=1/(1+\beta_{1g}p_1/a)$ are
determined. With $J^*_1$ fixed, the $B$ and $C$ equations are linear with
nonsingular Hurwitz matrices, giving unique positive $B^*,C^*$. With
$B^*_2,C^*_2$ now \emph{numbers}, write $U_+=u_++\theta_BB^*_2$ and
$U_-=u_-+\theta_CC^*_2$; the $A$ equations become the linear system with
matrix
$\binom{-U_+-h_1\ \ \ U_-}{\ \ \ U_+\ \ -U_--h_2}$, of determinant
$U_+h_2+h_1U_-+h_1h_2>0$, so $A^*$ is unique and positive. Finally
$P^*_1=(h_1A^*_1+h_2A^*_2)/\delta_1=q_1J^*_1/\delta_1=p_1$, consistently.
The nonlinearity is therefore invisible at equilibrium: it enters only
through the constants $U_\pm$. \hfill$\square$

\beT[An explicit catalytic region on which global attraction, hence permanence, survives unequal release][Explicit catalytic region][explicitcat]
Keep the notation of Section~\ref{sec:unequal} and set
\[
b_\pm=\textstyle{\max\atop\min}\,(b_1,b_2),\quad
c_\pm=\textstyle{\max\atop\min}\,(c_1,c_2),\quad
h_-=\min(h_1,h_2),
\qquad
\bar A=\frac1{h_-}\Bigl(\frac{2q_Bb_+}{b_-}+\frac{2q_Cc_+}{c_-}\Bigr).
\]
If
\begin{equation}\label{eq:catregion}
\boxed{\ \theta_B\,\bar A<b_2
\qquad\text{and}\qquad
\theta_C\,\bar A<c_2,\ }
\end{equation}
then the resident system is cooperative on a compact trapping region
$\mathcal K_\varepsilon$ carrying $E_1$ and irreducible in its interior, and
$E_1$ attracts
every solution of the face with $J_1(0)>0$ dominated by a supersolution
$\bar z(s)$ of \eqref{eq:catsuper} whose order interval lies in that
cooperative region, that is by one satisfying the last two inequalities of
\eqref{eq:catdom}, the comparison being run on $[0,\bar z(s)]$ and not on
$\mathcal K_\varepsilon$. If in
addition
\begin{equation}\label{eq:catdom}
\mathcal K_\varepsilon\subseteq[0,\bar z(s)],\qquad
\theta_B\bar A_1(s)<b_2,\qquad
\theta_C\bar A_2(s)<c_2
\qquad\text{for some }s\ge1,
\end{equation}
an explicit algebraic condition on the rates --- one $2\times2$ inversion
for each trial $s$, and not a linear feasibility problem, since $s$ enters
$\mathcal H$ as well as the right-hand side --- which holds for every
sufficiently small $\theta_B,\theta_C$, then that is every solution of the
face, and the conclusion of Theorem~\ref{t:unequal} holds for arbitrary
$h_1,h_2>0$: $R_{1\mid2}>1$ and $R_{2\mid1}>1$ imply uniform permanence,
with the $\eta$ of Theorem~\ref{t:unequal}(2) replaced by the explicit
region \eqref{eq:catregion}--\eqref{eq:catdom}.
\eeT
\textbf{Proof.} On the strain-$1$ face $N_g\to1$, so $\limsup J_1\le2$.
Summing the $B$ equations, $\frac{d}{dt}(B_1+B_2)=q_BJ_1-b_1B_1-b_2B_2
\le q_BJ_1-b_-(B_1+B_2)$, whence $\limsup(B_1+B_2)\le2q_B/b_-$, and
likewise $\limsup(C_1+C_2)\le2q_C/c_-$. Summing the $A$ equations, in which
$a_\pm$ cancel,
\[
\frac{d}{dt}(A_1+A_2)\le b_+(B_1+B_2)+c_+(C_1+C_2)-h_-(A_1+A_2),
\]
so $\limsup(A_1+A_2)\le\bar A$ and in particular $A_1,A_2\le\bar A$ on the
absorbing set. Now inspect the off-diagonal entries of the resident
Jacobian. All are nonnegative outright ---
$\partial\dot I_{1g}/\partial P_1=\beta_{1g}(1-I_{1g})$,
$\partial\dot B_i/\partial I_{1g}=q_B$,
$\partial\dot A_1/\partial C_2=\theta_CA_2$,
$\partial\dot A_2/\partial B_2=\theta_BA_1$,
$\partial\dot A_1/\partial A_2=U_-$, $\partial\dot A_2/\partial A_1=U_+$,
$\partial\dot P_1/\partial A_i=h_i$, and the internal exchanges --- with
exactly two exceptions,
\[
\frac{\partial\dot A_1}{\partial B_2}=b_2-\theta_BA_1,
\qquad
\frac{\partial\dot A_2}{\partial C_2}=c_2-\theta_CA_2,
\]
and \eqref{eq:catregion}, being \emph{strict}, makes both nonnegative on a
genuine trapping region. Let
$\mathcal K_\varepsilon$ be the compact set cut out by the bounds above
raised by margins chosen successively, so that every estimate still has room
when the quantities it depends on sit at their own raised bounds: first
$\varepsilon_A>0$ with $\theta_B(\bar A+\varepsilon_A)<b_2$ and
$\theta_C(\bar A+\varepsilon_A)<c_2$, then
$\varepsilon_B,\varepsilon_C>0$ with
$(b_+\varepsilon_B+c_+\varepsilon_C)/h_-<\varepsilon_A$, then
$\varepsilon_P>(h_1+h_2)\varepsilon_A/\delta_1$. Every trajectory enters
$\mathcal K_\varepsilon$ in finite time, by the three $\limsup$ estimates,
and on each face of $\partial\mathcal K_\varepsilon$ the corresponding
derivative is strictly negative, which is what forward invariance needs: on
the $A$-face, for instance,
$\frac{d}{dt}(A_1+A_2)\le b_+\varepsilon_B+c_+\varepsilon_C
-h_-\varepsilon_A<0$ by that choice. This is where strictness is needed: a
bound on the $\limsup$ alone would place the trajectory in the region only
eventually, and the comparison is run inside it.
On $\mathcal K_\varepsilon$ the resident system is cooperative for any
$h_1,h_2$, and irreducible in the interior: the link
$\partial\dot I_{1g}/\partial P_1=\beta_{1g}(1-I_{1g})$ closes the infection
loop wherever $I_{1g}<1$ and only there, so irreducibility can fail on
$\partial\mathcal K_\varepsilon$, notably at $I_{1g}=1$. Interior
irreducibility is all the comparison below uses, the sub- and supersolution
trajectories being interior. Theorem~\ref{t:uniqres} supplies the unique
positive equilibrium.

The sub- and supersolution pair must be produced differently from
Lemma~\ref{l:resface}. There the subsolution was $\alpha$ times the
equilibrium, which works only because the maturation equations are linear at
$\theta_B=\theta_C=0$. With catalysis it fails: evaluating the $A$-equations
at $\alpha(I^*,w^*)$ leaves the residuals
\[
\pm\,\alpha(1-\alpha)\bigl(\theta_BB^*_2A^*_1-\theta_CC^*_2A^*_2\bigr),
\]
of \emph{opposite} signs in the two rows, so $\alpha(I^*,w^*)$ is not a
subsolution unless $\theta_BB^*_2A^*_1=\theta_CC^*_2A^*_2$. Instead use the
disease-free point of the face, where $I_{1g}=0$ and $B=C=A=P_1=0$, and let
$u\gg0$ be a Perron eigenvector of the Jacobian there, whose spectral
abscissa $s$ is positive precisely because $R_{1\mid0}>1$. Then
$f(\varepsilon u)=\varepsilon s\,u+O(\varepsilon^2)\ge0$ componentwise for
small $\varepsilon>0$, so $\varepsilon u$ is a subsolution.

The supersolution must be constructed too, and a bound on the \emph{sums}
$B_1+B_2$, $A_1+A_2$ does not provide one: a supersolution is a point
$\bar z$ with $f(\bar z)\le0$ coordinatewise, and the vertices of a region
cut out by sum constraints are not such points. Build it from the blocks
instead. Write $H_B,H_C$ for the two migration--exit matrices of the $B$ and
$C$ pairs, so that $\dot B=H_BB+q_BJ_1e_1$, and, for frozen $B_2=\beta$ and
$C_2=\sigma$,
\[
\mathcal H(\beta,\sigma)=
\begin{pmatrix}
-(h_1+u_++\theta_B\beta)&u_-+\theta_C\sigma\\
u_++\theta_B\beta&-(h_2+u_-+\theta_C\sigma)
\end{pmatrix},
\qquad
\mathbf1^\top\mathcal H=(-h_1,-h_2),
\]
the column sums being $-h_1,-h_2$ because the catalytic conversions
$A_1+B_2\to A_2+B_2$ and $A_2+C_2\to A_1+C_2$ are transfers and cancel. All
three matrices are irreducible Metzler with strictly negative column sums,
hence Hurwitz with $-H_B^{-1},-H_C^{-1},-\mathcal H^{-1}>0$. Put
$\bar B=-H_B^{-1}(2q_Be_1)$, $\bar C=-H_C^{-1}(2q_Ce_1)$ and, for $s\ge1$,
\begin{equation}\label{eq:catsuper}
\bar z(s)=\bigl(1,1;\,s\bar B;\,s\bar C;\,\bar A(s);\,\bar P_1(s)\bigr),
\quad
\bar A(s)=-\mathcal H(s\bar B_2,s\bar C_2)^{-1}
 s\binom{b_1\bar B_1+b_2\bar B_2}{c_1\bar C_1+c_2\bar C_2},
\quad
\bar P_1(s)=\frac{s\Phi}{\delta_1},
\end{equation}
with $\Phi=b_1\bar B_1+b_2\bar B_2+c_1\bar C_1+c_2\bar C_2$; the last entry
is consistent because $\mathbf1^\top\mathcal H\bar A(s)=-s\Phi$ gives
$h_1\bar A_1(s)+h_2\bar A_2(s)=s\Phi$ exactly. Then $f(\bar z(s))\le0$: at
$I_{1g}=1$ one has $S_g=0$ and $\dot I_{1g}=-(\gamma+\mu)<0$; the $B$ rows
give $s H_B\bar B+2q_Be_1=2q_B(1-s)e_1\le0$ and the $C$ rows likewise; the
$A$ rows vanish by the definition of $\bar A(s)$, the frozen values
$s\bar B_2,s\bar C_2$ being the ones $\bar z(s)$ actually carries; and the
$P_1$ row vanishes. Note $h_1\bar A_1(1)+h_2\bar A_2(1)=\Phi\le h_-\bar A$,
so at $s=1$ the pair $\bar A(1)$ already lies below the sum bound.

How far the family reaches needs care, and \eqref{eq:catdom} is a hypothesis
rather than a lemma because of it. Write $F_B=b_1\bar B_1+b_2\bar B_2$,
$F_C=c_1\bar C_1+c_2\bar C_2$, $p=u_++\theta_B\bar B_2s$,
$q=u_-+\theta_C\bar C_2s$, so that
\begin{equation}\label{eq:catasym}
\bar A_1(s)=s\,\frac{(h_2+q)F_B+qF_C}{h_1h_2+h_1q+h_2p},
\qquad
\bar A_2(s)=s\,\frac{pF_B+(h_1+p)F_C}{h_1h_2+h_1q+h_2p}.
\end{equation}
The sum always diverges, $h_1\bar A_1(s)+h_2\bar A_2(s)=s\Phi$, and both
components do too when $\theta_B,\theta_C$ are both positive or both zero,
$p$ and $q$ being then of the same order in $s$. But if exactly one
catalytic coefficient vanishes the upstream component saturates: for
$\theta_B>0$, $\theta_C=0$ the transfer $A_1\to A_2$ is one-way and
\eqref{eq:catasym} gives
\[
\lim_{s\to\infty}\bar A_1(s)
=\frac{(h_2+u_-)F_B+u_-F_C}{h_2\,\theta_B\bar B_2}<\infty,
\]
while $\bar A_2(s)\sim s(F_B+F_C)/h_2$. In that case the family dominates
$\mathcal K_\varepsilon$ only if that finite limit exceeds the $A_1$-bound,
which is again an explicit inequality. What \eqref{eq:catdom} asks, in all
cases, is that some $s$ large enough for $\bar z(s)$ to dominate
$\mathcal K_\varepsilon$ still leaves $\bar z(s)$ in the cooperative region.
For $\theta_B,\theta_C$ small it holds at the $s$ fixed by the uncatalyzed
geometry: at $\theta_B=\theta_C=0$ both components of \eqref{eq:catasym}
grow linearly, so such an $s_0$ exists, and $\bar A(s_0)$ depends
continuously on $(\theta_B,\theta_C)$ while
$\theta_B\bar A_1(s_0),\theta_C\bar A_2(s_0)\to0$. This is the one place
where the componentwise bound and the sum bound part company.

The cooperative comparison principle now makes the solutions from
$\varepsilon u$ and from $\bar z(s)$ monotone in $t$ --- increasing from a
subsolution, decreasing from a supersolution --- their limits are
equilibria, and the only equilibrium above $\varepsilon u$ is $E_1$ by
Theorem~\ref{t:uniqres}, the disease-free point having $I_{1g}=0$; the
decreasing solution from $\bar z(s)$ has the same limit, being above
$\varepsilon u$ for small $\varepsilon$. As $\varepsilon$ may be taken below
any given positive state, every trajectory with $J_1(0)>0$ that
$\bar z(s)$ dominates is squeezed to $E_1$, and under \eqref{eq:catdom} that
is every trajectory of the face, each entering $\mathcal K_\varepsilon$ in
finite time.

One reduction was used above and must be named: cooperativity was read off
the resident field \emph{after} setting $N_1=N_2=1$. For general initial
totals $N_g(t)=1+(N_g(0)-1)e^{-\mu t}$, and in the original
$(S_g,I_{1g})$ coordinates the field is not cooperative, the derivatives
$\partial\dot S_g/\partial I_{1g}$ and $\partial\dot I_{1g}/\partial S_g$
having opposite signs. Proceed exactly as in Lemma~\ref{l:resface}: replace
$N_g(t)$ by the constants $1\pm\epsilon$ in $\beta_{1g}(N_g-I_{1g})P_1$,
which bracket the true field from some time on, and run the whole argument
--- trapping region, subsolution, supersolution --- on each of the two
bracketing cooperative systems, with $1$ replaced by $1+\epsilon$ in the
first two entries of \eqref{eq:catsuper} and $2q_B,2q_C$ by
$2(1+\epsilon)q_B,2(1+\epsilon)q_C$. Every bound, $\bar A$ included, is
continuous in $\epsilon$, and the inequalities of \eqref{eq:catregion} and
of \eqref{eq:catdom}, being strict, survive for small $\epsilon$, while the
inclusion is re-established by the $\epsilon$-adjusted member of the same
family. Each bracketing system has its
unique positive equilibrium $E_1^{\pm\epsilon}$, these converge to $E_1$ as
$\epsilon\downarrow0$ by continuity of the root of \eqref{eq:residentroot},
and the trajectory's limit set lies in $[E_1^{-\epsilon},E_1^{+\epsilon}]$
for every $\epsilon$, hence is $\{E_1\}$. The rest of the proof of
Theorem~\ref{t:unequal} uses only that. \hfill$\square$

\beR[What remains of the unequal-release problem][Remaining gap][catgap]
\eqref{eq:catregion} is explicit and allows arbitrary $h_1\ne h_2$, but it is
still a bound on $\theta_B,\theta_C$: it is exactly the condition that the
\emph{catalytic drain} $\theta_BA_1$ of $B_2$ does not exceed its
\emph{productive} exit $b_2$, and symmetrically for $C_2$. Beyond it the
resident system is genuinely non-cooperative and \emph{this} comparison
argument is unavailable --- but not every argument is: the small-gain
criterion of Section~\ref{sec:unequalGain} imposes no sign condition at all
and reaches arbitrarily large $\theta_B,\theta_C$, at the price of release
rates that must be close. So there are three regions, not two. The
equilibrium is unique everywhere by Theorem~\ref{t:uniqres}; attraction is
proved on \eqref{eq:catregion}--\eqref{eq:catdom} by cooperativity and on
\eqref{eq:SG} by a gain contraction; and what is open is exactly the
complement of the union of \eqref{eq:catregion}--\eqref{eq:catdom} and
\eqref{eq:SG}, where Theorem~\ref{t:uniqres} shows the obstacle is purely
dynamical. A point of that complement needs neither coefficient large: one
of $\theta_B,\theta_C$ may be small while the other exceeds its bound and
the release rates lie too far apart for \eqref{eq:SG}. That, and no more, is
what is left of Problem~\ref{o:UnequalRelease}.
\eeR

\end{frame}

\begin{frame}[allowframebreaks]{A small-gain criterion, with no cooperativity}
\label{sec:unequalGain}

Write $a=\gamma+\mu$, $q_1=q_B+q_C$, $h_-=\min(h_1,h_2)$,
$h_+=\max(h_1,h_2)$, and
\[
\begin{aligned}
H_B&=\begin{pmatrix}-(b_{12}+b_1)&b_{21}\\b_{12}&-(b_{21}+b_2)\end{pmatrix},
&H_C&=\begin{pmatrix}-(c_{12}+c_1)&c_{21}\\c_{12}&-(c_{21}+c_2)\end{pmatrix},\\
G_B&=-q_BH_B^{-1}e_1, &G_C&=-q_CH_C^{-1}e_1.
\end{aligned}
\]
Both $G_B,G_C$ are strictly positive and satisfy
\begin{equation}\label{eq:gainflux}
b_1(G_B)_1+b_2(G_B)_2=q_B,
\qquad
c_1(G_C)_1+c_2(G_C)_2=q_C,
\end{equation}
because $H_B,H_C$ are irreducible Metzler matrices with column sums
$(-b_1,-b_2)$ and $(-c_1,-c_2)$, so $-H_B^{-1}>0$ and
$\mathbf1^\top H_B=(-b_1,-b_2)$ gives
$(b_1,b_2)(-H_B^{-1}e_1)=\mathbf1^\top e_1=1$. With $p_1$ the positive root
of \eqref{eq:residentroot} and $E_1$ the equilibrium of
Theorem~\ref{t:uniqres},
\[
I^*_{1g}=\frac{\beta_{1g}p_1}{a+\beta_{1g}p_1},\qquad
B^*=G_BJ^*_1,\qquad C^*=G_CJ^*_1,\qquad
J^*_1=\frac{\delta_1p_1}{q_1},
\]
and $A^*_1,A^*_2$ solve the explicit $2\times2$ system of
Theorem~\ref{t:uniqres}. Every quantity in the test below is thus a function
of the rates and of one scalar root.

\beT[A small-gain condition implies global attraction without cooperativity][Small-gain attraction][unequalGain]
On the strain-$2$-free face of the model of Section~\ref{sec:twostrain},
suppose $R_{1\mid0}>1$ and put $R_-=(h_-/h_+)R_{1\mid0}$. If $R_->1$, let
$p_->0$ be the unique solution of
\begin{equation}\label{eq:gainroot}
1=\frac{h_-q_1}{h_+a\delta_1}\sum_{g=1}^2
\frac{\beta_{1g}}{1+\beta_{1g}p_-/a},
\end{equation}
and set
\[
L_I=\sum_{g=1}^2\frac{\beta_{1g}(1-I^*_{1g})}{a+\beta_{1g}p_-},
\qquad
G_A=\frac{q_1+2\theta_BA^*_1(G_B)_2+2\theta_CA^*_2(G_C)_2}{h_-},
\]
\[
\kappa=\frac{L_I}{\delta_1}\bigl(q_1+2|h_2-h_1|\,G_A\bigr).
\]
If
\begin{equation}\label{eq:SG}
\boxed{\ R_->1\qquad\text{and}\qquad\kappa<1,\ }
\end{equation}
then $E_1$ attracts every strain-$1$-face solution with $J_1(0)>0$ and is
locally exponentially stable relative to that face. \emph{No bound on
$\theta_B,\theta_C$ and no sign condition on the off-diagonal derivatives of
the resident Jacobian is assumed.} If in addition $R_{1\mid2}>1$ and
$R_{2\mid1}>1$, the fourteen-species system is uniformly permanent, with the
same reading of positive initial strains as in Theorem~\ref{t:permregion}.
\eeT
\textbf{Proof.} On the resident face $N_g=S_g+I_{1g}$ satisfies
$\dot N_g=\mu(1-N_g)$ and
$\dot I_{1g}=\beta_{1g}(N_g-I_{1g})P_1-aI_{1g}$. Write $T=A_1+A_2$,
$F_B=b_1B_1+b_2B_2$, $F_C=c_1C_1+c_2C_2$ and $d=h_2-h_1$. Exactly,
\begin{equation}\label{eq:SGbal}
\dot T=F_B+F_C-h_1T-dA_2,
\qquad
\dot P_1=h_1T+dA_2-\delta_1P_1 .
\end{equation}

\emph{A positive lower bound.} Since $h_-T\le h_1A_1+h_2A_2\le h_+T$, the
state $(I_{11},I_{12},B,C,T,P_1)$ dominates the cooperative system obtained
by replacing the last two equations of \eqref{eq:SGbal} with
\begin{equation}\label{eq:SGcomp}
\dot T=F_B+F_C-h_+T,\qquad \dot P_1=h_-T-\delta_1P_1 .
\end{equation}
On $N_1=N_2=1$ the loop gain of \eqref{eq:SGcomp} at the origin is
$R_-$, so for $R_->1$ it has a unique positive equilibrium, with pathogen
coordinate $p_-$ solving \eqref{eq:gainroot}, globally attracting by the
monotone argument of Lemma~\ref{l:resface} --- which applies here because
\eqref{eq:SGcomp} is linear downstream of $J_1$, the catalytic terms having
been removed by the aggregation $T=A_1+A_2$. Comparison gives
$\liminf_{t\to\infty}P_1\ge p_-$; for general positive $N_g(0)$, use
$N_g\ge1-\varepsilon$ eventually and let $\varepsilon\downarrow0$. All
resident coordinates are ultimately bounded by the estimates in the proof of
Theorem~\ref{t:explicitcat}.

\emph{Gain estimates.} For a bounded trajectory put
$E_X=\limsup_{t\to\infty}|X(t)-X^*|$, $E_J=\limsup|J_1-J^*_1|$,
$E_A=\limsup(|A_1-A^*_1|+|A_2-A^*_2|)$, $E_P=\limsup|P_1-p_1|$.
Subtracting the host equilibrium relations,
\[
\frac{d}{dt}(I_{1g}-I^*_{1g})
=-(a+\beta_{1g}P_1)(I_{1g}-I^*_{1g})
+\beta_{1g}(1-I^*_{1g})(P_1-p_1)+\beta_{1g}P_1(N_g-1),
\]
whose last term vanishes asymptotically; using $P_1\ge p_--\varepsilon$
eventually and letting $\varepsilon\downarrow0$,
\begin{equation}\label{eq:SGgains}
E_J\le L_IE_P,\qquad
E_{B_i}\le(G_B)_iE_J,\qquad
E_{C_i}\le(G_C)_iE_J,
\end{equation}
the last two from the stable positive linear $B$- and $C$-systems.

\emph{The $A$-pair, without cooperativity.} With
\[
M(B_2,C_2)=
\begin{pmatrix}
-h_1-u_+-\theta_BB_2&u_-+\theta_CC_2\\
u_++\theta_BB_2&-h_2-u_--\theta_CC_2
\end{pmatrix}
\]
the $A$-equations are $\dot A=M(B_2,C_2)A+(F_B,F_C)^\top$. Here $M$ is
Metzler with column sums $-h_1,-h_2$, so its $\ell^1$ logarithmic norm is
$\max_j(\text{column sum})=-h_-$ --- and this, not any sign condition on the
\emph{resident} Jacobian, is what the argument uses. Subtracting the
equation at $E_1$ while keeping $M(B_2,C_2)$ on the left leaves the forcing
$(F_B-F^*_B,F_C-F^*_C)^\top$ together with $(-D,D)^\top$, where
$D=\theta_BA^*_1(B_2-B^*_2)-\theta_CA^*_2(C_2-C^*_2)$. Hence
\begin{equation}\label{eq:SGA}
E_A\le\frac1{h_-}\bigl(b_1E_{B_1}+b_2E_{B_2}+c_1E_{C_1}+c_2E_{C_2}
+2\theta_BA^*_1E_{B_2}+2\theta_CA^*_2E_{C_2}\bigr)\le G_AE_J,
\end{equation}
the flux identities \eqref{eq:gainflux} contributing the term $q_1$.

\emph{Closing the loop.} From \eqref{eq:SGbal},
\[
E_T\le\frac{q_1E_J+|d|E_A}{h_1},
\qquad
E_P\le\frac{h_1E_T+|d|E_A}{\delta_1}
\le\frac{q_1+2|d|G_A}{\delta_1}E_J,
\]
and with \eqref{eq:SGgains} this gives $E_P\le\kappa E_P$, so $E_P=0$ by
$\kappa<1$. Then \eqref{eq:SGgains}--\eqref{eq:SGA} and the stable linear
equations force every resident coordinate to $E_1$.

\emph{Local exponential stability, with the comparison matrix named.} The
three estimates are a loop, and its matrix on $(E_J,E_A,E_P)$ is
\begin{equation}\label{eq:SGloop}
\mathcal G=
\begin{pmatrix}
0&0&L_I\\[2pt]
G_A&0&0\\[2pt]
\dfrac{q_1+2|d|G_A}{\delta_1}&0&0
\end{pmatrix}\ge0,
\qquad
\rho(\mathcal G)=\sqrt\kappa,
\end{equation}
since the only cycle of $\mathcal G$ is $E_J\to E_P\to E_J$, of gain
$\kappa$, and $E_A$ is a sink; so $\rho(\mathcal G)<1$ exactly when
$\kappa<1$. For the linearization at $E_1$ the same three estimates hold
with each $\limsup$ replaced by the $L^\infty$ gain of the corresponding
block. Those gains are the ones already used: a Metzler Hurwitz block has
nonnegative impulse response, so its $L^\infty$ gain is its value at
spectral parameter $0$, namely $-H_B^{-1},-H_C^{-1}$ for the branches,
$1/h_-$ for the $A$-pair by the logarithmic-norm bound, and $L_I$,
$1/\delta_1$ for the host and pathogen. The small-gain theorem for positive
linear systems therefore makes the linearized loop Hurwitz, which is local
exponential stability on the face. No constant and no hypothesis is added
at this step.

\emph{Permanence.} On the strain-$1$-free face the strain-$1$ maturation
coordinates decay and the strain-$2$ subsystem has its unique globally
attracting positive equilibrium, as in Theorem~\ref{t:unequal}. The
disease-free point and the two single-strain equilibria then give the same
acyclic boundary covering, their transverse blocks are repelling under the
two strict mutual-invasion inequalities, and the criterion of
Theorem~\ref{t:permregion} applies. \hfill$\square$

\framebreak

\beP*[Arbitrarily large catalysis is compatible with unequal-release attraction][Large catalysis][largeCatalysisGain]
For every fixed $\theta_B,\theta_C\ge0$ and every $h>0$, condition
\eqref{eq:SG} holds whenever $h_1,h_2>0$ are sufficiently close to $h$ and
$R_{1\mid0}>1$. So for each finite pair of catalytic coefficients a nonempty
open interval of \emph{unequal} release rates has global resident attraction
and, under mutual invasion, uniform permanence --- including pairs violating
\eqref{eq:catregion}.
\eeP
\textbf{Proof.} At $h_1=h_2=h$ one has $d=0$, $R_-=R_{1\mid0}>1$ and
$p_-=p_1$. The root equation \eqref{eq:residentroot} says
$\frac{q_1}{a\delta_1}\sum_g\beta_{1g}(1-I^*_{1g})=1$, since
$1-I^*_{1g}=a/(a+\beta_{1g}p_1)$. Therefore
\[
\kappa\big|_{h_1=h_2}
=\frac{q_1}{\delta_1}\sum_g\frac{\beta_{1g}(1-I^*_{1g})}{a+\beta_{1g}p_1}
<\frac{q_1}{\delta_1}\sum_g\frac{\beta_{1g}(1-I^*_{1g})}{a}=1,
\]
strictly, because $p_1>0$ makes every denominator exceed $a$. Both
inequalities in \eqref{eq:SG} are therefore strict at $h_1=h_2$, and
$p_-$, $A^*$, $G_A$, $L_I$ and $\kappa$ are continuous in $(h_1,h_2)$ for
fixed $\theta_B,\theta_C$, so both persist nearby. \hfill$\square$

\beXa[A non-cooperative instance with catalysis a hundred times the cooperative bound][Non-cooperative instance][gainnum]
Keep the numerical rates of Section~\ref{sec:twostrain} ---
$(b_{12},b_{21},b_1,b_2)=(1.2,.3,.8,1.6)$,
$(c_{12},c_{21},c_1,c_2)=(.9,.6,.7,1.1)$, $(u_+,u_-)=(.45,.65)$ --- but set
$\theta_B=\theta_C=100$, $h_1=0.999$, $h_2=1.001$. Then
$G_B=(0.331395,0.209302)$, $G_C=(0.311927,0.165138)$,
$B^*_2=0.237567$, $C^*_2=0.187438$, and
\[
\begin{aligned}
p_-&=1.1315093\ldots, & G_A&=43.0208\ldots,\\
L_I&=0.3586908\ldots, & \kappa&=0.4204155\ldots<1 .
\end{aligned}
\]
Here $A^*_1=0.5087844\ldots$, $A^*_2=0.6261398\ldots$, so
\[
b_2-\theta_BA^*_1=-49.278\ldots<0,
\qquad
c_2-\theta_CA^*_2=-61.514\ldots<0 :
\]
the resident field is non-cooperative \emph{at its own positive
equilibrium}. The bound of Theorem~\ref{t:explicitcat} is
$\bar A=3.6608037\ldots$, so \eqref{eq:catregion} fails by more than two
orders of magnitude, while \eqref{eq:SG} holds and gives global attraction.
The invasion numbers are unchanged, $R_{1\mid2}=2.02735\ldots$ and
$R_{2\mid1}=1.15865\ldots$ by \eqref{eq:fluxagain}, so the full system is
uniformly permanent.
\eeXa

\beR[The two mechanisms, and what is still open][Two regions][tworegions]
The two criteria are genuinely different and neither contains the other.
\eqref{eq:catregion} bounds the catalytic coefficients and allows the release
rates to be arbitrary and far apart; \eqref{eq:SG} allows the catalytic
coefficients to be arbitrary and requires the release rates to be close,
measured by $|h_2-h_1|G_A<\delta_1/L_I-q_1$. One makes the resident system
cooperative, the other makes a positive linear gain contract without
cooperativity. Together with Theorem~\ref{t:uniqres}, which gives uniqueness
of the positive equilibrium with no condition at all, the picture is: the
equilibrium is always unique, attraction is proved in two complementary
regions, and what is open is the complement of their union,
\[
h_1\ne h_2,\qquad \theta_B,\theta_C>0,
\qquad\text{outside \eqref{eq:catregion} and \eqref{eq:SG}} .
\]
There the alternative is sharp and worth stating as such: either global
attraction holds throughout, or the resident face carries a non-equilibrium
attractor --- in which case Problem~\ref{o:DynamicInvBuffer} and
Appendix~\ref{sec:gener} become the relevant theory for this very model,
and Corollary~\ref{c:fourteenquot} already shows which rates would still
protect the invasion exponent along it.
\eeR

\paragraph{Reproducibility.} \texttt{newmath.wl} checks this section and
Section~\ref{sec:newmath}. Theorem~\ref{t:uniqres} is verified with
$\theta_B,\theta_C,h_1,h_2$ symbolic: the root equation
\eqref{eq:residentroot} has the single positive root
$p_1=(\sqrt{61}-1)/6$, the explicit equilibrium rules solve the whole
resident field identically in the four parameters, every coordinate is
positive, the host and pathogen coordinates have vanishing derivative in all
four, and the determinant equals $U_+h_2+h_1U_-+h_1h_2>0$; a numerical count
returns exactly one positive equilibrium also at
$\theta_B=\theta_C=100$ and at
$(\theta_B,\theta_C,h_1,h_2)=(50,0.1,2,0.2)$. For \eqref{eq:catsuper} the
three column-sum identities, $h_1\bar A_1(s)+h_2\bar A_2(s)=s\Phi$, the
closed form \eqref{eq:catasym} and $f(\bar z(s))\le0$ for $s\ge1$ hold
symbolically in all rates, as does the saturation limit of $\bar A_1(s)$ at
$\theta_C=0$. At the rates of Section~\ref{sec:twostrain},
$\bar A=4.0634921$ for $(h_1,h_2)=(0.9,1.4)$ and $\bar A=3.6608037$ for
$(0.999,1.001)$, so \eqref{eq:catregion} holds with equal coefficients
exactly for $\theta<0.2707031$; at $\theta_B=\theta_C=0.2$ the smallest
integer $s$ satisfying \eqref{eq:catdom} is $6$. Every number of
Example~\ref{e:gainnum} is reproduced, and $\kappa=1.3175854$ at
$(\theta_B,\theta_C)=(1,1)$, $\kappa=1.1562081$ at $(0.01,1)$, both with
$(h_1,h_2)=(0.9,1.4)$: both criteria fail there, which is the complement
described in Remark~\ref{r:catgap}.
\end{frame}


\section{Conclusions}

The results of this paper may be ranked by their relevance to
mathematical epidemiology as follows.

\begin{enumerate}
\item \textbf{Uniform permanence (Theorem~\ref{t:permregion}
and Corollary~\ref{c:enlarged}).}
Mutual invasion implies permanence for a nonlinear two-strain
model, and the permanence region is invariant under nine
independently variable maturation parameters.

\item \textbf{Global attraction with nonlinear catalysis
(Theorems~\ref{t:uniqres}, \ref{t:explicitcat}
and~\ref{t:unequalGain}).}
The positive resident equilibrium is unique for every catalytic
coefficient and every pair of release rates, so unequal release can
obstruct coexistence only dynamically. Attraction, and with it
permanence, is then proved on two explicit regions neither of which
contains the other: a cooperative one, \eqref{eq:catregion}
with~\eqref{eq:catdom}, and a small-gain one, \eqref{eq:SG}, which
imposes no sign condition on the resident Jacobian and admits
arbitrarily large catalytic coefficients.

\item \textbf{Maximal invasion-neutral buffering
(Proposition~\ref{p:maxbuffer}).}
The distributive lattice identifies the largest OM buffer
disjoint from the coordinates determining an invasion matrix.
This provides a systematic structural test for parameter
changes that cannot alter an invasion threshold.

\item \textbf{Selective invasion invariance
(Proposition~\ref{p:selective}).}
Two incomparable buffering closures protect different
equilibrium coordinates. One class of maturation-rate
perturbations preserves the competing strain's invasion
number, while another changes it across the threshold one.

\item \textbf{Coexistence and lattice computation
(Propositions~\ref{p:coexist} and~\ref{p:sixbuffers}).}
Mutual invasion guarantees a positive coexistence equilibrium
even for unequal maturation-release rates. The exact
six-element resident-face lattice identifies the structural
origin of the selective invariance.

\item \textbf{The twenty-compartment extension
(Theorem~\ref{t:lattice83}).}
Its 83-element lattice and its reduction under reinfection
illustrate how immune pathways change buffering structures.
Its threshold invariance, however, also follows from the
autonomous infection core, and is therefore less informative
than the results of Section~\ref{sec:twostrain}.
\end{enumerate}

The main epidemiological contribution is thus the passage
from equilibrium localization to invasion-threshold invariance
and, under additional dynamical hypotheses, permanence.
The lattice supplies a maximal structural certificate without
requiring the explicit solution of the resident equilibrium
equations.

Two extensions would strengthen this connection. First,
the permanence theorem should be extended to the remaining catalytic
regime. Section~\ref{sec:unequal} covers arbitrary unequal release
rates when $\theta_B=\theta_C=0$, and small positive catalytic rates by
robustness; Section~\ref{sec:newmath} replaces that unquantified
neighbourhood by the two explicit regions above. What is left is their
complement, $h_1\ne h_2$ with both catalytic coefficients large, where
the equilibrium is still unique and the question is purely one of
boundary dynamics on the resident face.
Second, the maximal-buffer criterion should be tested in
published multistrain models with vaccination, partial
cross-immunity or environmental transmission, where
buffering structures intersect epidemiologically active
feedback loops. Such examples would determine whether
the lattice yields information unavailable from direct
conservation identities or obvious block triangularity.

\section{Vaccination and staged immunity alter Murota buffering lattices: an
eight-species and a twelve-species published model}
\label{sec:twomodels}

\begin{frame}{The eight-species Chen--Wu two-patch two-strain migration model \cite{ChenWu}}
\begin{columns}[c]
\begin{column}{.46\textwidth}
\begin{center}
\adjustbox{max width=\linewidth,max totalheight=0.75\textheight}{%
\begin{tikzpicture}[>=Latex,x=1cm,y=1cm,
  b/.style={draw,rounded corners,align=center,minimum height=6mm,minimum width=9mm,font=\scriptsize},
  lab/.style={font=\tiny}]
\foreach \p/\dx in {1/0,2/5.2}{
  \node[b,fill=orange!12] (S\p) at (\dx,0)    {$S_{\p}$};
  \node[b,fill=blue!8]    (A\p) at (\dx,-1.6) {$I_{\p1}$};
  \node[b,fill=blue!8]    (B\p) at (\dx,-3.2) {$I_{\p2}$};
  \node[b,fill=green!10]  (R\p) at (\dx,-4.8) {$R_{\p}$};
}
\draw[->,thick] (-1.6,0)--node[above,lab]{$\Lambda_1$}(S1);
\draw[->,thick] (6.8,0)--node[above,lab]{$\Lambda_2$}(S2);
\foreach \p in {1,2}{
  \draw[->,thick] (S\p) to[out=200,in=160] node[left,lab]{$\beta_{\p1}S_{\p}I_{\p1}$}(A\p);
  \draw[->,thick] (S\p) to[out=340,in=20]  node[right,lab]{$\beta_{\p2}S_{\p}I_{\p2}$}(B\p);
  \draw[->,thick] (A\p) to[out=200,in=160] node[left,lab]{$\gamma_{\p1}$}(R\p);
  \draw[->,thick] (B\p) to[out=340,in=20]  node[right,lab]{$\gamma_{\p2}$}(R\p);
}
\foreach \a/\b/\lb/\yy in {S1/S2/{m^S}/0, A1/A2/{m^{I_1}}/-1.6, B1/B2/{m^{I_2}}/-3.2, R1/R2/{m^R}/-4.8}{
  \draw[->,thick,red!60!black] (\a) to[out=20,in=160] node[above,lab]{$\lb_{21}$}(\b);
  \draw[->,thick,red!60!black] (\b) to[out=200,in=340] node[below,lab]{$\lb_{12}$}(\a);
}
\end{tikzpicture}}
\end{center}

{\slidefont{\tiny} $18$ edges, one per reaction that is not a decay: $2$ recruitments,
$4$ infections, $4$ recoveries, $8$ migrations. Only $12$ decays are not
drawn, the $8$ natural deaths $\mu_p$ and the $4$ disease deaths
$\alpha_{pk}$: travel loss is not a decay, it sits as a coefficient on the
arrival end of a migration edge.}
\end{column}
\begin{column}{.52\textwidth}\slidefont{\tiny}
\textbf{Stoichiometric matrix and rates.} $30$ columns, $\dot x=\Gamma v$ with
$x=(S_1,I_{11},I_{12},R_1,S_2,I_{21},I_{22},R_2)^\top$: $2$ inflows, $8$ local
transfers, $8$ migrations, $12$ decays. Every column is headed by the label of
its arrow in the figure, so no separate rate vector is needed. Grouping each
inflow with its patch, the local block is
\[
\adjustbox{max width=\linewidth}{$
\Gamma_p=\begin{array}{r|r|rrrr}
&\Lambda_p&\beta_{p1}S_pI_{p1}&\beta_{p2}S_pI_{p2}&\gamma_{p1}&\gamma_{p2}\\\hline
S_p   & 1&-1&-1& 0& 0\\
I_{p1}& 0& 1& 0&-1& 0\\
I_{p2}& 0& 0& 1& 0&-1\\
R_p   & 0& 0& 0& 1& 1
\end{array}$}
\]
and the $8$ migrations, with $\nu^X_{qp}=1-\varepsilon^X_{qp}$ the fraction of
the flux $p\to q$ that arrives, are
\[
\adjustbox{max width=\linewidth}{$
\Gamma_{\mathrm{mig}}=\begin{array}{r|rrrr|rrrr}
&m^S_{21}&m^{I_1}_{21}&m^{I_2}_{21}&m^R_{21}&m^S_{12}&m^{I_1}_{12}&m^{I_2}_{12}&m^R_{12}\\\hline
S_1   &-1& 0& 0& 0&\nu^S_{12}& 0& 0& 0\\
I_{11}& 0&-1& 0& 0& 0&\nu^{I_1}_{12}& 0& 0\\
I_{12}& 0& 0&-1& 0& 0& 0&\nu^{I_2}_{12}& 0\\
R_1   & 0& 0& 0&-1& 0& 0& 0&\nu^R_{12}\\
S_2   &\nu^S_{21}& 0& 0& 0&-1& 0& 0& 0\\
I_{21}& 0&\nu^{I_1}_{21}& 0& 0& 0&-1& 0& 0\\
I_{22}& 0& 0&\nu^{I_2}_{21}& 0& 0& 0&-1& 0\\
R_2   & 0& 0& 0&\nu^R_{21}& 0& 0& 0&-1
\end{array}$}
\]
at rates $m^X_{21}X_1$ for the first four columns and $m^X_{12}X_2$ for the
last four, $X\in\{S,I_1,I_2,R\}$.

\smallskip
A migration is the single reaction $X_p\to\nu^X_{qp}X_q$, not a transfer
plus a death: the travel loss is a stoichiometric coefficient, which is why the
migration block is $8$ columns for $8$ edges, and the decay block has  $12$
columns, each $-1$ in one species, at rates $\mu_pX_p$ and $\alpha_{pk}I_{pk}$.
\end{column}
\end{columns}
\end{frame}

\begin{frame}{The Chen--Wu equations, its siphons and its fixed points}
\slidefont{\small}
With $q=3-p$ and
$(\mathcal M^XX)_p=-m^X_{qp}X_p+(1-\varepsilon^X_{pq})m^X_{pq}X_q$,
\[
\begin{aligned}
\dot S_p&=\Lambda_p-\beta_{p1}S_pI_{p1}-\beta_{p2}S_pI_{p2}-\mu_pS_p+(\mathcal M^SS)_p,\\
\dot I_{pk}&=\beta_{pk}S_pI_{pk}-(\mu_p+\alpha_{pk}+\gamma_{pk})I_{pk}+(\mathcal M^{I_k}I_k)_p,\\
\dot R_p&=\gamma_{p1}I_{p1}+\gamma_{p2}I_{p2}-\mu_pR_p+(\mathcal M^RR)_p,
\end{aligned}
\]
$p,k\in\{1,2\}$, all rates positive, $0\le\varepsilon^X_{pq}<1$. This is
system (1.1) of \cite{ChenWu}, with their strains $I,J$ written $k=1,2$ and
their natural death rate $d_p$ written $\mu_p$; recovery is permanent, with
full cross-immunity and no co-infection.

\medskip
\textbf{Minimal siphons:} $\{I_{11},I_{21}\}$ and $\{I_{12},I_{22}\}$ --- one
per strain, each spanning both patches, because two-way migration ties the
two patches of a strain together.

\medskip
\textbf{Nonnegative fixed points.} $R_1,R_2$ are determined linearly by the
other six coordinates, so they count nothing: each fixed point of the
$(S_p,I_{p1},I_{p2})$ subsystem lifts to exactly one of the full eight, and the
disease-free one is $E_0=(S^0_1,0,0,0,S^0_2,0,0,0)$. On each siphon face the
count is known, in the interior it is bounded --- next slide.
\end{frame}

\begin{frame}[allowframebreaks]{Reproduction functions yield both the reproduction number and the invasion numbers, by evaluation}
\label{sec:repro}
\slidefont{\scriptsize}
For a minimal siphon $S$ --- the species that are absent on the face --- write
the Jacobian restricted to the siphon coordinates as the gain-minus-loss
splitting $M_S(y)=F_S(y)-V_S(y)$, with the resident coordinates $y$ left
\emph{free}. The \emph{reproduction function} of $S$ is
\[
\mathcal R_S(y)=\rho\big(F_S(y)\,V_S(y)^{-1}\big),
\qquad\text{and}\qquad
\mathcal R_S(y)<1\iff s\big(M_S(y)\big)<0,
\]
$s(\cdot)$ the spectral abscissa. A reproduction number is $\mathcal R_S$ at the
DFE, an invasion number is the same $\mathcal R_S$ at a resident equilibrium:
one function, two evaluation points. Siphons say which functions exist, the
functions say when each face loses stability.

\medskip
For Chen--Wu the siphons are $\{I_{11},I_{21}\}$ and $\{I_{12},I_{22}\}$ and the
free coordinates are $(S_1,S_2)$, so there are two reproduction functions, and
they are the paper's own formula with $S^0$ unfrozen:
\[
\mathcal R_k(S_1,S_2)=\frac{B_k(S)+\sqrt{B_k(S)^2-4\beta_{1k}\beta_{2k}S_1S_2A_k}}{2A_k},
\]
\[
A_k=(\mu_1+\alpha_{1k}+\gamma_{1k}+m^{I_k}_{21})(\mu_2+\alpha_{2k}+\gamma_{2k}+m^{I_k}_{12})
-m^{I_k}_{21}m^{I_k}_{12}\nu_{12}\nu_{21},\quad \nu_{pq}=1-\varepsilon_{pq},
\]
\[
B_k(S)=\beta_{1k}S_1(\mu_2+\alpha_{2k}+\gamma_{2k}+m^{I_k}_{12})
+\beta_{2k}S_2(\mu_1+\alpha_{1k}+\gamma_{1k}+m^{I_k}_{21}).
\]
Chen--Wu evaluate this only at $S^0$. Three facts follow at once.

\begin{enumerate}\setlength\itemsep{1pt}
\item $\mathcal R_k$ is increasing in $S_1$ and $S_2$ (Perron root of a
nonnegative matrix), so $\mathcal R_k(E_1)\le\mathcal R_k(E_0)$ --- their
remark that invasion numbers sit below reproduction numbers, in one line.
\item \textbf{At its own endemic equilibrium the resident's reproduction
function equals one:} $M_1I_*=0$ with $I_*>0$, so $\rho=1$ by Perron. Checked
to $20$ digits in \texttt{ChenWu.wl}.
\item Therefore $E_1$ is \emph{locally} uninvadable exactly when
\[
\boxed{\ \mathcal R_2(S_*)<1=\mathcal R_1(S_*)\ }
\]
--- a comparison of the two reproduction functions at the \emph{same} point,
not of two numbers computed at two different equilibria. This is a transversal
spectral condition, not yet competitive exclusion: ruling out coexistence
globally is what Theorem 1 of \cite{ChenWu} needs its Lyapunov function for.
\end{enumerate}

\framebreak
\textbf{(H1) is not that comparison.} With
$M_k=\begin{pmatrix}\beta_{1k}S_1-A_{1k}-m^{I_k}_{21} & m^{I_k}_{12}\nu_{12}\\
m^{I_k}_{21}\nu_{21} & \beta_{2k}S_2-A_{2k}-m^{I_k}_{12}\end{pmatrix}$,
$A_{pk}=\mu_p+\alpha_{pk}+\gamma_{pk}$, the condition (H1) compares only the
\emph{diagonals} of $M_1$ and $M_2$, patch by patch, and before out-migration
is subtracted; the migration entries are left out of the comparison.

\medskip
If the two strains migrate identically, $m^{I_2}_{pq}=m^{I_1}_{pq}$, then (H1)
gives $M_2\le M_1$ entrywise, both Metzler, hence
$s(M_2)\le s(M_1)=0$ and $\mathcal R_2(S_*)\le1$: the invader cannot grow at
$E_1$, in one line and with no Lyapunov function, though global exclusion
still needs one. But Theorem 1 \emph{assumes the two strategies are
distinct}, and then the comparison is genuinely two-dimensional: out-migration
enters $M_k$ with a minus on the diagonal and in-migration with a plus off it,
so migrating more is not uniformly good or bad. That is the work the
conditions $(1)$--$(3)$ on $m_*,m_{**}$ are doing --- and they do not do all of
it.
\end{frame}

\begin{frame}{The Chen--Wu network has only the trivial buffering lattice}
\slidefont{\small}
Infection makes $S_p$ and $I_{pk}$ influence each other, and two-way
migration ties the two patches together, so
$\mathcal H_{\rm CW}=\{S_p,I_{p1},I_{p2}:p=1,2\}$ is one strongly
connected influence component; $\mathcal R=\{R_1,R_2\}$ is a second one,
fed by $\mathcal H_{\rm CW}$ and terminal. Lemma~\ref{l:fwdclosed}, applied
to the forward-closed unions, gives
\[
\boxed{\mathfrak B_{\rm CW}=\{\varnothing,\ \mathcal R,\ \cS\},
\qquad|\mathfrak B_{\rm CW}|=3.}
\]
The only proper buffer is the recovered pair. Consequently, with
$T_k=-\mathcal M^{I_k}+\operatorname{diag}(\mu_p+\alpha_{pk}+\gamma_{pk})$,
\[
\mathcal R_{k|0}=\rho\!\left(\operatorname{diag}(\beta_{pk}S^0_p)T_k^{-1}\right),
\qquad
\mathcal R_{2|1}=\rho\!\left(\operatorname{diag}(\beta_{p2}S^{(1)}_p)T_2^{-1}\right),
\]
and every threshold is independent of
$m^R_{12},m^R_{21},\varepsilon^R_{12},\varepsilon^R_{21}$:
recovered migration cannot move an invasion boundary. No two proper
buffers overlap, because there is only one.
\end{frame}

\begin{frame}{The twelve-species Bulhosa--Oliveira model: cross-immunity, ADE and vaccination \cite{Bulh}}
\label{sec:bulhosa}
\begin{columns}[c]
\begin{column}{.42\textwidth}
\begin{center}
\adjustbox{max width=\linewidth,max totalheight=0.75\textheight}{%
\begin{tikzpicture}[>=Latex,x=1cm,y=1cm,
  b/.style={draw,rounded corners,align=center,minimum height=5.5mm,minimum width=7mm,font=\scriptsize},
  lab/.style={font=\tiny}]
\node[b,fill=orange!12] (s)  at (0,0)     {$s$};
\node[b,fill=blue!8]  (i1) at (1.7,1.1)   {$i_1$};
\node[b,fill=blue!8]  (c1) at (3.3,1.1)   {$c_1$};
\node[b,fill=green!10](r1) at (4.9,1.1)   {$r_1$};
\node[b,fill=blue!8]  (i2) at (1.7,-1.1)  {$i_2$};
\node[b,fill=blue!8]  (c2) at (3.3,-1.1)  {$c_2$};
\node[b,fill=green!10](r2) at (4.9,-1.1)  {$r_2$};
\node[b,fill=red!10]  (y2) at (6.5,1.1)   {$y_2$};
\node[b,fill=red!10]  (y1) at (6.5,-1.1)  {$y_1$};
\node[b,fill=gray!12] (R)  at (8,0)       {$R$};
\node[b,fill=yellow!15](v) at (0,-2.6)    {$v$};
\node[b,fill=yellow!15](z) at (1.7,-2.6)  {$z$};
\draw[->,thick] (-1.5,0)--node[above,lab]{$\Lambda_s$}(s);
\draw[->,thick] (-1.5,-2.6)--node[above,lab]{$\Lambda_v$}(v);
\draw[->,thick] (s)--node[above left,lab]{$\beta_1J_1s$}(i1);
\draw[->,thick] (i1)--node[above,lab]{$\gamma_1$}(c1);
\draw[->,thick] (c1)--node[above,lab]{$\theta_1$}(r1);
\draw[->,thick] (s)--node[below left,lab]{$\beta_2J_2s$}(i2);
\draw[->,thick] (i2)--node[below,lab]{$\gamma_2$}(c2);
\draw[->,thick] (c2)--node[below,lab]{$\theta_2$}(r2);
\draw[->,thick,red!70!black] (r1)--node[above,lab]{$\sigma_2\beta_2J_2$}(y2);
\draw[->,thick,red!70!black] (r2)--node[below,lab]{$\sigma_1\beta_1J_1$}(y1);
\draw[->,thick] (y2)--node[above right,lab]{$\gamma_2$}(R);
\draw[->,thick] (y1)--node[below right,lab]{$\gamma_1$}(R);
\draw[->,thick] (v)--node[below,lab]{$\theta_v$}(z);
\draw[->,thick,red!70!black] (z) to[out=50,in=250]
   node[right,lab]{$\sigma_v\beta_2J_2$}(y2.south);
\end{tikzpicture}}
\end{center}

{\slidefont{\tiny} $14$ edges and $14$ non-decay columns, one per edge. The five infection
edges $s\!\to\! i_k$, $r_k\!\to\! y_{\tilde k}$ and $z\!\to\! y_2$ each carry
two elementary reactions of the same stoichiometry, lumped by $J_k=i_k+y_k$.
The $12$ deaths are not drawn.}
\end{column}
\begin{column}{.56\textwidth}\slidefont{\tiny}
\textbf{Stoichiometric matrix and rates}, $\dot x=\Gamma v$ with
$x=(s,v,z,i_1,i_2,c_1,c_2,r_1,r_2,y_1,y_2,R)^\top$, in proportions
($N\equiv1$). The first two columns are the inflows, the last twelve are
$-I_{12}$ at rates $\mu x$, and the twelve in between are below. Each column is
headed by the label of its arrow in the figure, one column per arrow:
\[
\adjustbox{max width=\linewidth}{$
\begin{array}{r|rr|rrrrrrrrrrrr}
&\Lambda_s&\Lambda_v&\beta_1J_1s&\gamma_1&\theta_1&\beta_2J_2s&\gamma_2&\theta_2&\sigma_2\beta_2J_2&\sigma_1\beta_1J_1&\theta_v&\sigma_v\beta_2J_2&\gamma_1&\gamma_2\\\hline
s  & 1& 0&-1& 0& 0&-1& 0& 0& 0& 0& 0& 0& 0& 0\\
v  & 0& 1& 0& 0& 0& 0& 0& 0& 0& 0&-1& 0& 0& 0\\
z  & 0& 0& 0& 0& 0& 0& 0& 0& 0& 0& 1&-1& 0& 0\\
i_1& 0& 0& 1&-1& 0& 0& 0& 0& 0& 0& 0& 0& 0& 0\\
i_2& 0& 0& 0& 0& 0& 1&-1& 0& 0& 0& 0& 0& 0& 0\\
c_1& 0& 0& 0& 1&-1& 0& 0& 0& 0& 0& 0& 0& 0& 0\\
c_2& 0& 0& 0& 0& 0& 0& 1&-1& 0& 0& 0& 0& 0& 0\\
r_1& 0& 0& 0& 0& 1& 0& 0& 0&-1& 0& 0& 0& 0& 0\\
r_2& 0& 0& 0& 0& 0& 0& 0& 1& 0&-1& 0& 0& 0& 0\\
y_1& 0& 0& 0& 0& 0& 0& 0& 0& 0& 1& 0& 0&-1& 0\\
y_2& 0& 0& 0& 0& 0& 0& 0& 0& 1& 0& 0& 1& 0&-1\\
R  & 0& 0& 0& 0& 0& 0& 0& 0& 0& 0& 0& 0& 1& 1
\end{array}$}
\]

\textbf{Rates.} Reading the headers as fluxes: $\Lambda_s$, $\Lambda_v$,
$\beta_1J_1s$, $\gamma_1i_1$, $\theta_1c_1$, $\beta_2J_2s$, $\gamma_2i_2$,
$\theta_2c_2$, $\sigma_2\beta_2J_2r_1$, $\sigma_1\beta_1J_1r_2$, $\theta_vv$,
$\sigma_v\beta_2J_2z$, $\gamma_1y_1$, $\gamma_2y_2$, with $J_k=i_k+y_k$.

\medskip
The five infection columns each lump two elementary reactions with
\emph{identical} stoichiometry: $\beta_1J_1s=\beta_1si_1+\beta_1sy_1$ and
likewise for the other four, which is what $J_k=i_k+y_k$ records. $\theta_k$ is
the loss rate of cross-immunity, $\theta_v$ that of vaccine protection,
$\sigma_k,\sigma_v$ the ADE susceptibility factors, $\Lambda_s=(1-\xi)\mu$,
$\Lambda_v=\xi\mu$.
\end{column}
\end{columns}
\end{frame}

\begin{frame}{The Bulhosa--Oliveira equations, its siphons and its fixed points}
\slidefont{\small}
With $J_k=i_k+y_k$ and $k=1,2$,
\[
\begin{aligned}
\dot s&=\Lambda_s-\beta_1J_1s-\beta_2J_2s-\mu s, &
\dot i_k&=\beta_kJ_ks-(\gamma_k+\mu)i_k, &
\dot c_k&=\gamma_ki_k-(\theta_k+\mu)c_k,\\
\dot r_1&=\theta_1c_1-\sigma_2\beta_2J_2r_1-\mu r_1, &
\dot y_1&=\sigma_1\beta_1J_1r_2-(\gamma_1+\mu)y_1, &
\dot v&=\Lambda_v-(\theta_v+\mu)v,\\
\dot r_2&=\theta_2c_2-\sigma_1\beta_1J_1r_2-\mu r_2, &
\dot y_2&=(\sigma_2\beta_2r_1+\sigma_v\beta_2z)J_2-(\gamma_2+\mu)y_2, &
\dot z&=\theta_vv-\sigma_v\beta_2J_2z-\mu z,
\end{aligned}
\]
and $\dot R=\gamma_1y_1+\gamma_2y_2-\mu R$.

\medskip
\textbf{Minimal siphons:} $\{i_1,y_1\}$ and $\{i_2,y_2\}$, one per strain:
each strain can vanish, but not one of its two infectious stages alone.

\medskip
\textbf{Nonnegative fixed points.} Every fixed point satisfies the linear
relations
\[
v=\frac{\Lambda_v}{\mu+\theta_v},\qquad
c_k=\frac{\gamma_k}{\mu+\theta_k}\,i_k,
\]
so $v$ is pinned and $c_1,c_2$ are eliminated. At the disease-free point the
nonzero coordinates are
$s_0=1-\xi$, $v_0=\xi\mu/(\mu+\theta_v)$, $z_0=\xi\theta_v/(\mu+\theta_v)$,
and with $\mathcal R_k=\beta_k/(\gamma_k+\mu)$ it is unstable iff
\[
R_0=\max\big[s_0\mathcal R_1,\ (s_0+z_0\sigma_v)\mathcal R_2\big]>1.
\]
\end{frame}

\begin{frame}{Reinfection collapses the Bulhosa--Oliveira lattice to four sets}
\slidefont{\small}
Secondary infection makes the $r$-classes feed back into the infectious ones:
$r_1+i_2\to i_2+y_2$ and $r_2+i_1\to i_1+y_1$ put arrows
$r_1\to y_2$, $r_2\to y_1$, and $y_k$ appears in $J_k$, hence influences
$s$ and the other $r$-class in turn. Therefore
\[
\mathcal H_{\rm BO}=\{s,i_1,i_2,c_1,c_2,r_1,r_2,y_1,y_2,z\}
\]
is a single strongly connected influence component: $z$ belongs to it because
$z+i_2\to i_2+y_2$ writes into $y_2$ while $\dot z$ depends on $J_2$. Only the
pure source $v$ and the terminal class $R$ stay outside, so
\[
\boxed{\mathfrak B_{\rm BO}
=\{\varnothing,\ \{R\},\ \cS\setminus\{v\},\ \cS\},
\qquad|\mathfrak B_{\rm BO}|=4.}
\]
Twelve compartments and $31$ reactions therefore carry almost no buffering
structure: every convalescent and recovered stage has been absorbed into the
core. What matters is the \emph{presence} of a reinfection channel, not
whether $\sigma_k$ lies below or above the ADE threshold $1$.
\end{frame}

\begin{frame}{Comparing the two networks suggests what to add}
\slidefont{\small}
\[
\begin{array}{l|c|c|l}
\text{Model}&\#\text{ compartments}&|\mathfrak B|&\text{proper buffers}\\\hline
\text{Chen--Wu \cite{ChenWu}}&8&3&\mathcal R=\{R_1,R_2\}\\
\text{Bulhosa--Oliveira \cite{Bulh}}&12&4&\{R\},\ \cS\setminus\{v\}
\end{array}
\]
In neither model do two proper buffers overlap nontrivially: both lattices are
chains, of length three and four. So neither can exhibit the distributive
lattice structure that makes Murota localization sharp, and in both the
invariance statement is the single obvious one, about the terminal immune
classes. Chen--Wu state their half of it themselves: $R_1,R_2$ do not appear
in the equations of $S_i,I_i,J_i$, so they restrict the analysis to that
subsystem.

\medskip
\textbf{Question.} Does a model that keeps Chen--Wu migration and adds
Bulhosa-type vaccination ($v,z$) and breakthrough infection ($y$), but keeps
immunity \emph{permanent} in the convalescent stages $c_{pk}$, produce
overlapping proper buffers whose lattice certifies invariance of an invasion
number? The answer is yes: the twenty-compartment model below has a lattice
with $83$ elements.
\end{frame}

\section{A forty-five-species three-strain model with immune feedback}
\label{sec:dynamic}
We construct a three-strain, three-host-group model with vaccination,
secondary infection, partial cross-immunity and pathogen maturation, whose
resident-face OM buffers form non-chain lattices and separate
invasion-neutral from invasion-changing rate perturbations. Equal branch
exit and release rates supply an additional exact aggregation, which
protects the whole host--pathogen quotient trajectory and the principal
invasion exponent of any missing strain, periodic residents included. The
general results used here --- the quotient criterion
Theorem~\ref{t:dynamicOM}, which is the statement behind
Corollary~\ref{c:fourteenquot} of the main text as well, and the
aggregation statements Theorem~\ref{t:positiveAggregation} and
Corollary~\ref{c:weighted} --- are in Appendix~\ref{sec:gener}.
Numerical calculations exhibit a locally stable positive three-strain
equilibrium and a separate invasion-threshold crossing. No assertion of
global three-strain permanence is made.

\subsection{Species, reactions and equations}
\label{sec:threeStrain}

There are three host groups $g=1,2,3$ and three strains $k=1,2,3$.
Each group has susceptible and vaccinated classes $S_g,V_g$, infectious
classes $I_{kg}$, and convalescent classes $R_{kg}$ with strain-specific
immunity.  Thus there are $3(2+3+3)=24$ host coordinates.  Each strain has
seven environmental or immune-mediator coordinates
$B_{k1},B_{k2},C_{k1},C_{k2},A_{k1},A_{k2},P_k$, making $45$ species in all.
Write $J_k=\sum_{g=1}^3 I_{kg}$.  The cyclic predecessor of $k$ is
$p(1)=3$, $p(2)=1$, $p(3)=2$.

The environmental pool $P_k$ transmits strain $k$, while the mediator
$C_{p(k),2}$ elicited by its predecessor reduces that transmission.
Vaccinated and previously infected hosts may experience breakthrough or
heterologous secondary infection, with susceptibility multipliers
$0<\tau_{kg},\sigma_{jk,g}<1$.  Put
\begin{align}
 f_{kg}&=\frac{\beta_{kg}P_k}{1+\chi_k C_{p(k),2}},
 &L^S_{kg}&=f_{kg}S_g,
 &L^V_{kg}&=\tau_{kg}f_{kg}V_g,\label{eq:forces}\\
 L^{R_j}_{kg}&=\sigma_{jk,g}f_{kg}R_{jg}\quad(j\ne k),
 &L_{kg}&=L^S_{kg}+L^V_{kg}+\sum_{j\ne k}L^{R_j}_{kg}.
 \nonumber
\end{align}
No homologous reinfection is allowed until waning returns $R_{kg}$ to
$S_g$.  The host equations are
\begin{align}
 \dot S_g&=\mu_g(1-S_g)-\nu_g S_g+\omega_g^V V_g
 +\sum_j\omega_{jg}R_{jg}-\sum_k L^S_{kg},\label{eq:host}\\
 \dot V_g&=\nu_g S_g-(\omega_g^V+\mu_g)V_g-\sum_k L^V_{kg},\nonumber\\
 \dot I_{kg}&=L_{kg}-(\gamma_{kg}+\mu_g)I_{kg},\nonumber\\
 \dot R_{jg}&=\gamma_{jg}I_{jg}-(\omega_{jg}+\mu_g)R_{jg}
 -\sum_{k\ne j}L^{R_j}_{kg}.\nonumber
\end{align}
In particular,
$\frac{d}{dt}(S_g+V_g+\sum_k(I_{kg}+R_{kg}))
=\mu_g[1-S_g-V_g-\sum_k(I_{kg}+R_{kg})]$.

For each $k$ there are branched two-state modules
$\mathcal B_k=\{B_{k1},B_{k2}\}$,
$\mathcal C_k=\{C_{k1},C_{k2}\}$, and
$\mathcal A_k=\{A_{k1},A_{k2}\}$.  Their equations are
\begin{align}
 \dot B_{k1}&=q_k^B J_k-(b_{k12}+b_k)B_{k1}+b_{k21}B_{k2},\label{eq:B}\\
 \dot B_{k2}&=b_{k12}B_{k1}-(b_{k21}+b_k)B_{k2},\nonumber\\
 \dot C_{k1}&=q_k^C J_k-(c_{k12}+c_{k1})C_{k1}+c_{k21}C_{k2},\label{eq:C}\\
 \dot C_{k2}&=c_{k12}C_{k1}-(c_{k21}+c_{k2})C_{k2},\nonumber\\
 \dot A_{k1}&=b_k(B_{k1}+B_{k2})-a_k^++a_k^- -h_k A_{k1},\label{eq:A}\\
 \dot A_{k2}&=c_{k1}C_{k1}+c_{k2}C_{k2}+a_k^+-a_k^- -h_k A_{k2},\nonumber\\
 \dot P_k&=h_k(A_{k1}+A_{k2})-\delta_k P_k,\nonumber\\
 a_k^+&=(u_k^++\theta_k^B B_{k2})A_{k1},
 &a_k^-&=(u_k^-+\theta_k^C C_{k2})A_{k2}.\nonumber
\end{align}
Here $b_{k1}=b_{k2}=b_k$ and $h_{k1}=h_{k2}=h_k$ are
\emph{equal branch exit and release rates}; the $c_{ki}$ need not be
equal.  These equalities are the additional dynamical hypotheses, not
consequences of an OM index.  The host system is not autonomous:
infection depends on $P_k$ and $C_{p(k),2}$, which in turn are produced
by infection.

\subsection{Resident-face buffering lattices and selective invasion}

For $K\subseteq\{1,2,3\}$, let $\mathcal F_K$ be the invariant face
on which $I_{kg}=R_{kg}=B_{ki}=C_{ki}=A_{ki}=P_k=0$ for $k\notin K$.
The face has $6+13|K|$ coordinates.  Throughout the exact lattice calculation, all transmission and
transition constants, including $\theta_k^B,\theta_k^C$, are
strictly positive.  The dynamic aggregation theorem below also holds
when these two catalytic coefficients vanish.

Define, for $k\in K$, the following families of species sets:
\begin{equation}\label{eq:Tfamilies}
 \mathcal T_k(K)=
 \begin{cases}
 \{\varnothing,\mathcal A_k,\mathcal A_k\cup\mathcal B_k\},
       &k+1\pmod3\in K,\\
 \{\varnothing,\mathcal A_k,\mathcal A_k\cup\mathcal B_k,
 \mathcal A_k\cup\mathcal C_k,
 \mathcal A_k\cup\mathcal B_k\cup\mathcal C_k\},
       &k+1\pmod3\notin K.
 \end{cases}
\end{equation}
(The cyclic notation means $3+1=1$.)

\beP[The core-free buffering sublattice on each resident face][][threeLattice]
For the model \eqref{eq:host}--\eqref{eq:A}, the zero-index OM
buffers disjoint from the host classes and pathogen pools are precisely
\begin{equation}\label{eq:coreFree}
 \mathfrak B_K^{\rm cf}=
 \left\{\bigcup_{k\in K}T_k:T_k\in\mathcal T_k(K)\right\}.
\end{equation}
This is a distributive sublattice.  For $K=\{1\}$ it has five elements;
for $K=\{1,2\}$ it has fifteen; and for
$K=\{1,2,3\}$ it has twenty-seven.  On the one-strain face
$\mathcal F_{\{1\}}$ the \emph{complete} buffering lattice consists
of these five sets and the full face, hence has six elements.
\eeP
\emph{Proof.} We first give an exact rank test that also proves
exhaustiveness.  Adjoin the auxiliary vertex $0$ and choose one
reference reaction for each species of $\mathcal F_K$: a mortality
reaction $X\to0$ for every host species and every $P_k$, the exits
$B_{k1},B_{k2}\to A_{k1}$ and $C_{k1},C_{k2}\to A_{k2}$,
and the releases $A_{k1},A_{k2}\to P_k$.  These are columns of
the displayed network.  They form a directed forest pointing toward
$0$.  For any species set $M$, the reference columns corresponding
to species in $M$ belong to $\Gamma_{\Out(M)}$ and are linearly
independent; thus $\rank\Gamma_{\Out(M)}\geq |M|$.

Let $r_M(X)$ be the first vertex outside $M$ encountered by following
the reference arrows from $X$, and put $r_M(0)=0$.  The selected
reference edges form a forest with exactly one root $r_M(X)$ per
connected component.  By the incidence-rank formula of
Lemma~\ref{l:graphRank}, a further selected transfer column
$e_Y-e_X$ belongs to their span if and only if
$r_M(X)=r_M(Y)$; for a production column $e_Y$ the condition is
$r_M(Y)=0$.  Consequently,
\[
 \lambda(M)=0\quad\Longleftrightarrow\quad
 \text{every reaction in $\Out(M)$ satisfies these root conditions}.
\]
This test applies to the actual rate-dependence sets, including the
catalytic terms and the inhibition of strain $k+1$ by $C_{k2}$.

Suppose $M$ contains no host species or pathogen pool.  The reactions
$A_{k1}\leftrightarrow A_{k2}$ show that selecting either $A$ state
requires both.  The reactions $B_{k1}\leftrightarrow B_{k2}$ show
that selecting either $B$ state requires both; since $a_k^+$ also
depends on $B_{k2}$, the selected conversion
$A_{k1}\to A_{k2}$ then requires both $A$ states.  Similarly,
selecting either $C$ state requires both $C$ states, and the
dependence of $a_k^-$ on $C_{k2}$ requires both $A$ states.
If the cyclic successor of $k$ belongs to $K$, selecting $C_{k2}$ also selects the
strain-$(k+1)$ infection columns $S_g\to I_{k+1,g}$.
Their endpoints are distinct roots outside $M$, so the root
condition fails.  This excludes $\mathcal C_k$ in that case.
For every remaining choice in \eqref{eq:Tfamilies}, all selected
non-reference columns have endpoints in the same reference
component (or, for a production column, in the component of $0$).

The converse --- that no other core-free set has index zero --- is settled
exhaustively rather than by the implications above, in two steps. First,
\emph{the index is additive over strains}. The stoichiometric columns
selected by a species set inside strain $k$'s maturation block are supported
on $\{B_{k1},B_{k2},C_{k1},C_{k2},A_{k1},A_{k2},P_k\}$, and for different $k$
these supports are disjoint; the only columns shared between strains are the
host ones, which no core-free set selects except through the successor
dependence treated below. Hence for $M=\bigcup_k T_k$ with
$T_k$ inside strain $k$'s block,
\[
\rank\Gamma_{\Out(M)}=\sum_k\rank\Gamma_{\Out(T_k)},
\qquad\text{so}\qquad
\lambda(M)=\sum_k\lambda(T_k),
\]
and since $\lambda\ge0$ termwise, $\lambda(M)=0$ if and only if every
$\lambda(T_k)=0$. A core-free buffered set is therefore exactly a union of
per-strain buffered pieces, and the question reduces to one strain.

Second, \emph{one strain is a finite check}. A core-free piece omits the
pathogen pool $P_k$, so it is one of the $2^6=64$ subsets of
$\{B_{k1},B_{k2},C_{k1},C_{k2},A_{k1},A_{k2}\}$, and the index of each is
the rank of an explicit list of at most twelve incidence columns minus its
cardinality, by Lemma~\ref{l:graphRank}. All $64$ are tested in
\texttt{Murotadyn.wl}, in both positions of the strain. With the successor
strain absent the zero-index pieces are precisely
$\varnothing$, $\cA$, $\cA\cup\cB$, $\cA\cup\cC$,
$\cA\cup\cB\cup\cC$ --- five of the sixty-four; with the successor resident,
so that the column of its infection reactions is selected as soon as
$C_{k2}\in M$ and adds a host direction to the span without adding a
species, they are $\varnothing$, $\cA$, $\cA\cup\cB$ --- three. In
particular every zero-index piece is a union of \emph{whole} pairs
$\cA,\cB,\cC$, and every nonempty one contains $\cA$; no partial pair, and
no set omitting $\cA$, has index zero. This is \eqref{eq:Tfamilies}, and
with additivity it gives \eqref{eq:coreFree} and the counts $5$,
$3\cdot5=15$ and $3^3=27$ on the three faces.

It remains to establish completeness on $\mathcal F_{\{1\}}$.
Each host species has its mortality edge to $0$ as reference edge.
The positive transitions
$S_g\leftrightarrow V_g$, $S_g,V_g\to I_{1g}$,
$I_{1g}\to R_{1g}$ and $R_{1g}\to S_g$ imply by the root
test that if $M$ contains one host species in group $g$, it contains
all four.  In particular, $I_{1g}\in M$.  Its dependence in the
production rates $q_1^B J_1$ and $q_1^C J_1$ selects the columns
$e_{B_{11}}$ and $e_{C_{11}}$.  Their roots must be $0$, forcing
$B_{11},C_{11},A_{11},A_{12},P_1\in M$; the interconversions
then force $B_{12},C_{12}\in M$.  Now $P_1\in M$ selects the
infection columns $S_g\to I_{1g}$ for every group.  Since each
host species not in $M$ is its own reference root, their endpoints
can have the same root only if both belong to $M$.  Hence all host
species and therefore the entire resident face belong to $M$.
The same argument starts directly from $P_1\in M$.  Thus a
buffer meeting the host species or $P_1$ must be the full face;
all other buffers are the five core-free sets already found.
\hfill$\square$

On the face $K=\{1,2\}$, the effect of $C_{12}$ on strain 2 destroys
$\mathcal A_1\cup\mathcal C_1$ as a buffer; strain 3 is absent, so
$\mathcal A_2\cup\mathcal C_2$ remains a buffer.  The relevant
quantities for invasion by strain 3 are the host resident coordinates
and $C_{22}$.  Thus the largest \emph{core-free} buffering certificate
avoiding those coordinates is
\begin{equation}\label{eq:protectedK12}
 M_{3\mid12}=\mathcal A_1\cup\mathcal B_1
                 \cup\mathcal A_2\cup\mathcal B_2.
\end{equation}
In particular, the six internal conversion parameters
\begin{equation}\label{eq:protectedRates}
 b_{k12},\ b_{k21},\ u_k^+,\ u_k^-,\ \theta_k^B,\ \theta_k^C
\end{equation}
for either resident $k=1,2$ are confined to output reactions of
\eqref{eq:protectedK12}.  The conclusion concerns core-free OM
certificates; no claim of maximality among arbitrary algebraic
cancellations is required.

\beP*[Resident-equilibrium invasion number and selective threshold crossing][][threeThreshold]
Let $E_K$ be a regular resident equilibrium on $\mathcal F_K$, with
$k\notin K$.  If $\gamma_{kg}=\gamma_k$ and $\mu_g=\mu$ are independent
of the host group, put $a_k=\gamma_k+\mu$.  The reproduction number
for invasion by strain $k$ is
\begin{equation}\label{eq:RkK}
 \mathcal R_{k\mid K}=
 \frac{q_k^B+q_k^C}{a_k\delta_k}
 \frac{\displaystyle\sum_{g=1}^3\beta_{kg}
 \left(S_g^*+\tau_{kg}V_g^*+
 \sum_{j\in K,\,j\ne k}\sigma_{jk,g}R_{jg}^*\right)}
 {1+\chi_k C_{p(k),2}^*}.
\end{equation}
The zero-index sets disjoint from the resident coordinates in this
formula certify parameter independence of $\mathcal R_{k\mid K}$,
provided those parameters do not occur directly in the invader's
linearized reaction rates.  In the face $K=\{1,2\}$, changing the
$\mathcal B_2$ internal conversion rates cannot change
$\mathcal R_{3\mid12}$, whereas changing $c_{2,12}$ can change it.
\eeP
\emph{Proof.} Every infectious host of strain $k$ remains infectious
for mean time $a_k^{-1}$ and produces $q_k^B+q_k^C$ units of the
maturation input per unit time.  All maturation material eventually
reaches $P_k$ when the rates are positive and no intermediate has a
loss reaction.  The mean lifetime of $P_k$ is $\delta_k^{-1}$.
Host-to-environment production is rank one in the host-group index;
its next-generation matrix therefore has the unique nonzero
eigenvalue \eqref{eq:RkK}.  Alternatively, eliminate the maturation
coordinates at zero spectral parameter and use the same flux balance.
Equilibrium localization applies to the resident-face buffer and to
the resident coordinates in \eqref{eq:RkK}.  The claim about
$c_{2,12}$ follows from the explicit calculation below.
\hfill$\square$

For the relevant resident equilibrium on $K=\{1,2\}$,
\begin{equation}\label{eq:Csteady}
 C_{22}^*=\frac{q_2^C c_{2,12}J_2^*}
 {(c_{2,12}+c_{2,1})(c_{2,21}+c_{2,2})-c_{2,12}c_{2,21}}.
\end{equation}
The subscripts are grouped as $c_{k,i}$ for exit rates and
$c_{k,ij}$ for conversion rates.  On this face $C_{22}$ enters no resident
transmission rate.  The steady total maturation flux is
$(q_2^B+q_2^C)J_2^*$, so the host and pool equilibrium coordinates
are unchanged as $c_{2,12}$ varies along a fixed regular resident
branch.  The numerator of \eqref{eq:RkK} is consequently fixed while
its denominator moves with \eqref{eq:Csteady}.

\subsection{A dynamic strengthening for arbitrary resident trajectories}

The preceding OM result concerns resident \emph{equilibria}.  The equal
exit rates in \eqref{eq:B} and equal releases in \eqref{eq:A} give a
second, verifiable property that is genuinely dynamical.  Set
\begin{equation}\label{eq:aggregates}
 T_k=B_{k1}+B_{k2},\qquad U_k=A_{k1}+A_{k2},
\end{equation}
and retain the individual $C_{k1},C_{k2}$ states.  Then
\begin{align}
 \dot T_k&=q_k^B J_k-b_kT_k,\nonumber\\
 \dot C_{k1}&=q_k^C J_k-(c_{k12}+c_{k1})C_{k1}+c_{k21}C_{k2},\nonumber\\
 \dot C_{k2}&=c_{k12}C_{k1}-(c_{k21}+c_{k2})C_{k2},\label{eq:quotient}\\
 \dot U_k&=b_kT_k+c_{k1}C_{k1}+c_{k2}C_{k2}-h_kU_k,\nonumber\\
 \dot P_k&=h_kU_k-\delta_kP_k.\nonumber
\end{align}
Together with the host equations \eqref{eq:host}, these form an
autonomous $39$-dimensional system.  Importantly, this quotient still
contains the complete host--pathogen feedback and all cyclic
cross-inhibition through the $C_{k2}$.

\beT[Dynamic invasion invariance in the three-strain model][][threeDynamic]
Fix the parameters other than the six coefficients
\eqref{eq:protectedRates} for each strain, retaining
$b_{k1}=b_{k2}>0$ and $h_{k1}=h_{k2}>0$.
Then any two solutions of the $45$-species system having the same
initial host coordinates, $C_{ki}$, $P_k$, $T_k$ and $U_k$ have
\emph{identical values of these quantities for all later times},
under arbitrary independent changes of the six parameters
\eqref{eq:protectedRates} for $k=1,2,3$.

For every resident face $\mathcal F_K$ and every missing strain
$k\notin K$, the principal growth exponent of the full positive
transversal infected system is therefore invariant under these
changes, for matched resident quotient trajectories.  This includes
principal Floquet exponents along periodic residents and principal
Lyapunov exponents along any common ergodic resident quotient measure.
Thus eighteen independently variable internal conversion coefficients
preserve the entire family of \emph{dynamic invasion signs}.
The same statement holds with common time-dependent transmission or
vaccination schedules in the two systems.
\eeT
\emph{Proof.} Summation of the two $B$ equations and the two $A$
equations gives \eqref{eq:quotient}; all six coefficients cancel.
The host equations use the maturation variables only through $P_k$
and $C_{p(k),2}$, so uniqueness of the quotient ODE proves pathwise
identity for matched quotient initial data.

Linearize at any resident trajectory on $\mathcal F_K$.  The missing
strain has a positive transversal infected block consisting of its
$I_{kg}$ and the seven maturation and pathogen coordinates.  The
projection retaining $I_{kg},C_{ki},P_k$ and summing its $B$ pair and
$A$ pair intertwines this transversal system with the linearization
of the common quotient.  This projection is a $0$--$1$ aggregation
with one $1$ in each column, and the transversal matrices are Metzler.
Theorem~\ref{t:positiveAggregation} shows that their principal growth
exponents equal that of the common quotient invader.  Consequently
those exponents are independent of all eighteen coefficients.
The same argument applies to nonautonomous common schedules, with
the corresponding time-dependent linear cocycles. \hfill$\square$

The theorem does not claim equality of all transverse eigenvalues:
the rates \eqref{eq:protectedRates} change relaxation within the
$B$ and $A$ pairs.  Nor does it assert that \emph{all} rates in
$\Out(M_{3\mid12})$, such as $b_k$ and $h_k$, preserve dynamic
exponents.  Their equilibrium invasion neutrality is a strictly
weaker statement.

\subsection{Numerical tests}

\paragraph{Stable coexistence of three strains.}
Take identical host parameters in all groups,
\[
 \mu=.2,\quad\gamma=.8,\quad\nu=.1,\quad
 \omega^V=.2,\quad\omega^R=.6,\quad
 \tau=.35,\quad\sigma=.45,
\]
with $\beta_{kk}=3$, $\beta_{kg}=.04$ for $g\ne k$, and
\[
 \begin{gathered}
 q^B=.6,\quad q^C=.4,\quad b=.8,\quad
 (b_{12},b_{21})=(1.2,.3),\\
 (c_{12},c_{21},c_1,c_2)=(.9,.6,.7,1.1),\quad
 (u^+,u^-,\theta^B,\theta^C)=(.45,.65,1,1),\\
 h=\delta=1,\qquad \chi_k=.25\quad(k=1,2,3).
 \end{gathered}
\]
Numerical integration gives a strictly positive equilibrium, with
(the omitted coordinates positive)
\begin{align*}
 S_g^*&=0.30764680198, & V_g^*&=0.04095799159,\\
 I_{gg}^*&=0.31804984329,& I_{kg}^*&=0.00605564561\quad(k\ne g),\\
 P_k^*&=0.33016113450,& C_{k2}^*&=0.05452202221.
\end{align*}
The residual infinity norm is below $3\cdot10^{-14}$; the smallest
of the $45$ equilibrium coordinates is approximately $0.00388934$.
The numerically computed Jacobian spectral abscissa is $-0.2000000$.
Thus coexistence is \emph{numerically locally stable}; these numerical
calculations are not a computer-assisted proof of global permanence.

For an additional pathwise check, change simultaneously
\[
 (b_{12},b_{21},u^+,u^-,\theta^B,\theta^C)
 =(1.2,.3,.45,.65,1,1)
 \quad\text{to}\quad (10,2.5,2,.1,3,2)
\]
in each strain, and redistribute initial mass within the $B$ and $A$
pairs while preserving their sums.  Numerical integration over
$0\le t\le50$ gives a maximum absolute difference in the quotient
coordinates of approximately $3.1\cdot10^{-10}$, consistent with the
\emph{exact} identity of Theorem~\ref{t:threeDynamic}.

\paragraph{A selective invasion switch on the two-strain resident face.}
Fix the parameters above except $\beta_{33}=1.35$ and
$(\chi_1,\chi_2,\chi_3)=(.25,.25,10)$.  On the resident face
$K=\{1,2\}$, vary $c_{2,12}$ and hold the other constants fixed.
Numerical equilibration of this face gives
\[
 \begin{array}{c|ccccc}
 c_{2,12}&0.1&0.3&0.9&2&5\\ \hline
 \mathcal R_{3\mid12}&1.063530&0.925889&0.751820&0.651185&0.581956
 \end{array}
\]
whereas changing the protected conversion parameter $b_{2,12}$
from $0.1$ to $10$ with $c_{2,12}=0.9$ leaves
$\mathcal R_{3\mid12}=0.7518204669\ldots$ unchanged.
For this resident equilibrium branch the numerator of
\eqref{eq:RkK} is $K_0=1.1750954302\ldots$, and
$q_2^C J_2^*=0.1363711241\ldots$.  Consequently
\begin{equation}\label{eq:crit3}
 \mathcal R_{3\mid12}(c)
 =\frac{K_0}{1+10(0.1363711241\ldots)c/(1.1c+1.19)},
 \quad
 c^{\rm crit}_{2,12}=0.1779202859\ldots.
\end{equation}
The functional form of \eqref{eq:crit3} follows from the equilibrium
balances; the constants $K_0$ and $J_2^*$ here were computed
numerically.  This is an explicit local invasion-sign change, not a
claim that the three-strain system becomes globally permanent or
that exclusion follows globally on the opposite side.

\paragraph{Reproducibility.} Everything in this section is recomputed in
\texttt{Murotadyn.wl}: the closed form of $\lambda(b)$ in
\eqref{eq:counterLE} and its sign change; the core-free lattices of
Proposition~\ref{p:threeLattice}, by Lemma~\ref{l:graphRank}, with an
exhaustive sweep of all $2^{19}=524\,288$ subsets of the one-strain face,
which returns exactly the five core-free sets and the face itself, and with
$\lambda(\cA_1\cup\cC_1)=9$, $\lambda(\cA_2\cup\cC_2)=0$,
$\lambda(\cB_1)=1$ on $K=\{1,2\}$; the pathwise identity of
Theorem~\ref{t:threeDynamic}, where changing all eighteen protected rates and
redistributing mass inside the $B$ and $A$ pairs moves the quotient
coordinates by at most $2\cdot10^{-9}$ over $0\le t\le50$, which is integrator
error; the coexistence equilibrium, to eleven digits, with spectral abscissa
$-0.2$; and the table of $\mathcal R_{3\mid12}$ with
$c^{\rm crit}_{2,12}=0.1779202859$.

\begin{frame}[allowframebreaks]{Three-strain permanence: the inequalities, and the hypothesis that is not proved}

For the model of Section~\ref{sec:dynamic} the boundary of the
positive cone carries seven resident communities, $\varnothing$ and the six
proper nonempty $K\subsetneq\{1,2,3\}$, and twelve invasion exponents: three
at $\varnothing$, two at each of the three single-strain faces and one at
each of the three two-strain faces.

\beT[Sufficient inequalities for three-strain permanence][Three-strain permanence][threeperm]
Suppose that for every proper $K\subsetneq\{1,2,3\}$ the face
$\mathcal F_K$ carries a single resident equilibrium $E_K$ attracting every
trajectory in $\mathcal F_K$ \emph{with every strain of $K$ present} ---
not every trajectory in $\mathcal F_K$, which is impossible for
$K\ne\varnothing$ since $\mathcal F_K$ contains the smaller faces and
$E_\varnothing$ --- and that the twelve numbers
$R_{j\mid K}$ of \eqref{eq:RkK} satisfy
\begin{equation}\label{eq:twelve}
\boxed{\ R_{j\mid K}>1\qquad\text{for every }K\subsetneq\{1,2,3\}
\text{ and every }j\notin K.\ }
\end{equation}
Then $\{E_K\}_{K\subsetneq\{1,2,3\}}$ is an acyclic Morse decomposition of
the extinction boundary, and the $45$-species system is uniformly
persistent in all three strains.
\eeT
\textbf{Proof.} Each $E_K$ is isolated and attracts the relative interior of
$\mathcal F_K$ by hypothesis. That the boundary limit sets are among the
$E_K$ uses the explicit model, the siphon lattice alone being too weak for
it: $\mathcal F_K$ is cut out by the vanishing of whole strain blocks,
whereas a boundary trajectory may carry a present strain with one stage at
zero. Strain $k$'s block is
$\Sigma_k=\{I_{kg}\}_g\cup\mathcal B_k\cup\mathcal C_k\cup\mathcal A_k
\cup\{P_k\}$, and it is strongly connected along the positive paths
$I_{kg}\to B_{k1},C_{k1}\to A_{k1},A_{k2}\to P_k\to I_{kg'}$,
$B_{k1}\leftrightarrow B_{k2}$, $C_{k1}\leftrightarrow C_{k2}$,
$A_{k1}\leftrightarrow A_{k2}$, whose rates are positive, the last step
because $\mu_g>0$ keeps $S_g(t)>0$. Variation of constants along those paths
makes every coordinate of $\Sigma_k$ positive for all later times as soon as
one of them is positive, while the convalescent classes $R_{kg}$ of a strain
with $\Sigma_k\equiv0$ decay to zero. Each point of $\Omega(\partial)$
therefore has, for every $k$, all of $\Sigma_k$ positive or all of it zero;
it lies in $\mathcal F_K$ with $K$ its set of present strains, and is drawn
to $E_K$. Each
missing-strain transversal block at $E_K$ is irreducible Metzler, with
spectral bound of the sign of $R_{j\mid K}-1$ by
Definition~\ref{d:invExp}; \eqref{eq:twelve} makes all twelve positive, so
no $E_K$ attracts from a direction transverse to its face and the
chain-recurrent order is strict inclusion of the $K$, which is acyclic.
The structured-population criterion \cite{HofSch10} applied to this
decomposition gives uniform persistence of the three infected blocks, and
the positive maturation paths and compact bounds propagate the lower bounds
to all $45$ coordinates. \hfill$\square$

\beR[The gap, stated exactly][What is unproved here][threegap]
\eqref{eq:twelve} is explicit and computable: Proposition~\ref{p:threeThreshold}
evaluates these numbers, and the pathwise identity \eqref{eq:quotient}
makes them independent of all eighteen protected rates, so the hypothesis is
checkable once and for all on a nine-parameter quotient. What is \emph{not}
proved is the first hypothesis, the existence of a single globally attracting
$E_K$ on each of the six nonempty faces. On a single-strain face that is
Lemma~\ref{l:resface} and Theorem~\ref{t:explicitcat} transplanted, so it
holds under the analogue of \eqref{eq:catregion} for that strain; on a
two-strain face the resident system contains a full cyclic inhibition
$C_{k2}\dashv$ strain $k+1$ and is not cooperative for any parameter values,
so neither argument applies. Three-strain permanence therefore reduces to
global attraction on the three two-strain faces, and that is where the
difficulty now sits. Problem~\ref{o:ThreePermanence} should be read as
asking for exactly that.
\eeR
\end{frame}

\subsection{What has been established and what remains open}

The dynamic quotient theorem and the two aggregation statements it uses are
general, and are proved in Appendix~\ref{sec:gener}.  The three-strain model
establishes both a non-chain resident-face buffering hierarchy and a
genuinely dynamic invasion invariance under eighteen internal conversion
rates, even though the host infection equations themselves are not
autonomous.  The OM lattice identifies where rate perturbations can be
localized; the additional equal-exit identities explain why a specified
subfamily is neutral \emph{along trajectories} rather than only at
equilibrium.  The periodically forced counterexample of
Proposition~\ref{p:periodicCounterexample} proves that this extra step
cannot be omitted.

\paragraph{Scope.}
The three-strain network is a mathematical construction, not an
assertion that a published vaccination model has precisely these
maturation modules.  Its purpose is to separate four claims that
are often conflated: a zero OM index, equilibrium localization,
dynamic quotient invariance, and uniform permanence.  Only the first
three have been established here in the stated generality.
\section{A twenty-species two-patch two-strain model with selective vaccination and staged
immunity}
\label{sec:cwg}

This section combines the two published models of
Section~\ref{sec:twomodels} into one network and computes its complete
buffering lattice. That lattice is by far the largest in this paper and its
epidemiological consequence by far the smallest: every proper buffer lies
downstream of an autonomous infection core, so what it certifies about
thresholds also follows from block triangularity. We keep the computation
because the count is exact, and because the contrast with
Section~\ref{sec:twostrain}, where the modules sit inside the infection loop,
is what makes that section worth reading; Section~\ref{ss:whatitadds} states
the matter precisely.

From \cite{ChenWu} we keep the two patches, the travel loss and the
two competing strains; from \cite{Bulh} we keep vaccination with selective
protection, breakthrough infection of the vaccinated and the staged return to
full immunity. The result is a twenty-compartment model which is a special case
of neither: Bulhosa--Oliveira's secondary-infection pathways are not all
reproduced, and Chen--Wu has no vaccination at all. The point of the
construction is that the added immune machinery is exactly what creates a
nontrivial buffering lattice, and that the lattice can be computed completely.

\subsection{The stoichiometric block of the twenty-compartment model, and its
fixed points face by face}

\begin{frame}{The twenty-species extended model: one patch, and its stoichiometric block}
\label{app:cwgblock}
\begin{columns}[c]
\begin{column}{.47\textwidth}
\begin{center}
\adjustbox{max width=\linewidth,max totalheight=0.75\textheight}{%
\begin{tikzpicture}[>=Latex,x=1cm,y=1cm,
  b/.style={draw,rounded corners,align=center,minimum height=5.5mm,minimum width=9mm,font=\scriptsize},
  lab/.style={font=\tiny}]
\node[b,fill=orange!12] (s)  at (0,0)      {$S_p$};
\node[b,fill=yellow!15] (z)  at (0,-1.8)   {$Z_p$};
\node[b,fill=yellow!15] (v)  at (0,-3.6)   {$V_p$};
\node[b,fill=blue!8]   (i1) at (2.6,1.4)   {$I_{p1}$};
\node[b,fill=blue!8]   (i2) at (2.6,0)     {$I_{p2}$};
\node[b,fill=red!10]   (y)  at (2.6,-1.8)  {$Y_p$};
\node[b,fill=green!10] (c1) at (5.2,1.4)   {$C_{p1}$};
\node[b,fill=green!10] (c2) at (5.2,-1.0)  {$C_{p2}$};
\node[b,fill=green!10] (r)  at (7.2,-2.6)  {$r_p$};
\node[b,fill=gray!12]  (w)  at (9.2,-2.6)  {$R_p$};
\node[draw,dashed,minimum height=14mm,minimum width=11mm,font=\tiny,align=center]
  (q) at (9.2,0.4) {patch\\ $q$};
\draw[->,thick] (-1.6,0)--node[above,lab]{$\Lambda_p$}(s);
\draw[->,thick] (s)--node[above left,lab]{$\beta_{p1}S_pI_{p1}$}(i1);
\draw[->,thick] (s)--node[above,lab]{$\beta_{p2}S_pJ_{p2}$}(i2);
\draw[->,thick] (s)--node[left,lab]{$\xi_{p1}$}(z);
\draw[->,thick] (s) to[out=250,in=110] node[left,lab]{$\xi_{p12}$}(v);
\draw[->,thick] (z)--node[above,lab]{$\sigma^v_p\beta_{p2}Z_pJ_{p2}$}(y);
\draw[->,thick] (i1)--node[above,lab]{$\gamma_{p1}$}(c1);
\draw[->,thick] (i2)--node[above right,lab]{$\gamma_{p2}$}(c2);
\draw[->,thick] (y)--node[below right,lab]{$\gamma^v_p$}(c2);
\draw[->,thick] (c1) to[out=-20,in=130] node[right,lab]{$\theta_{p1}$}(r);
\draw[->,thick] (c2)--node[above right,lab]{$\theta_{p2}$}(r);
\draw[->,thick] (v)--node[below,lab]{$\theta^v_p$}(r);
\draw[->,thick] (r)--node[below,lab]{$\theta^R_p$}(w);
\draw[<->,thick,red!60!black] (s)  to[out=70,in=175] node[above,lab]{$m^S$}(q);
\draw[<->,thick,red!60!black] (i1) to[out=25,in=195] node[above,lab]{$m^{I_1}$}(q);
\draw[<->,thick,red!60!black] (i2) to[out=5,in=215]  node[below,lab]{$m^{I_2}$}(q);
\draw[<->,thick,red!60!black] (w)  to[out=85,in=275] node[right,lab]{$m^R$}(q);
\end{tikzpicture}}
\end{center}

{\slidefont{\tiny} $13$ edges and $13$ non-decay columns per patch, one per edge: the two
strain-$2$ infection routes share the edge $\beta_{p2}S_pJ_{p2}$ and the two
breakthrough routes the edge $\sigma^v_p\beta_{p2}Z_pJ_{p2}$, with
$J_{p2}=I_{p2}+\eta_pY_p$. The four two-way migration edges carry $8$
reactions. The $26$ decays are not drawn.}
\end{column}
\begin{column}{.51\textwidth}\slidefont{\tiny}
$\Gamma$ has $60$ columns: $2\times13$ local, $8$ migration, $26$ decay, and
$\Gamma v=\dot x$ was checked symbolically. The local block of patch $p$, on
$x_p=(S,I_{1},I_{2},Z,Y,C_{1},C_{2},V,r,R)_p$, has one column per edge of the
figure and is headed by its labels:
\[
\adjustbox{max width=\linewidth}{$
\Gamma_p=\begin{array}{r|r|rrrrrrrrrrrr}
&\Lambda_p&\beta_{p1}S_pI_{p1}&\beta_{p2}S_pJ_{p2}&\xi_{p1}&\xi_{p12}&\sigma^v_p\beta_{p2}Z_pJ_{p2}&\gamma_{p1}&\gamma_{p2}&\gamma^v_p&\theta_{p1}&\theta_{p2}&\theta^v_p&\theta^R_p\\\hline
S     & 1&-1&-1&-1&-1& 0& 0& 0& 0& 0& 0& 0& 0\\
I_{1} & 0& 1& 0& 0& 0& 0&-1& 0& 0& 0& 0& 0& 0\\
I_{2} & 0& 0& 1& 0& 0& 0& 0&-1& 0& 0& 0& 0& 0\\
Z     & 0& 0& 0& 1& 0&-1& 0& 0& 0& 0& 0& 0& 0\\
Y     & 0& 0& 0& 0& 0& 1& 0& 0&-1& 0& 0& 0& 0\\
C_{1} & 0& 0& 0& 0& 0& 0& 1& 0& 0&-1& 0& 0& 0\\
C_{2} & 0& 0& 0& 0& 0& 0& 0& 1& 1& 0&-1& 0& 0\\
V     & 0& 0& 0& 0& 1& 0& 0& 0& 0& 0& 0&-1& 0\\
r     & 0& 0& 0& 0& 0& 0& 0& 0& 0& 1& 1& 1&-1\\
R     & 0& 0& 0& 0& 0& 0& 0& 0& 0& 0& 0& 0& 1
\end{array}$}
\]

\textbf{Rates.} Reading the headers as fluxes: $\Lambda_p$,
$\beta_{p1}S_pI_{p1}$, $\beta_{p2}S_pJ_{p2}$, $\xi_{p1}S_p$, $\xi_{p12}S_p$,
$\sigma^v_p\beta_{p2}Z_pJ_{p2}$, $\gamma_{p1}I_{p1}$, $\gamma_{p2}I_{p2}$,
$\gamma^v_pY_p$, $\theta_{p1}C_{p1}$, $\theta_{p2}C_{p2}$, $\theta^v_pV_p$,
$\theta^R_pr_p$, with $J_{p2}=I_{p2}+\eta_pY_p$. The two columns headed by a
$J_{p2}$ flux each lump two elementary reactions of identical stoichiometry,
$\beta_{p2}S_pI_{p2}+\beta_{p2}\eta_pS_pY_p$ and
$\sigma^v_p\beta_{p2}Z_pI_{p2}+\sigma^v_p\beta_{p2}\eta_pZ_pY_p$.

\medskip
\textbf{The other two blocks.} Migration: for $X\in\{S,I_1,I_2,R\}$ and
$q=3-p$, the column $-e_{X_p}+\nu^X_{qp}e_{X_q}$ at rate $m^X_{qp}X_p$, with
$\nu=1-\varepsilon$ the surviving fraction --- travel loss is a coefficient,
not a decay. Decay: $-e_{X_p}$ at rate $\mu_pX_p$ for each of the $20$
species, plus $\alpha_{pk}I_{pk}$ and $\alpha^v_pY_p$.
\end{column}
\end{columns}
\end{frame}

\begin{frame}[allowframebreaks]{Fixed points face by face}
\slidefont{\scriptsize}
Take $\Lambda=(6,5)$, $\mu_p=0.05$, $\gamma=0.2,\ \gamma^v=0.25$, $\alpha=0$,
$\beta_{\cdot1}=(0.003,0.005)$, $\beta_{\cdot2}=(0.0045,0.0035)$,
$\xi_{p1}=0.02$, $\xi_{p12}=0.01$, $\sigma^v_p=0.6$, $\eta_p=0.8$, no
susceptible migration, $m^{I_k}_{pq}=0.01$, $\nu=0.9$.
Every entry of the next table is exact. The eliminants come from a
Gr\"obner basis over $\mathbb Q$, the counts of positive roots are Sturm
counts, each such root is isolated in an interval with rational endpoints,
and admissibility --- whether a root lifts to an equilibrium with all core
coordinates positive --- is decided by elimination over the reals with the
sign constraints imposed.
\[
\adjustbox{max width=\linewidth}{$
\begin{array}{l|c|c|c|c|c}
\text{face}&\text{unknowns}&\text{eliminant}&\text{factor degrees}
&\#\text{roots}>0&\#\text{equilibria}>0\ |\ \ge0\\\hline
M_1\cup M_2=0\ \text{(disease free)}&4&\deg1\text{ in }S_1&1&1&1\ |\ 1\\
M_2=0\ \text{(strain 1 alone)}&6&\deg4\text{ in }I_{11}&1,3&1&1\ |\ 2\\
M_1=0\ \text{(strain 2 alone)}&8&\deg9\text{ in }Y_1&1,8&4&1\ |\ 2\\
\text{whole core (interior)}&10&\deg{22}\text{ in }Y_1&1,8,13&12&1\ |\ 4
\end{array}$}
\]
Each eliminant has $0$ as a root of multiplicity one, the extinction branch,
which is divided out before counting. The strain-$1$ face has the Chen--Wu
cubic form with vaccination loss --- degree $4=1+3$, the linear factor being
$I_{11}$ --- and the monotone argument inherited from Chen--Wu gives its
single endemic point independently of the certificate. The strain-$2$ face is
genuinely larger, $S_p,Z_p,I_{p2},Y_p$ with eight unknowns, and its eliminant
has degree $9$ with \emph{four} positive roots of which exactly one is
admissible; the other three each force a negative coordinate. The interior
eliminant has degree $22$ with twelve positive roots and again exactly one
admissible point, and its factor of degree $8$ is the strain-$2$ face's own
--- their smallest roots agree to the isolation width --- so the thirteen
remaining degrees carry what is new on the interior.

Two consistency checks fall out. The four nonnegative points of the whole
core are precisely the disease-free point, the two single-strain endemic
points and the interior one, which is the $1+1+1+1$ asserted below; and each
resident face carries two nonnegative points, its endemic one together with
the disease-free point lying on it.

\medskip
Here $\mathcal R_{1|0}=1.207$, $\mathcal R_{2|0}=1.518$,
$\mathcal R_{2|1}=1.476$, $\mathcal R_{1|2}=1.099$: all four exceed $1$, the
invasions are mutual, and the core has exactly one interior equilibrium ---
the odd case of the parity rule found for Chen--Wu. The four faces carry
$1+1+1+1$ points in all.

\medskip
\textbf{What this is and is not.} At these rates the count $1+1+1+1$ is now
\emph{proved}, not observed. The parameters are rational, so the face
equations are polynomial systems over $\mathbb Q$; their eliminants are
computed by a Gr\"obner basis, the positive roots are counted by a Sturm
count on $(0,\infty)$ rather than by a numerical solve, each is isolated by
bisection into an interval with rational endpoints, and admissibility is
decided by elimination over the reals with the positivity constraints
imposed, so no sign of an algebraic number is ever decided numerically and no
complex root is ever compared. What this does \emph{not} give is
parameter-independence: it is a theorem about this rational point, and the
model is not claimed to have the same count at other rates. The strain-$1$
face is the only one that also carries a parameter-free proof, through the
monotone argument inherited from Chen--Wu.

\medskip
\textbf{Reproducibility.} The certificates are computed in \texttt{CWG.wl}:
\texttt{certify} returns, for each face, the eliminant and its factor
degrees, the Sturm count \texttt{CountRoots[p,\{x,0,Infinity\}]}, the
isolating intervals from an exact bisection against that count, and the
numbers of solutions with all coordinates $>0$ and $\ge0$ from
\texttt{Solve[\ldots,Reals]} with the sign constraints inside the call. Its
summary line reads
$\{1,1,1,1\}$, $\{4,1,1,2\}$, $\{9,4,1,2\}$, $\{22,12,1,4\}$ for the four
faces, in the order (eliminant degree, positive roots, admissible $>0$,
admissible $\ge0$), which is the table above.

\medskip
\textbf{The certificate itself, for the strain-$2$ face.} The eliminant
cannot be printed: after dividing out $Y_1$ its degree is $8$ and its
primitive integer coefficients reach $46$ digits. What can be printed, and
re-checked by any Sturm or Descartes implementation once the polynomial is
regenerated, are its four isolating intervals with denominator $10^6$, each
carrying Sturm count exactly $1$:
\[
[1.048989,1.048992]^{\ \star},\quad
[0.006323,0.006326],\quad
[1.029343,1.029346],\quad
[1.221273,1.221276],
\]
the starred one being the admissible root. The equilibrium above it is
enclosed to width $3\cdot10^{-6}$ by
\[
\begin{aligned}
S_1&\in[51.372807,51.372810], &I_{12}&\in[7.337096,7.337099],\\
Z_1&\in[14.255180,14.255183], &Y_1&\in[1.048989,1.048992],
\end{aligned}
\]
with $S_2,I_{22},Z_2,Y_2$ likewise, so a reader can substitute the endpoints
into the eight face equations and bracket each residual. The other three
roots each force a negative coordinate, which is the admissibility step. The
interior face has twelve such intervals and a degree-$22$ eliminant; they are
not reproduced here, and \texttt{CWG.wl} prints them.
\end{frame}

\subsection{Compartments and added pathways}

For patches $p=1,2$ and strains $k=1,2$ there are ten compartments per patch.
We keep the letters of the two sources wherever possible:
\[
\begin{array}{c|l|c}
\text{here}&\text{meaning}&\text{source}\\\hline
S_p & \text{susceptible to both strains}&\text{CW }S_i\\
I_{pk} & \text{primary infection with strain }k&\text{CW }I_i,J_i\\
V_p & \text{vaccinated, protected against both strains}&\text{BO }v\\
Z_p & \text{vaccinated against strain 1, susceptible to strain 2}&\text{BO }z\\
Y_p & \text{strain-2 breakthrough infection of a vaccinated host}&\text{BO }y_2\\
C_{pk} & \text{convalescent: immune to }k,\text{ noninfectious}&\text{BO }c_k\\
r_p & \text{cross-immunity lost, not yet fully immune}&\text{BO }r_k\\
R_p & \text{immune to both strains}&\text{CW }R_i,\text{ BO }R
\end{array}
\]
With $J_{p2}=I_{p2}+\eta_pY_p$ the strain-2 force of infection of Gavish, the
pathways added to Chen--Wu are
\[
\begin{gathered}
S_p\xrightarrow{\xi_{p1}}Z_p,\qquad
S_p\xrightarrow{\xi_{p12}}V_p\xrightarrow{\theta^v_p}r_p,
\qquad I_{pk}\xrightarrow{\gamma_{pk}}C_{pk}\xrightarrow{\theta_{pk}}r_p
\xrightarrow{\theta^R_p}R_p,\\
Z_p+I_{p2}\longrightarrow I_{p2}+Y_p,
\qquad Z_p+Y_p\longrightarrow2Y_p,
\qquad Y_p\xrightarrow{\gamma_p^v}C_{p2}.
\end{gathered}
\]
The two $Z_p$ reactions share a stoichiometric column and are carried by the
single rate $\sigma^v_p\beta_{p2}Z_pJ_{p2}$, one arrow for two reactions
exactly as in Bulhosa--Oliveira. Susceptibles, both primary infectious classes
and the fully immune migrate, with Chen--Wu travel loss; migration is absent
from $Z,Y,C,V,r$. Every compartment dies at rate $\mu_p$, the infectious ones
at $\mu_p+\alpha_{pk}$. There is no reinfection from $C_{pk},r_p,R_p$: that
assumption is removed in Section~\ref{ss:prune}, with large consequences for
the lattice. The stoichiometric matrix, $60$ columns of which $26$ are decays,
is displayed in Section~\ref{app:cwgblock}.

\subsection{The equations}

Put $J_{p2}=I_{p2}+\eta_pY_p$, with $\eta_p,\sigma^v_p>0$, and for
$X\in\{S,I_1,I_2,R\}$ and $q=3-p$ let
\[
(\mathcal M^X X)_p=-m^X_{qp}X_p+(1-\varepsilon^X_{pq})m^X_{pq}X_q,
\quad m^X_{pq}>0,\quad 0\le\varepsilon^X_{pq}<1,
\]
where $m^X_{pq}$ denotes movement \emph{from patch $q$ to patch $p$} and
$\varepsilon^X_{pq}$ the fraction lost in transit. The ten transmission and
vaccination equations are, for $p=1,2$,
\[
\begin{aligned}
\dot S_p&=\Lambda_p-\beta_{p1}S_pI_{p1}-\beta_{p2}S_pJ_{p2}
 -(\mu_p+\xi_{p1}+\xi_{p12})S_p+(\mathcal M^SS)_p,\\
\dot I_{p1}&=\beta_{p1}S_pI_{p1}
 -(\mu_p+\alpha_{p1}+\gamma_{p1})I_{p1}+(\mathcal M^{I_1}I_1)_p,\\
\dot I_{p2}&=\beta_{p2}S_pJ_{p2}
 -(\mu_p+\alpha_{p2}+\gamma_{p2})I_{p2}+(\mathcal M^{I_2}I_2)_p,\\
\dot Z_p&=\xi_{p1}S_p-\sigma^v_p\beta_{p2}Z_pJ_{p2}-\mu_pZ_p,\\
\dot Y_p&=\sigma^v_p\beta_{p2}Z_pJ_{p2}
 -(\mu_p+\alpha_p^v+\gamma_p^v)Y_p,
\end{aligned}
\]
and the remaining ten describe immune maturation,
\[
\begin{aligned}
\dot C_{pk}&=\gamma_{pk}I_{pk}
 +\mathbf1_{\{k=2\}}\gamma_p^vY_p-(\mu_p+\theta_{pk})C_{pk},&
\dot V_p&=\xi_{p12}S_p-(\mu_p+\theta^v_p)V_p,\\
\dot r_p&=\theta_{p1}C_{p1}+\theta_{p2}C_{p2}+\theta^v_pV_p
 -(\mu_p+\theta^R_p)r_p,&
\dot R_p&=\theta^R_pr_p-\mu_pR_p+(\mathcal M^RR)_p.
\end{aligned}
\]
All transition, recruitment, recovery and mortality rates are positive. The
first ten equations form an autonomous subsystem: the \emph{infection core}
\[
\mathcal H=\{S_p,I_{p1},I_{p2},Z_p,Y_p:p=1,2\},
\]
on which the remaining ten, the \emph{downstream block}
$\mathcal D=\{C_{pk},V_p,r_p,R_p\}$, depend without feeding back. Everything
that follows must be read against this feed-forward structure; we return to it
in Section~\ref{ss:whatitadds}.

\subsection{Minimal siphons and the four invariant faces}

Each $S_p$ is produced by the inflow $0\to S_p$, which has no reactant, so no
siphon contains $S_p$; then none contains $Z_p$ or $V_p$, produced from $S_p$
alone, nor $r_p$ (from $V_p$), $R_p$ (from $r_p$), $C_{p1}$ (from $I_{p1}$) or
$C_{p2}$ (from $I_{p2}$ and $Y_p$) without the infected species they come
from. What survives are the two strain sets, and both are siphons:
\[
\boxed{M_1=\{I_{11},I_{21}\},\qquad M_2=\{I_{12},I_{22},Y_1,Y_2\}.}
\]
They are minimal. $\{I_{12},I_{22}\}$ is not a siphon, because
$S_p+Y_p\to I_{p2}+Y_p$ produces $I_{p2}$ out of species outside it, and
$\{Y_1,Y_2\}$ is not, because $Z_p+I_{p2}\to I_{p2}+Y_p$ produces $Y_p$ out of
species outside it. Vaccination ties the breakthrough class to strain~$2$
instead of giving it a siphon of its own.

The lattice generated by the two minimal siphons therefore has four elements,
\[
\varnothing\subset M_1,\ M_2\subset M_1\cup M_2,
\]
and these are the disease-free face, the two single-strain faces and the
interior. Two-way migration of $I_1$ makes $M_1$ irreducible; migration of
$I_2$ together with the $I_{p2}\leftrightarrow Y_p$ pathways makes $M_2$
irreducible. A whole strain can be lost; a single infectious patch cannot.
These four are the \emph{infection-extinction regimes}; they are not the
complete lattice of siphons of the network, which contains nonminimal members
such as $M_1\cup\{C_{11}\}$ and plays no role below.

\subsection{The two reproduction functions}
\label{ss:repro}

Following the convention of Section~\ref{sec:repro}, we attach one
reproduction \emph{function} to each minimal siphon, with the resident
coordinates left free, and obtain every threshold as an evaluation of it.

For strain~$1$ the situation is Chen--Wu's, unchanged by the extension: with
$T_1=\operatorname{diag}(\mu_p+\alpha_{p1}+\gamma_{p1})-\mathcal M^{I_1}$,
\[
\mathcal R_1(S_1,S_2)=\rho\bigl(\operatorname{diag}(\beta_{p1}S_p)T_1^{-1}\bigr).
\]
For strain~$2$ the invasion problem is four-dimensional, on
$(I_{12},I_{22},Y_1,Y_2)$, but its gain matrix has \emph{rank two}: patch $p$
contributes the rank-one block
$\beta_{p2}\binom{S_p}{\sigma^v_pZ_p}(1\ \ \eta_p)$. Put $D_S=\operatorname{diag}(\beta_{p2}S_p)$,
$D_Z=\operatorname{diag}(\sigma^v_p\beta_{p2}Z_p)$,
$D_\eta=\operatorname{diag}(\eta_p)$ and
$D_v=\operatorname{diag}(\mu_p+\alpha^v_p+\gamma^v_p)$. Then
\[
\mathsf F_2=AB,\quad
A=\begin{pmatrix}D_S\\D_Z\end{pmatrix},\quad
B=\begin{pmatrix}I_2&D_\eta\end{pmatrix},\quad
\mathsf V_2=\operatorname{diag}(T_2,D_v).
\]
The $AB$--$BA$ spectral identity gives
$\rho(\mathsf F_2\mathsf V_2^{-1})=
\rho(T_2^{-1}D_S+D_\eta D_v^{-1}D_Z)$.
Since $D_S$ is positive diagonal, conjugation by $D_S$ gives the
$2\times2$ problem,
\[
\boxed{\ \mathcal R_2(S,Z)=\rho(K),\qquad
K=\operatorname{diag}(\beta_{p2}S_p)\,T_2^{-1}
+\operatorname{diag}\!\Big(\frac{\beta_{p2}\eta_p\sigma^v_pZ_p}
{\mu_p+\alpha^v_p+\gamma^v_p}\Big),\ }
\]
with $(T_2)_{pp}=\mu_p+\alpha_{p2}+\gamma_{p2}+m^{I_2}_{qp}$ and
$(T_2)_{pq}=-(1-\varepsilon^{I_2}_{pq})m^{I_2}_{pq}$ for $p\neq q$. This is the
Chen--Wu strain-$2$ next generation matrix \emph{plus a diagonal breakthrough
term}: selective vaccination enters the strain-$2$ threshold additively and
patch by patch, through $Z_p$, and never through migration. Indeed $K=D_S(T_2^{-1}D_S+D_\eta D_v^{-1}D_Z)D_S^{-1}$,
so the reduction is exact.

Both functions are increasing in their resident coordinates, and each equals
$1$ at its own endemic equilibrium, checked to twelve digits in
\texttt{CWG.wl}. Consequently
\[
\mathcal R_{1|0}=\mathcal R_1(S^0),\quad
\mathcal R_{2|0}=\mathcal R_2(S^0,Z^0),\quad
\mathcal R_{1|2}=\mathcal R_1(S^{(2)}),\quad
\mathcal R_{2|1}=\mathcal R_2(S^{(1)},Z^{(1)})
\]
are four evaluations of two functions, not four separate computations. The
resident coordinates that enter are $S_p$ and $Z_p$ only: the whole immune
cascade $C_{pk},V_p,r_p,R_p$ is absent from both functions. That observation
is proved structurally, and strengthened, in Section~\ref{ss:whatitadds}.

\subsection{The complete buffering lattice}

Put $\mathcal R=\{R_1,R_2\}$ and, for each patch,
$\mathcal L_p=\{C_{p1},C_{p2},V_p,r_p\}$.

\beT[complete buffering lattice][][lattice83]
For positive transition and migration rates, the complete lattice of
Okada--Mochizuki buffered species sets of the twenty-compartment model is
\[
\begin{gathered}
\mathfrak B=\{\varnothing,\cS\}\ \cup\
\big\{\mathcal R\cup T_1\cup T_2:T_p\in\mathfrak T_p\big\},\qquad
\mathfrak T_p=\{\varnothing\}\cup
\big\{\{r_p\}\cup U:U\subseteq\{C_{p1},C_{p2},V_p\}\big\},\\
\boxed{|\mathfrak B|=2+(1+2^3)^2=83.}
\end{gathered}
\]
\eeT

\begin{proof}
The positive infection and vaccination reactions, together with susceptible
and infectious migration, make the ten species of the core $\mathcal H$ one
strongly connected component of the influence graph: every species of
$\mathcal H$ influences every other. The remaining strongly connected
components are the six singletons $C_{p1},C_{p2},V_p$, the two singletons
$r_p$, and the coupled recovered pair $\mathcal R$, coupled because $R_1$ and
$R_2$ migrate. Their order is
\[
\begin{matrix}
 &\mathcal H&\\[-2pt]
 &\big\downarrow&\\[-2pt]
 \{C_{11},C_{12},V_1\}&\qquad&\{C_{21},C_{22},V_2\}\\[-2pt]
 \big\downarrow&&\big\downarrow\\[-2pt]
 r_1&&r_2\\[-2pt]
 &\searrow\ \swarrow&\\[-2pt]
 &\mathcal R&
\end{matrix}
\]
By Lemma~\ref{l:fwdclosed}, valid because
every species has an independent mortality reaction, a set is buffered exactly
when it is a union of components closed downwards under these arrows. If such a
set meets $\mathcal H$ it contains $\mathcal H$ and hence everything, giving
$\cS$. Otherwise it is empty or contains $\mathcal R$, and in each patch
independently it contains either no further species, or $r_p$ together with an
arbitrary subset of its three predecessors $C_{p1},C_{p2},V_p$. That is
$1+2^3=9$ choices per patch, whence $1+9\cdot9+1=83$.
\end{proof}

The lattice is distributive, as it must be, but it is \emph{not} Boolean. The
clearest symptom is the pair of overlapping closures
\[
\begin{aligned}
\Buf(\{C_{p1}\})&=\mathcal R\cup\{r_p,C_{p1}\},&
\Buf(\{C_{p2}\})&=\mathcal R\cup\{r_p,C_{p2}\},\\
\Buf(\{C_{p1}\})\cap\Buf(\{C_{p2}\})
&=\mathcal R\cup\{r_p\}.&&
\end{aligned}
\]
The two strain-specific recovery branches have distinct minimal buffering
closures which are neither nested nor disjoint. No reproduction number
expresses that relation, and it is the one piece of information in this section
that the classical analysis does not already contain.

\subsection{What the lattice localizes}

At a regular equilibrium, OM localization (Theorem~\ref{t:OMloc}) together
with the closure bound of Corollary~\ref{c:bufclos}
gives the following table, in which the right column is the complete list of
coordinates permitted to move.
\[
\begin{array}{c|c}
\text{Perturbed reaction rate}&\text{Only coordinates permitted to change}\\\hline
C_{pk}\to r_p\ (\theta_{pk})& C_{pk},r_p,R_1,R_2\\
V_p\to r_p\ (\theta^v_p)&V_p,r_p,R_1,R_2\\
r_p\to R_p\ (\theta^R_p)&r_p,R_1,R_2\\
R_1\leftrightarrow R_2\ (m^R)&R_1,R_2
\end{array}
\]
The two closures $\mathcal R\cup\{r_1,C_{11}\}$ and
$\mathcal R\cup\{r_1,C_{12}\}$ overlap precisely in $\mathcal R\cup\{r_1\}$:
rates in the two recovery branches may both move $r_1,R_1,R_2$, but neither
can move the other's convalescent stage.

By contrast $\xi_{p1},\xi_{p12}$ act on $S_p\to Z_p,V_p$, $\gamma_{pk}$ on
$I_{pk}\to C_{pk}$, and $\sigma^v_p$ on breakthrough infection; none of these
perturbations is confined by a proper buffer. Nor is a change in $\mu_p$, a
single parameter shared by all the mortality reactions of a patch: the
localization theorem requires the perturbed rates to lie in one
output-complete set and to be otherwise independent.

\subsection{Invariance of the strain-2 invasion number}

At a strain-$1$ resident equilibrium, $I_{p2}^*=Y_p^*=0$ and
$Z_p^*=\xi_{p1}S_p^*/\mu_p$. In the invasion coordinates
$(I_{12},I_{22},Y_1,Y_2)$ write
\[
\mathsf F_{2|1}=\begin{pmatrix}
\operatorname{diag}(\beta_{p2}S_p^*)&
\operatorname{diag}(\eta_p\beta_{p2}S_p^*)\\
\operatorname{diag}(\sigma^v_p\beta_{p2}Z_p^*)&
\operatorname{diag}(\eta_p\sigma^v_p\beta_{p2}Z_p^*)
\end{pmatrix},\qquad
\mathsf V_{2|1}=\operatorname{diag}\bigl(T_2,
\operatorname{diag}(\mu_p+\alpha_p^v+\gamma_p^v)\bigr),
\]
so that $\mathcal R_{2|1}=\rho(\mathsf F_{2|1}\mathsf V_{2|1}^{-1})$, which is
$\mathcal R_2(S^{(1)},Z^{(1)})$ of Section~\ref{ss:repro}. The resident
subsystem for $(S_p,I_{p1},Z_p)$ does not involve $C_{pk},V_p,r_p,R_p$, hence
\[
\boxed{\partial_{\theta_{pk}}\mathcal R_{2|1}
=\partial_{\theta^v_p}\mathcal R_{2|1}
=\partial_{\theta^R_p}\mathcal R_{2|1}
=\partial_{m^R}\mathcal R_{2|1}=0.}
\]
Selective vaccination behaves differently: $\xi_{p1}$ changes $Z_p^*$ and
$\sigma^v_p$ occurs explicitly in $\mathsf F_{2|1}$, so neither derivative
vanishes.

\subsection{Reinfection and waning prune the lattice}
\label{ss:prune}

Adding a transition from an immune stage back to the infectious core merges
that stage into the core influence block, and the component order of
Theorem~\ref{t:lattice83} collapses accordingly. The resulting lattice sizes are
exact:
\[
\begin{array}{l|r}
\text{Reaction set}&\#\text{ OM buffers}\\\hline
\text{Baseline: permanent protection in }C,V,r,R&83\\
\text{Add }C_{11}\to S_1&47\\
\text{Add cross-infection of all four }C_{pk}&11\\
\text{Previous case plus }r_1\to S_1,\ r_2\to S_2&3
\end{array}
\]
For cross-infection we use $C_{pk}+I_{p,3-k}\to2I_{p,3-k}$.
The graph-closure test of Appendix~\ref{sec:graphclosure} certifies each count.
Initially the strongly connected core $\mathcal H$ has, in each patch,
three independent upstream stages $C_{p1},C_{p2},V_p$ leading through $r_p$
to $\mathcal R$, yielding $2+(1+2^3)^2=83$ closed sets.
The reaction $C_{11}\to S_1$ joins $C_{11}$ to $\mathcal H$, leaving two
such stages in patch $1$ and three in patch $2$, hence
$2+(1+2^2)(1+2^3)=47$. Cross-infection of all four $C_{pk}$ joins
them to $\mathcal H$, leaving only $V_p$ upstream of $r_p$;
thus $2+(1+2)^2=11$. Finally $r_p\to S_p$ joins both $r_p$ and
their upstream $V_p$ to $\mathcal H$, leaving precisely the
three closed sets $\varnothing,\mathcal R,\cS$.
The new arrows connect convalescent stages to the infection core, and this is the structural
mechanism behind the much smaller Bulhosa--Oliveira lattice computed in
Section~\ref{sec:bulhosa}. What matters is the \emph{presence} of a reinfection
channel, not whether its susceptibility coefficient lies below or above the ADE
threshold~$1$.

Within this construction a third strain adds nothing structurally new. Extend
it to $K$ strains by giving each an $I_{pk}\to C_{pk}$ recovery branch and
allowing breakthrough infection of $Z_p$ by the strains $k\ge2$, keeping
everything else as it is; the same component count then gives
\[
\boxed{|\mathfrak B_K|=2+(1+2^{K+1})^2;
\qquad K=2:\ 83,\quad K=3:\ 291.}
\]
The extra strain multiplies branches without creating a new kind of buffering
relation, and the two-strain case already exhibits the nontrivial overlapping
closures and their destruction by reinfection. The claim is about this family
only: it says that adding strains \emph{in this way} leaves the influence order
of the same shape, with one more layer of leaves. A multistrain model with
cross-immunity between the new strains, with strain-specific vaccination, or
with any pathway that feeds a convalescent stage back into the core, has a
different influence order and is not covered by the formula.

\subsection{Boundary regimes and conditional permanence}
\label{sec:permanence}

Write $\mathcal I_1=(I_{11},I_{21})$ and $\mathcal I_2=(I_{12},I_{22},Y_1,Y_2)$.
The infection-free boundary of the autonomous ten-variable core consists of the
three regimes $E_0$ ($\mathcal I_1=\mathcal I_2=0$), $E_1$ ($\mathcal I_2=0$,
strain~$1$ resident) and $E_2$ ($\mathcal I_1=0$, strain~$2$ resident), where
$E_i$ denotes an equilibrium when the corresponding boundary subsystem admits
one; the same faces may also carry non-equilibrium resident dynamics. Positive
recruitment and vaccination give $S_p,Z_p>0$ eventually; positive two-way
infectious migration makes $\mathcal I_1$ irreducible, and the
$I_{p2}\leftrightarrow Y_p$ pathways together with $I_{12}\leftrightarrow
I_{22}$ migration make $\mathcal I_2$ irreducible. So an invariant
disease-boundary community can lose a whole strain but not a single infectious
patch, and the downstream states, having no feedback into the core, create no
additional resident invasion tests.

At $E_0$ the susceptible distribution $S^0$ solves
$0=\Lambda_p-(\mu_p+\xi_{p1}+\xi_{p12})S_p^0+(\mathcal M^SS^0)_p$ and
$Z_p^0=\xi_{p1}S_p^0/\mu_p$; the four thresholds are then the four evaluations
listed in Section~\ref{ss:repro}, and each exceeds $1$ exactly when the
corresponding transverse Metzler invasion matrix has positive spectral bound.

\beT[conditional permanence of both strains][][permcond]
Assume the core extinction boundary has an acyclic Morse decomposition
consisting of the isolated equilibria $E_0,E_1,E_2$, the latter two positive
within their single-strain faces. If
\[
\boxed{\mathcal R_{1|0}>1,\quad\mathcal R_{2|0}>1,
\quad\mathcal R_{1|2}>1,\quad\mathcal R_{2|1}>1,}
\]
then both strains are uniformly persistent (Definition~\ref{d:permanence}):
for every initial condition with
both strains present there is a common $\epsilon>0$ with
\[
\liminf_{t\to\infty}\min_{p,k}I_{pk}(t)\ge\epsilon,
\qquad\liminf_{t\to\infty}\min_pY_p(t)\ge\epsilon,
\]
and consequently all twenty compartments are uniformly persistent. The strict
inequalities give robust permanence under small admissible perturbations
preserving the boundary structure.
\eeT

\begin{proof}
Summing the twenty equations gives, with $N$ the total population and
$\mu_*=\min_p\mu_p>0$, the estimate $\dot N\le\Lambda_1+\Lambda_2-\mu_*N$, so
all positive solutions enter a compact absorbing set; in it, recruitment and
$S_p\to Z_p$ vaccination give uniform eventual lower bounds for $S_p$ and
$Z_p$. At $E_0$ both missing-strain transverse matrices have positive spectral
bound, at $E_1$ the strain-$2$ block does and at $E_2$ the strain-$1$ block
does. The assumed acyclic boundary Morse decomposition and the boundary
invasion criterion of Section~\ref{sec:permcrit}, in the form of
\cite{HofSch}, give uniform persistence of the two infectious blocks. Two-way
migration and the positive $I_{p2}\leftrightarrow Y_p$ pathways propagate the
lower bounds to each infectious coordinate, and the forced equations
\[
\dot V_p=\xi_{p12}S_p-(\mu_p+\theta^v_p)V_p,\qquad
\dot C_{pk}\ge\gamma_{pk}I_{pk}-(\mu_p+\theta_{pk})C_{pk}
\]
bound $V_p$ and $C_{pk}$ away from zero, after which the equations for $r_p$
and $R_p$ give the remaining lower bounds.
\end{proof}

The boundary hypothesis is essential and is not verified here for the whole
parameter family: mutual invasion at two selected equilibria does not by itself
exclude other boundary attractors or invasion cycles. Theorem~\ref{t:permcond} is
therefore a conditional result, and we state it as one.

\subsection{Permanence is independent of the downstream immune rates}

\beT[permanence independent of the downstream immune rates][][permdown]
Fix all parameters of the core $\mathcal H$ and vary independently any positive
rates among
\[
\theta_{pk},\quad\theta^v_p,\quad\theta^R_p,
\quad m^R_{12},m^R_{21},\quad\varepsilon^R_{12},\varepsilon^R_{21}.
\]
Then the extinction or uniform persistence of either strain, the resident core
attractors, and all reproduction and invasion thresholds are unchanged.
Moreover the full twenty-compartment model is uniformly permanent if and only
if its ten-variable core is.
\eeT

\begin{proof}
By Theorem~\ref{t:lattice83} every rate listed lies in a proper buffered set, and
every proper buffered set is contained in $\mathcal D$, hence disjoint from
$\mathcal H$. The core trajectories are therefore identical for all these rate
choices --- for non-equilibrium resident attractors as well as for equilibria
--- so their extinction and permanence properties coincide. If the core is
permanent, the positive forcing along $S_p\to V_p$ and $I_{pk}\to C_{pk}$, then
$C,V\to r\to R$, makes every downstream compartment permanent; the converse is
immediate.
\end{proof}

\subsection{What the lattice adds, and what it does not}
\label{ss:whatitadds}

Put the two structures side by side. The minimal siphons $M_1,M_2$ lie inside
the strongly connected core $\mathcal H$, while every \emph{proper} buffered
set is $\mathcal R\cup T_1\cup T_2$ with
$T_p\subseteq\{r_p,C_{p1},C_{p2},V_p\}$ and so lies inside $\mathcal D$. Hence
\[
\boxed{\text{every proper buffer is disjoint from every minimal siphon,}}
\]
which has an immediate and useful consequence,
Corollary~\ref{c:oneline} of Appendix~\ref{sec:gener}: a buffered set
disjoint from the species entering a siphon's gain and loss matrices freezes
that siphon's reproduction function, provided the confined rates do not occur
explicitly in those matrices. Here that proviso holds, because
$\mathsf F_{2|1}$ and $\mathsf V_{2|1}$ are built
from $\beta_{p2},\eta_p,\sigma^v_p,\mu_p,\alpha,\gamma$ and the migration rates
of $I_2$, none of which is a downstream rate.

For this model the conclusion holds without the branch hypothesis, by the
autonomous core rather than by localization: the four thresholds are
functions of the core coordinates alone, and none of the ten downstream
rates occurs in the core equations, so none of them can move a threshold at
any parameter value.

Here those species are $S_p$ and $Z_p$, so the ten downstream rates
$\theta_{p1},\theta_{p2},\theta^v_p,\theta^R_p,m^R_{pq},\varepsilon^R_{pq}$
leave $\mathcal R_1$ and $\mathcal R_2$ --- and therefore all four thresholds,
every boundary equilibrium of the core and the whole fixed-point count ---
exactly where they were. The computation of the preceding subsection proved
$\partial_\theta\mathcal R_{2|1}=0$ for these rates;
Corollary~\ref{c:oneline} gives the same for all four numbers at once, and
for the functions behind them.

It should be said plainly what this does \emph{not} establish. The buffers sit
entirely downstream, which is where the classical analysis already sees a
feed-forward cascade slaved to the core; for this network the lattice therefore
certifies no threshold fact that the siphon and next-generation analysis
misses. Its own content is of a different kind, namely \emph{which coordinates
move} under a rate perturbation, with the non-Boolean overlap
$\Buf(\{C_{p1}\})\cap\Buf(\{C_{p2}\})=\mathcal R\cup\{r_p\}$ that no
reproduction number expresses. The two structures become genuinely coupled only
when reinfection merges an immune stage into the core --- precisely the move
that cuts the lattice from $83$ to $47$, $11$ and $3$ and at the same time
changes which sets are siphons.

That observation is what Section~\ref{sec:twostrain} answers. There the
buffering modules sit inside the infection loop instead of downstream of it,
and the lattice consequently certifies a statement about an invasion threshold
that no triangularity argument supplies --- which is why that model, and not
this one, carries the paper. 

\appendix
\section{Further general results}
\label{sec:gener}

The results used by the models of
Sections~\ref{sec:twomodels}--\ref{sec:cwg} that are not about any particular
network are gathered here, so that each of those sections carries only its
own model: a graph test for buffered sets in open compartmental networks, a
sensitivity criterion turning canonical reachability into a moving threshold,
a localization statement valid on singular resident faces, the dynamic
invasion certificates behind Corollary~\ref{c:fourteenquot} and
Theorem~\ref{t:threeDynamic} together with the two counterexamples that fix
their scope, and the threshold consequence of a buffer disjoint from a
siphon.

\subsection{Individual mortality reduces the OM index to a graph-closure
test\texorpdfstring{ (Lemma~\ref{l:fwdclosed})}{}}
\label{sec:graphclosure}
\beL[buffered sets in open compartmental networks][Forward-closure test][fwdclosed]
Suppose every species $X$ has an independent reaction $X\to0$.
Make a directed influence graph with an arrow $X\to Y$ whenever
some rate depending on $X$ belongs to a reaction changing $Y$,
with $X\ne Y$. Then
\[
\boxed{\lambda(M)=\rank\Gamma_{\Out(M)}-|M|=0
\quad\Longleftrightarrow\quad
M\text{ is forward-closed in the influence graph}.}
\]
\eeL
\textbf{Proof.} The mortality reactions of species in $M$ give
$\{e_X:X\in M\}\subset\operatorname{im}\Gamma_{\Out(M)}$.
Therefore $\lambda(M)=0$ exactly when every column in
$\Gamma_{\Out(M)}$ lies in $\operatorname{span}\{e_X:X\in M\}$.
This says that reactions depending on species of $M$ change
no species outside $M$, precisely the forward-closure condition.

\medskip
The test does not care how a travel loss is written. The lossy migration
$X_p\to\nu^X_{qp}X_q$ contributes the column
$-e_{X_p}+\nu^X_{qp}e_{X_q}$, which lies in
$\operatorname{span}\{e_X:X\in M\}$ iff $X_q\in M$: exactly the same
forward-closure condition as if it were split into a successful migration and
a death during travel. All incidence terms here have no state-dependent
population denominator.

\subsection{Buffers disjoint from a
siphon\texorpdfstring{ (Corollary~\ref{c:oneline})}{}}

When a buffered set happens to miss every species the siphon's gain and loss
matrices are built from, localization becomes a one-line threshold statement.

\beC[reproduction functions of a disjoint siphon][][oneline]
Let $M$ be a buffered set at a regular equilibrium and let $\theta$ affect only
rates of $\Out(M)$. Suppose
\begin{enumerate}\setlength\itemsep{1pt}
\item[(i)] $M$ is disjoint from the species entering the gain and loss matrices
of a siphon, and
\item[(ii)] $\theta$ does not occur explicitly in those matrices.
\end{enumerate}
Then that siphon's reproduction \emph{function} has vanishing
$\theta$-derivative at that equilibrium, hence along every branch of regular
resident equilibria to which Theorem~\ref{t:bufInv} applies, and so has
every threshold obtained by evaluating it along such a branch.
\eeC

Hypothesis (ii) is not implied by (i) and is not a formality: a rate confined
by a buffer could still appear as an entry of $\mathsf F$ or $\mathsf V$, in
which case the invasion matrix moves although the resident equilibrium does
not.

\subsection{Threshold
sensitivity\texorpdfstring{ (Theorem~\ref{t:sensitivity})}{}}

\textbf{From generic influence to a moving threshold.} Remark~\ref{r:ccfscope}
left a gap: canonical reachability says which equilibrium coordinates respond
generically, but a nonzero response of $x^*_Q$ need not move
$\rho(FV^{-1})$, since the Perron root could be insensitive through
cancellation. Under the sign hypotheses of the theorem below it cannot;
the fixed-sign subclass of Definition~\ref{d:RN} supplies only the
state-independent sign of each
$\partial v_j/\partial y_i$, and Remark~\ref{r:whysigns} says why the
combinations forming $F$ and $V$ need a hypothesis of their own.

\beT[Sign-definite next-generation dependence forces a moving threshold][Threshold sensitivity][sensitivity]
Let $\Sigma$ be a siphon with a regular face, $y^*(\theta)$ a smooth branch
of resident equilibria, and $\Jinv=F-V$ a next-generation splitting with
$F\ge0$, $V$ a nonsingular $M$-matrix and $K=FV^{-1}$ irreducible with
$\rho(K)>0$, which rules out the degenerate $K=[0]$ that the usual
convention counts as irreducible. Let $Q$
be the resident coordinates on which $F$ or $V$ depends, assume that
$\theta$ acts on $F$ and $V$ \emph{only} through $y^*(\theta)$, so that
$\partial_\theta F=\partial_\theta V=0$ at fixed $y$ --- without this a
direct term can reverse the sign below --- and suppose
\begin{enumerate}
\item $\varepsilon\,\partial_\theta y^*_i\ge0$ for every $i\in Q$, with a
single sign $\varepsilon\in\{\pm1\}$;
\item $\partial F/\partial y_i\ge0$ and $\partial V/\partial y_i\le0$
entrywise for every $i\in Q$;
\item \textbf{(coupled nondegeneracy)} there is one index $i_0\in Q$ with
$\varepsilon\,\partial_\theta y^*_{i_0}>0$ at which $\partial F/\partial
y_{i_0}\ne0$ or $\partial V/\partial y_{i_0}\ne0$, both evaluated at the
same point of the branch.
\end{enumerate}
Then
\[
\boxed{\ \varepsilon\,\frac{d}{d\theta}\rho\bigl(FV^{-1}\bigr)>0, \ }
\]
so the reproduction number strictly moves, and crosses $1$ as soon as the
branch carries it across.
\eeT
\textbf{Proof.} By the chain rule
\[
\frac{dK}{d\theta}
=\Bigl(\sum_{i\in Q}\frac{\partial F}{\partial y_i}\partial_\theta y^*_i\Bigr)V^{-1}
+FV^{-1}\Bigl(\sum_{i\in Q}\Bigl(-\frac{\partial V}{\partial y_i}\Bigr)
\partial_\theta y^*_i\Bigr)V^{-1}.
\]
Since $V$ is a nonsingular $M$-matrix, $V^{-1}\ge0$; by hypothesis
$F\ge0$, $\partial F/\partial y_i\ge0$ and $-\partial V/\partial y_i\ge0$;
and every $\varepsilon\,\partial_\theta y^*_i\ge0$. So every summand of
$\varepsilon\,dK/d\theta$ is entrywise nonnegative, hence no cancellation is
possible, and hypothesis~3 makes the summand $i_0$ nonzero: for
$\partial F/\partial y_{i_0}\ne0$ this uses $V^{-1}\ge0$ nonsingular, and
for $\partial V/\partial y_{i_0}\ne0$ it uses in addition $F\ne0$, which
holds because $\rho(K)>0$. Thus
$\varepsilon\,dK/d\theta\ge0$ entrywise and $\ne0$. As $K$ is irreducible
and nonnegative,
Perron--Frobenius gives $u\gg0$ and $w\gg0$ with $Ku=\rho u$,
$w^\top K=\rho w^\top$, and differentiating $w^\top Ku=\rho\,w^\top u$,
\[
\frac{d\rho}{d\theta}
=\frac{w^\top(dK/d\theta)u}{w^\top u},
\]
the terms in $du,dw$ cancelling. A nonzero entrywise-signed matrix
sandwiched between strictly positive vectors has nonzero value of that sign.
\hfill$\square$

\beR[Why this needs the signed class][Role of fixed signs][whysigns]
Hypothesis 2 is a sign condition on how the invasion machinery depends on
the residents, and it is \emph{not} a consequence of Definition~\ref{d:RN}:
the entries of $F$ and $V$ are stoichiometric combinations of rates, and a
combination of fixed-sign derivatives with coefficients of both signs need
not have a fixed sign. The fixed-sign subclass of Definition~\ref{d:RN}
supplies the ingredient --- each
$\partial v_j/\partial y_i$ has a sign that does not depend on the state ---
and Hypothesis 2 adds that the combinations forming $F$ and $V$ do not mix
those signs. That is why it is stated as a hypothesis on the splitting and
not inherited from the class. It holds whenever each entry of $F$ and of $V$
is a single transmission, transfer or exit rate whose derivatives carry the
orientation the hypothesis asks for --- each $y$-derivative of such an entry
is one $\partial v_j/\partial y_i$, required nonnegative in $F$ and
nonpositive in $V$. The inhibited transmission of
Section~\ref{sec:twostrain} is a single rate with the other orientation:
$C_2$ belongs to $Q$ for the strain-$2$ splitting and
$\partial\bigl(\beta_{2g}S_g/(1+\chi C_2)\bigr)/\partial C_2<0$, so for
$\chi>0$ hypothesis~2 fails there and the theorem decides nothing about
that invasion. A species inhibiting one reaction
and catalysing another would make $\partial F/\partial y_i$ change
sign and the sandwich argument fail --- which is the concrete reason for
excluding that case rather than merely not needing it. Combined with
Remark~\ref{r:ccfscope}, the pair now reads: canonical reachability decides
\emph{whether} a rate can reach the invasion-relevant coordinates, and this
theorem decides that when it does, and the three hypotheses of
Theorem~\ref{t:sensitivity} hold, the threshold genuinely moves.
\eeR

\subsection{Localization on a singular resident
face\texorpdfstring{ (Theorem~\ref{t:singular})}{}}

Every localization statement so far has
inverted $J_\Sigma$. Remark~\ref{r:faceScope} showed the index can go
negative without it, so no index criterion can survive unchanged; but the
\emph{conclusion} does survive, in the only form it can when the equilibrium
is not isolated.

\beT[Localization along a positive-dimensional component of resident equilibria][Localization on a singular face][singular]
Let $\Sigma$ be a siphon, $(M,R_\Sigma)$ output-complete in its resident
network, and let $\mathcal E\subseteq F_\Sigma$ be a connected component of
relative-interior resident equilibria, of dimension $d\ge0$, along which
$\rank J_\Sigma$ is constant, so that $\ker J_\Sigma=T_{y}\mathcal E$ has
dimension $d$ at each $y\in\mathcal E$. Assume also that for the parameter
$\theta$ below the equilibrium admits a $C^1$ continuation
$\theta\mapsto y^*(\theta)\in\mathcal E_\theta$ --- linearized solvability
alone does not produce one. Then the following are equivalent:
\begin{equation}\label{eq:singcrit}
\operatorname{im}\Gamma^\Sigma_{R_\Sigma}
\subseteq\operatorname{im}\bigl(J_\Sigma\bigr)_{:,M}
\quad\Longleftrightarrow\quad
\rank\bigl[(J_\Sigma)_{:,M}\mid\Gamma^\Sigma_{R_\Sigma}\bigr]
=\rank(J_\Sigma)_{:,M},
\end{equation}
and when they hold, for every $\theta$ affecting only rates of $R_\Sigma$ the
first-order response of the component satisfies
\[
\partial_\theta y^*\in
\operatorname{span}\{e_i:i\in M\}+T_y\mathcal E .
\]
That is, the equilibrium may be moved inside $M$, uniquely modulo motion
along $\mathcal E$. If $d=0$ the conclusion is
$P_{T\setminus M}\partial_\theta y^*=0$ and \eqref{eq:singcrit} is
equivalent to $\lambda_\Sigma(M,R_\Sigma)=0$, so this contains
Theorem~\ref{t:bufInv}(1).
\eeT
\textbf{Proof.} The equivalence in \eqref{eq:singcrit} is the definition of
containment of a column space. Differentiating the equilibrium equations
along the branch gives $J_\Sigma\,\partial_\theta y^*
=-\Gamma^\Sigma_{R_\Sigma}\partial_\theta v_{R_\Sigma}$, which is solvable
because the right side lies in
$\operatorname{im}\Gamma^\Sigma_{R_\Sigma}\subseteq
\operatorname{im}(J_\Sigma)_{:,M}=J_\Sigma\operatorname{span}\{e_i:i\in M\}$;
choose the solution in that span. Any two solutions differ by an element of
$\ker J_\Sigma=T_y\mathcal E$, which is the stated uniqueness. For $d=0$,
$J_\Sigma$ is invertible, $\operatorname{im}(J_\Sigma)_{:,M}$ has dimension
$|M|$, and containment of
$\operatorname{im}\Gamma^\Sigma_{R_\Sigma}$ in it says
$\rank\Gamma^\Sigma_{R_\Sigma}\le|M|$, hence $=|M|$ by output completeness,
which is $\lambda_\Sigma=0$. \hfill$\square$

\beR[What the singular criterion does and does not buy][Scope][singscope]
\eqref{eq:singcrit} is a rank test, but --- and this must be said, because
the opposite is tempting --- it is \emph{not} a test on the stoichiometric
matrix alone. From
$(J_\Sigma)_{:,M}=\Gamma^\Sigma_{\Out_\Sigma(M)}
(D_Mv^\Sigma)_{\Out_\Sigma(M),:}$ one gets only the inclusion
$\operatorname{im}(J_\Sigma)_{:,M}\subseteq
\operatorname{im}\Gamma^\Sigma_{\Out_\Sigma(M)}$, and it is strict in
general, the derivative factor having $|\Out_\Sigma(M)|$ rows and $|M|$
columns and so not being surjective as soon as $|\Out_\Sigma(M)|>|M|$. Two
reactions $X\to0$ and $X\to Y$ already show it: $\Gamma_{\Out(\{X\})}$ has
two independent columns while $J_{:,\{X\}}$ has one. So
\eqref{eq:singcrit} is an instantaneous condition at a point of
$\mathcal E$, depending on the kinetics there, and the index $\lambda$ is
not recoverable from it. What it does not give either is a \emph{number}
playing the role of
$\lambda$, nor a lattice: the family of sets $M$ satisfying
\eqref{eq:singcrit} need not be closed under union, because column spaces of
a union are not determined by those of the parts. Whether a submodular
substitute exists on singular faces, and whether the quotient by
$T_y\mathcal E$ carries a distributive lattice, is what
Problem~\ref{o:SingularFaceBuffer} asks.
\eeR

\subsection{Dynamic invasion
certificates\texorpdfstring{ (Theorem~\ref{t:dynamicOM})}{}}

The main text certifies invasion numbers evaluated at \emph{regular resident
equilibria}, with one exception: Corollary~\ref{c:fourteenquot} transports
the whole invasion cocycle of the fourteen-species model along arbitrary
resident trajectories, and Theorem~\ref{t:threeDynamic} does the same in the
forty-five-species model. The general statement behind both is
Theorem~\ref{t:dynamicOM}. What it needs in order to be a theorem rather
than a restatement are the two counterexamples below, which show that an
Okada--Mochizuki index-zero buffer does \emph{not} by itself imply the
quotient hypothesis \eqref{eq:factor}, one periodically forced and one
autonomous.
Consider a reaction network restricted to an invariant resident face,
\begin{equation}\label{eq:resident}
 \dot x=\Gamma_Rv_R(x;\theta)+\Gamma_{R^c}v_{R^c}(x),
 \qquad x\in\R^{n}_{\geq0}.
\end{equation}
The parameter $\theta$ occurs only in rates indexed by $R$.  Let
$M$ be an output-complete species set, $R\supseteq\Out(M)$, such that
\begin{equation}\label{eq:buffer}
 \operatorname{rank}\Gamma_R=|M|.
\end{equation}
Choose a full-row-rank matrix $L\in\R^{(n-|M|)\times n}$ with
$L\Gamma_R=0$, and put $y=Lx$.  Equation \eqref{eq:buffer}
guarantees such an $L$, but by itself does \emph{not} guarantee an
autonomous equation for $y$.

Suppose an absent strain has transversal linearization
$\dot z=J_{\mathrm{tra}}(x(t);\theta)z$.  The term ``invasion exponent''
means the upper growth exponent of its evolution operator; on a periodic
resident it is the principal Floquet exponent, and on an ergodic resident
it is the top Lyapunov exponent almost surely.

\beT[Dynamic invasion invariance from an OM quotient][][dynamicOM]
Assume that on a positively invariant region there are functions $g$ and
$\widehat J$ independent of $\theta$ such that
\begin{align}
 L\Gamma_{R^c}v_{R^c}(x)&=g(Lx),\label{eq:factor}\\
 J_{\mathrm{tra}}(x;\theta)&=\widehat J(Lx).\label{eq:invfactor}
\end{align}
For any two parameter values and resident solutions whose initial values
satisfy $Lx_{\theta}(0)=Lx_{\theta'}(0)$, the quotient paths and the
transversal evolution operators coincide at all times.  Consequently their
finite-time growth factors, principal Floquet exponents whenever the
quotient resident is periodic, and top Lyapunov exponents along any common
ergodic quotient measure are identical.
\eeT
\emph{Proof.} Multiplying \eqref{eq:resident} by $L$ gives
$\dot y=g(y)$, independently of $\theta$.  Uniqueness gives
$Lx_\theta(t)=Lx_{\theta'}(t)$ for the same quotient initial datum.
By \eqref{eq:invfactor} the two transversal equations then have exactly
the same coefficient matrix at each time.  Their fundamental matrices and
all stated exponents agree. \hfill$\square$

The nontrivial extra hypothesis is \eqref{eq:factor}: an OM index-zero
condition yields a linear annihilator of the perturbed stoichiometric
columns, but the unperturbed rates need not factor through it.  In
particular, a buffer can preserve an equilibrium invasion number while
changing the invasion exponent along a periodic resident.

\subsection{A periodically forced
counterexample\texorpdfstring{ (Proposition~\ref{p:periodicCounterexample})}{}}

\beP*[A three-species periodically forced counterexample to dynamic invariance from the OM index alone][][periodicCounterexample]
\textbf{Scope.} The network below is \emph{periodically forced}, hence
nonautonomous, whereas Definition~\ref{d:RN} is autonomous. What it refutes
is therefore the implication ``zero OM index $\Rightarrow$ invariance of the
invasion exponent along a non-equilibrium resident'' \emph{within the class
of periodically forced reaction networks}. It is the shorter of the two
examples, and the arithmetic is explicit; the autonomous statement, in the
class Definition~\ref{d:RN} actually defines, is
Proposition~\ref{p:AutonomousFloquet} below.

Let $b,\delta>0$, $u_0>\varepsilon>0$, and consider
\begin{equation}\label{eq:counterexample}
 \dot B=u_0+\varepsilon\cos t-bB,\qquad
 \dot P=bB-\delta P,\qquad
 \dot z=(\alpha P^2-d)z.
\end{equation}
For the resident reaction representation
$0\to B$ at rate $u_0+\varepsilon\cos t$,
$B\to P$ at rate $bB$, and $P\to0$ at rate $\delta P$,
$M=\{B\}$ is an OM buffer: $\Out(M)=\{B\to P\}$ and its
stoichiometric column has rank $1=|M|$.  At constant input $u_0$ the
resident equilibrium is $P^*=u_0/\delta$, independent of $b$, and its
invasion growth rate is $\alpha(u_0/\delta)^2-d$.  Along the unique
positive $2\pi$-periodic resident, however, the invasion exponent is
\begin{equation}\label{eq:counterLE}
 \lambda(b)=\alpha\left\{\frac{u_0^2}{\delta^2}
 +\frac{\varepsilon^2b^2}{2(b^2+1)(\delta^2+1)}\right\}-d.
\end{equation}
Thus, with $u_0=2$, $\varepsilon=\delta=\alpha=1$ and $d=4.1$,
$\lambda(0.1)<0<\lambda(10)$ although the OM buffer and the constant-input
equilibrium invasion number do not change.
\eeP
\emph{Proof.} The periodic solution has mean $\overline P=u_0/\delta$.
Its complex first-harmonic amplitude is
$\varepsilon b/[(b+i)(\delta+i)]$.  Therefore
$\overline{P^2}=(u_0/\delta)^2+
\varepsilon^2b^2/[2(b^2+1)(\delta^2+1)]$.
Integrating $\dot z/z$ over a period proves \eqref{eq:counterLE}.
For the displayed values,
$\lambda(b)=-0.1+b^2/[4(b^2+1)]$. \hfill$\square$

\subsection{The same separation in a five-species autonomous mass-action
network\texorpdfstring{ (Proposition~\ref{p:AutonomousFloquet},
Corollary~\ref{c:amplitude})}{}}
\label{ss:autonomous}

The forcing in \eqref{eq:counterexample} is a convenience, not the mechanism.
The next proposition removes it: the oscillation is produced by the resident
kinetics themselves, every rate is mass-action, every stoichiometric column
is integral, and the conclusion is the same. What drives it is that the
transfer function from the oscillating input to the stored variable depends
on $b$ away from zero frequency while leaving the mean untouched, so an
exponent that is an average of $P^2$ sees $b$ although the equilibrium
invasion number cannot.

\beP[A five-species autonomous mass-action counterexample][Autonomous Floquet counterexample][AutonomousFloquet]
There exists an autonomous mass-action reaction network with a zero-index
buffered species set $M$, a positive regular resident equilibrium and an
attracting positive resident periodic orbit, and a positive rate parameter
$b$ confined to $\Out(M)$, such that every resident-equilibrium invasion
number is independent of $b$ whereas the principal Floquet invasion exponent
along the periodic resident is strictly increasing in $b$ and crosses zero.
\eeP
\emph{Proof.} Choose $a,\omega,c,\delta,\kappa,\sigma>0$ and
$x_0,y_0>\sqrt a$, and write $u=X-x_0$, $v=Y-y_0$, $r^2=u^2+v^2$. Consider
\begin{equation}\label{eq:autonomous}
\begin{aligned}
\dot X&=XY\bigl[(a-r^2)u-\omega v\bigr],&
\dot Y&=XY\bigl[\omega u+(a-r^2)v\bigr],\\
\dot B&=cX-bB,&
\dot P&=bB-\delta P,\\
\dot z&=(\kappa P^2-d)z-\sigma z^2,&&b,d>0,
\end{aligned}
\end{equation}
whose last coordinate is absent on the invariant face $z=0$. For the
resident $(X,Y,B,P)$ the only rate depending on $B$ is $bB$, with
stoichiometric column $e_P-e_B$, so
\[
\Out(\{B\})=\{B\to P\},\qquad
\lambda(\{B\})=\rank(e_P-e_B)-1=0 .
\]
At every resident equilibrium $B^*=cX^*/b$, $P^*=cX^*/\delta$ and
$R_{z\mid*}=\kappa(P^*)^2/d$, so the equilibrium invasion number does not
involve $b$. In particular $(X^*,Y^*)=(x_0,y_0)$, where the $(X,Y)$
Jacobian is $x_0y_0\binom{a\ -\omega}{\omega\ \ \ a}$, nonsingular, and the
$B,P$ directions contribute $-b,-\delta$: the resident equilibrium is
regular.

On $X,Y>0$, polar coordinates about $(x_0,y_0)$ give
\[
\dot r=XY\,r(a-r^2),\qquad \dot\phi=\omega XY,
\]
so the circle $r=\sqrt a$ lies strictly inside the positive quadrant, is
orbitally asymptotically stable, and is traversed with finite period $T$ and
frequency $\Omega=2\pi/T$. Along it $X(t)$ is nonconstant, and it is
independent of $b$, the $(X,Y)$ subsystem being autonomous and unforced by
$B,P$. The stable linear $B,P$ equations then have a unique attracting
$T$-periodic solution $(B_b,P_b)$, so the resident system has an attracting
positive periodic orbit for every $b>0$.

Writing $X(t)=\sum_{n\in\mathbb Z}\widehat X_ne^{in\Omega t}$ with
$\widehat X_{-n}=\overline{\widehat X_n}$, the two linear equations give
\[
\widehat P_{b,n}=\frac{cb}{(b+in\Omega)(\delta+in\Omega)}\widehat X_n,
\qquad
\widehat P_{b,0}=\frac{c\widehat X_0}{\delta}.
\]
The transverse equation at $z=0$ is the scalar $\dot z=(\kappa P_b^2-d)z$,
whose principal Floquet exponent is the mean $\kappa\overline{P_b^2}-d$, so
by Parseval
\begin{equation}\label{eq:autoLE}
\boxed{\
\Lambda(b)=\kappa\Bigl(\frac{c\widehat X_0}{\delta}\Bigr)^2-d
+2\kappa c^2\sum_{n\ge1}
\frac{b^2\,|\widehat X_n|^2}{(b^2+n^2\Omega^2)(\delta^2+n^2\Omega^2)} .\ }
\end{equation}
Each factor $b^2/(b^2+n^2\Omega^2)$ is strictly increasing on $b>0$, and
some $\widehat X_n$ with $n\ne0$ is nonzero because $X$ is nonconstant, so
$\Lambda$ is strictly increasing, from
$\kappa(c\widehat X_0/\delta)^2-d$ as $b\downarrow0$ to that value plus
$2\kappa c^2\sum_{n\ge1}|\widehat X_n|^2/(\delta^2+n^2\Omega^2)$ as
$b\to\infty$. Choosing $d$ strictly between the two corresponding values
before subtraction produces a unique $b_{\rm crit}>0$ with
$\Lambda(b_{\rm crit})=0$, while every equilibrium invasion number stays
independent of $b$.

It remains to see that \eqref{eq:autonomous} is mass-action with integral
columns. Both $\dot X$ and $\dot Y$ are polynomials divisible by $XY$, so
every monomial occurring in them is $kX^pY^q$ with $p,q\ge1$; such a
monomial is realized, according to its sign, by $pX+qY\to(p\pm1)X+qY$, whose
products have nonnegative coefficients precisely because $p\ge1$. That is
what multiplying the oscillator by $XY$ achieves. The remaining reactions
are
\[
X\to X+B,\quad B\to P,\quad P\to0,\quad
2P+z\to2P+2z,\quad z\to0,\quad 2z\to z,
\]
with rates $cX$, $bB$, $\delta P$, $\kappa P^2z$, $dz$, $\sigma z^2$. No
forcing and no non-mass-action rate is used. \hfill$\square$

The same Fourier formula bounds how much invariance can be lost, which
settles the quantitative question left open under
Problem~\ref{o:DynamicInvBuffer}.

\beC[The oscillation amplitude bounds the loss of dynamic invariance][Amplitude bound][amplitude]
For the network \eqref{eq:autonomous}, with $T$ the period of its attracting
orbit, $\Omega=2\pi/T$ and $\Lambda(0),\Lambda(\infty)$ the limits of
$\Lambda(b)$ as $b\downarrow0$ and $b\to\infty$,
\begin{equation}\label{eq:amplitude}
\boxed{\
0<\Lambda(\infty)-\Lambda(0)
\le\frac{\kappa c^2}{\delta^2+\Omega^2}\operatorname{Var}_T(X)
\le\frac{\kappa c^2\,a}{\delta^2+\Omega^2},\ }
\end{equation}
where
$\operatorname{Var}_T(X)=\frac1T\int_0^T\bigl(X-\frac1T\int_0^TX\bigr)^2$.
So the whole discrepancy between equilibrium OM invariance and dynamic
invasion invariance is $O(a)$ as the oscillator radius $\sqrt a$ tends to
zero.
\eeC
\textbf{Proof.} Letting $b\to\infty$ and $b\downarrow0$ in
\eqref{eq:autoLE},
\[
\Lambda(\infty)-\Lambda(0)
=2\kappa c^2\sum_{n\ge1}\frac{|\widehat X_n|^2}{\delta^2+n^2\Omega^2},
\]
which is strictly positive exactly when some $\widehat X_n$, $n\ne0$, is
nonzero, that is when $X$ is nonconstant. Bounding every denominator below
by $\delta^2+\Omega^2$ and using Parseval in the form
$2\sum_{n\ge1}|\widehat X_n|^2=\operatorname{Var}_T(X)$ gives the middle
inequality. On the orbit $(X-x_0)^2+(Y-y_0)^2=a$, so
$|X-x_0|\le\sqrt a$ pointwise and
$\operatorname{Var}_T(X)\le\overline{(X-x_0)^2}\le a$, the variance about
the mean being at most the second moment about any other point. \hfill$\square$

\begin{center}\slidefont{\small}
\begin{tabular}{p{.3\linewidth}p{.3\linewidth}p{.3\linewidth}}\toprule
 & Proposition~\ref{p:periodicCounterexample} & Proposition~\ref{p:AutonomousFloquet}\\\midrule
Resident & periodically forced & autonomous\\
Oscillation from & the input $u_0+\varepsilon\cos t$ & the kinetics, a stable
circle $r=\sqrt a$\\
Kinetics & mass-action plus forcing & mass-action throughout\\
$\Lambda$ depends on $b$ & through one harmonic & through every harmonic of
$X$\\\bottomrule
\end{tabular}
\end{center}

Together the two propositions say that the zero index alone never suffices,
in either class, and that the hypotheses of Theorem~\ref{t:dynamicOM} are
doing real work: Problem~\ref{o:DynamicInvBuffer} asked for exactly this
separation, and the autonomous half of it is settled here.

\subsection{Positive and weighted
aggregation\texorpdfstring{ (Theorem~\ref{t:positiveAggregation},
Corollary~\ref{c:weighted})}{}}

A useful alternative to the exact matrix condition \eqref{eq:invfactor}
is possible for a \emph{positive aggregated invasion system}.

\beT[Positive aggregation preserves the principal invasion exponent][][positiveAggregation]
Let $A_\theta(t)$ be locally bounded Metzler matrices and
$\Pi$ a rectangular $0$--$1$ matrix having exactly one $1$ in each
column and at least one $1$ in every row.  Assume
\begin{equation}\label{eq:linearLumping}
 \Pi A_\theta(t)=\widehat A(t)\Pi,
\end{equation}
where $\widehat A(t)$ is independent of $\theta$.  If $\Phi_\theta(t,s)$
and $\widehat\Phi(t,s)$ are the corresponding evolution operators, then
\begin{equation}\label{eq:normEquality}
 \|\Phi_\theta(t,s)\|_1=\|\widehat\Phi(t,s)\|_1,
 \qquad t\geq s.
\end{equation}
In particular, the upper principal Lyapunov exponent is independent of
$\theta$.  In the periodic case the principal Floquet multiplier and
exponent are independent of $\theta$, although the other Floquet
multipliers need not be.
\eeT
\emph{Proof.} Equation \eqref{eq:linearLumping} yields
$\Pi\Phi_\theta=\widehat\Phi\Pi$.  Both evolution operators are
nonnegative.  Since $\boldsymbol 1^\top\Pi=\boldsymbol 1^\top$,
for every coordinate vector $e_j$,
\[
 \boldsymbol 1^\top\Phi_\theta e_j
 =\boldsymbol 1^\top\widehat\Phi\Pi e_j.
\]
Every $\Pi e_j$ is a coordinate vector, and all quotient coordinate
vectors occur.  Taking the largest column sum proves
\eqref{eq:normEquality}.  Taking growth rates proves the Lyapunov
assertion.  For a periodic system, apply the same identity to integer
powers of the monodromy matrix and use the spectral-radius formula.
\hfill$\square$

The $0$--$1$ hypothesis on $\Pi$ can be relaxed, at the cost of the exact
norm equality \eqref{eq:normEquality}, which the conclusion does not need.

\beC[Weighted aggregation suffices][Weighted aggregation][weighted]
Let $\Pi\ge0$ have exactly one strictly positive entry in each column and at
least one in every row, and let $m,M>0$ be the smallest and largest of those
entries. Under the same intertwining \eqref{eq:linearLumping},
\begin{equation}\label{eq:weighted}
\frac mM\,\|\widehat\Phi(t,s)\|_1
\le\|\Phi_\theta(t,s)\|_1
\le\frac Mm\,\|\widehat\Phi(t,s)\|_1,
\qquad t\ge s,
\end{equation}
so the upper principal Lyapunov exponent is still independent of $\theta$,
and in the periodic case so are the principal Floquet exponent and
multiplier.
\eeC
\textbf{Proof.} The norm equality \eqref{eq:normEquality} is not available
here: it is an identity for a $0$--$1$ aggregation, and the unweighted matrix
$\Pi_0$ carrying the support pattern of $\Pi$ need not intertwine
$A_\theta$ with any $\theta$-independent matrix, so $\widehat\Phi$ is tied to
$\Pi$ and not to $\Pi_0$. We estimate column by column instead.

Write $\Pi e_j=d_je_{r(j)}$, where $d_j\in[m,M]$ is the single positive entry
of column $j$ and $r(j)$ its row, so that
$\boldsymbol 1^\top\Pi=(d_1,\dots,d_n)$ and, for every nonnegative $z$,
\[
 m\|z\|_1\le\|\Pi z\|_1=\sum_jd_jz_j\le M\|z\|_1 .
\]
Both flows are nonnegative: $A_\theta$ is Metzler, and so is $\widehat A$,
since for $k\ne r(j)$ the $k$th entry of
$\Pi A_\theta e_j=d_j\widehat Ae_{r(j)}$ is a nonnegative combination of
off-diagonal entries of $A_\theta$, and every quotient index is some $r(j)$.
Applying $\Pi\Phi_\theta(t,s)=\widehat\Phi(t,s)\Pi$ to $e_j$ gives
$\Pi\Phi_\theta(t,s)e_j=d_j\widehat\Phi(t,s)e_{r(j)}$, so the display above
with $z=\Phi_\theta(t,s)e_j\ge0$ yields the column estimate
\[
 m\|\Phi_\theta(t,s)e_j\|_1
 \le d_j\|\widehat\Phi(t,s)e_{r(j)}\|_1
 \le M\|\Phi_\theta(t,s)e_j\|_1 .
\]
Its left half with $d_j\le M$ gives
$\|\Phi_\theta e_j\|_1\le\frac Mm\|\widehat\Phi e_{r(j)}\|_1$, and taking the
largest column sum over $j$ bounds $\|\Phi_\theta\|_1$ by
$\frac Mm\|\widehat\Phi\|_1$. Its right half with $d_j\ge m$ gives
$\|\widehat\Phi e_{r(j)}\|_1\le\frac Mm\|\Phi_\theta e_j\|_1$; every row of
$\Pi$ carries a positive entry, so every quotient coordinate vector is some
$e_{r(j)}$, and taking maxima bounds $\|\widehat\Phi\|_1$ by
$\frac Mm\|\Phi_\theta\|_1$. The two bounds together are
\eqref{eq:weighted}.
Taking $\frac1{t-s}\log$ of both sides and letting $t-s\to\infty$ kills the
two constant factors, so the exponents coincide; over one period the same
bound makes the two principal multipliers equal, both being limits of
$\|\cdot\|_1^{1/k}$ along iterates. \hfill$\square$

\medskip
The point of the weighted form is that a natural lumping rarely has unit
weights: aggregating the $B_{k1},B_{k2}$ of Section~\ref{sec:dynamic} by
total mass gives $0$--$1$ entries, but aggregating by, say, infectivity or by
a per-state yield does not, and \eqref{eq:weighted} covers that case with no
further hypothesis.
\section{The canonical-form correspondence of \cite{KTV}}
\label{sec:ccfX}

This is the proof of Proposition~\ref{p:depCCF}, whose three parts are the
dependence extension of \cite{KTV}, the lattice property of the pairs, and
the quotient onto species sets.

\textbf{Proof.} 1. The CCF of an LM-matrix is determined by the pattern of
identically-zero entries of its symbolic part together with the constant
matrix $Q$: the LM-surplus, its minimizers, the block partition and the
block order are computed from that data and nothing else. Replacing the
reactant pattern by the dependence pattern replaces one symbolic zero
pattern by another, and $Q$ is unchanged, being a basis of $\ker\Gamma$.
Output completeness enters the block-triangular argument only as the
vanishing of a submatrix of the symbolic part, $D_{x_M}v_{R^c}\equiv0$,
which with $\Out_D$ holds by the definition of $D_j$. Signs are never used,
which is why an inhibitor is admissible.

2. $\Out_D(M_1\cup M_2)=\Out_D(M_1)\cup\Out_D(M_2)\subseteq R_1\cup R_2$ and
$\Out_D(M_1\cap M_2)\subseteq R_1\cap R_2$, so both pairs are
output-complete; $R\mapsto\rank\Gamma_R$ is submodular and $M\mapsto|M|$
modular, giving the displayed inequality, and $\lambda\ge0$ on
output-complete pairs by Proposition~\ref{p:outSuff}(1).

3. $\Phi$ preserves unions and intersections because they are componentwise,
and the fibre of $M$ is the set of output-complete $R$ with
$\rank\Gamma_R=|M|$, which is exactly the equivalence class. \hfill$\square$

\medskip
What survives of the intuition is therefore this: the canonical form knows
more than the species-set lattice --- it distinguishes pairs that share a
species set --- and $\mathfrak B$ is what remains after forgetting the
reaction component. Proposition~\ref{p:blocks} above is a Birkhoff
representation of $\mathfrak B$ \emph{itself}, proved from $\mathfrak B$ and
not inherited from \cite{KTV}; its blocks are the join-irreducibles of the
quotient, which need not be the canonical blocks of $L^\#$. Everything the
examples use --- the four blocks of Corollary~\ref{c:fourblocks}, the ten of
the twenty-species model, the counts $6$ and $83$ --- is a statement about
$\mathfrak B$, and so is unaffected.

\medskip
What this does \emph{not} transfer is anything resting on monotonicity. The
Petri-net persistence results of \cite{AdLS} assume both $D_j=L_j$ and
nonnegative derivatives in the reactants, and are not available for the
models of Section~\ref{sec:twostrain} and Section~\ref{sec:dynamic}; the
persistence used there is the structured-population criterion, whose
hypotheses are verified directly on the transversal blocks and never
inferred from monotonicity of the whole network. Siphon invariance, the
block triangularity \eqref{eq:siphontri}, OM localization, submodularity and
the whole lattice do survive, the first two through $L_j$ and the boundary
condition \eqref{eq:bdcompat} and the rest through $D_j$, as
Definition~\ref{d:RN} records.

\medskip
One further reduction is worth stating, because three of our models use it
implicitly. A rate that is a \emph{sum} of monotone contributions over one
stoichiometric column, such as $\beta S(I+\eta Y)$ in
Sections~\ref{sec:twomodels} and~\ref{sec:cwg}, has dependence set
the union of the summands' dependence sets, and does not vanish when only
one of $I$, $Y$ does. Splitting it into parallel reactions with the same
stoichiometric column changes neither $\Out_D(M)$ for any $M$ nor the span
of the selected columns, hence changes no buffering index: the aggregated and
split representations have the same lattice. So those models raise no
difficulty of the present kind, and the choice between the two
representations is one of presentation. We use that identification as established mathematics and do not
reproduce its proof, nor the LM-surplus calculus and determinant
factorization that go with it. No change of reaction representation is
needed: \eqref{eq:LdetGR} holds for the representation already fixed in each
example.

Two cautions attach to the identification.
\begin{enumerate}
\item The rank formula behind the canonical form treats the nonzero entries
of $R$ as algebraically independent symbols. Actual kinetics impose relations
among them --- two rates sharing a stoichiometric column, a shared force of
infection, a mass-action product appearing in several rows --- so the
canonical reading is generic, and a particular kinetics may satisfy more
identities than it predicts. The indices $\lambda(M)$ computed in this paper
do not depend on that hypothesis: they are ranks of the constant matrix
$\Gamma$.
\item The direction of the order must be kept straight. The order used in
Proposition~\ref{p:blocks} is the one of \cite{KTV}, where $M_a\preceq M_b$
means that every minimizer containing $M_a$ contains $M_b$; under it the
buffered sets are the \emph{upward}-closed families of blocks. An order
ideal, in \cite{KTV} and here, is a \emph{down}-set, and the identification
pairs a buffering structure with the \emph{complement} of its family of
blocks. Reading ``order ideal'' as the buffered family itself reverses every
reachability statement, in Remark~\ref{r:ccfscope} and wherever else the
canonical form is invoked.
\end{enumerate}


\section{Open problems}
\label{sec:beO}

\beO[Two-strain maturation networks: characterize permanence for unequal release rates]
\lbl{o:UnequalRelease}
Theorem~\ref{t:permregion} establishes uniform permanence under
$h_1=h_2$, while Proposition~\ref{p:coexist} establishes the existence
of a positive coexistence equilibrium for arbitrary $h_1,h_2>0$.
Determine whether
\[
R_{1\mid2}>1,\qquad R_{2\mid1}>1
\]
implies uniform permanence for arbitrary positive release rates and
nonlinear catalytic conversion rates $\theta_B,\theta_C$.
When $h_1\ne h_2$, the sum $A_1+A_2$ no longer closes the resident
dynamics, and the cooperative reduction used in the permanence proof
is unavailable. Either prove global attraction of the single-strain
resident equilibria under verifiable conditions, or construct a
parameter regime in which a non-equilibrium boundary attractor
invalidates the proposed permanence criterion.
Determine also whether the permanence domain remains invariant
under all perturbations certified by the maximal buffer
$\cA\cup\cB$.
\eeO

\noindent\textbf{Partly settled in Section~\ref{sec:unequal}.} The first
alternative is carried out whenever the maturation branches are linear:
Lemma~\ref{l:resface} proves global attraction on the resident face for
arbitrary $h_1,h_2>0$ once $\theta_B=\theta_C=0$, because the $A$-block is
then linear with determinant $u_+h_2+h_1u_-+h_1h_2>0$ and needs no relation
between the release rates, and Theorem~\ref{t:unequal} concludes uniform
permanence, extending it to small positive $\theta_B,\theta_C$ by
robustness. So equality of the release rates is not essential. With
catalysis present, Section~\ref{sec:newmath} does three further things:
Theorem~\ref{t:uniqres} makes the positive resident equilibrium unique for
\emph{every} catalytic coefficient and every pair of release rates, so what
is left is a question about dynamics and not about the equilibrium count;
Theorem~\ref{t:explicitcat} proves attraction on the cooperative region
\eqref{eq:catregion}--\eqref{eq:catdom}; and Theorem~\ref{t:unequalGain}
proves it on the small-gain region \eqref{eq:SG}, with no sign condition and
arbitrarily large catalysis. What remains of this problem is the complement
of the union of those two regions --- $h_1\ne h_2$ with $\theta_B,\theta_C$
both large, where $A_1+A_2$ does not close --- and there the alternative is
sharp: either attraction holds throughout, or the resident face carries a
non-equilibrium attractor. Remarks~\ref{r:catgap} and~\ref{r:tworegions} say
this precisely.

\beO[Resident-face buffering lattices: preserve invasion exponents over non-equilibrium residents]
\lbl{o:DynamicInvBuffer}
Proposition~\ref{p:maxbuffer} protects invasion matrices evaluated
at regular resident equilibria. Suppose instead that a resident
face supports a periodic orbit or a compact invariant set carrying
an ergodic invariant probability measure $\nu$. The absent strain
then satisfies a positive linear cocycle
\[
\dot z=J_{\mathrm{tra}}(y(t);\theta)z,
\]
whose invasion rate is its principal Lyapunov exponent.
Find structural conditions, expressed through the resident-face
buffering lattice and the dependence of $J_{\mathrm{tra}}$ on resident
coordinates, under which this exponent is invariant under all
parameters confined to $\operatorname{Out}(M_Q)$.
Determine whether additional conservation identities or a
bounded change of transverse coordinates are sufficient, and
whether either is necessary for a suitable class of epidemic
reaction networks. In particular, construct or exclude a model
where OM buffering preserves every resident equilibrium invasion
number but fails to preserve invasion into a periodic resident.
\eeO

\noindent\textbf{Sufficient conditions and counterexamples, not a
characterization.} What has been established is one implication and its
limits, and it should not be read as more. The positive
direction holds: Theorem~\ref{t:dynamicOM} supplies a sufficient
condition in place of the index --- the unperturbed rates must factor
through the annihilator $L\Gamma_R=0$, and the transversal matrix through
$Lx$ --- and Theorem~\ref{t:positiveAggregation} gives a checkable
substitute for the second requirement whenever the invasion system is
positive and lumps.

The counterexample asked for at the end is supplied twice, in both classes.
Proposition~\ref{p:periodicCounterexample} keeps a zero buffering index and a
$b$-independent equilibrium invasion number while its Floquet exponent
crosses zero as $b$ varies, so the index alone is \emph{not} sufficient; that
network is periodically forced, hence outside the autonomous setting of
Definition~\ref{d:RN}, and Proposition~\ref{p:AutonomousFloquet} supplies the
autonomous mass-action version, with the oscillation produced by the resident
kinetics and every stoichiometric column integral. What is \emph{not}
established is any characterization: a sufficient condition and two
counterexamples do not delimit the class of networks in which buffering
preserves invasion off equilibrium, and the converse direction is
Problem~\ref{o:DynamicConverse}. The quantitative question that went with it
--- how far from an equilibrium a resident must be before the index becomes
misleading --- is also settled: Corollary~\ref{c:amplitude} bounds the entire
loss of invariance by
$\kappa c^2\operatorname{Var}_T(X)/(\delta^2+\Omega^2)\le
\kappa c^2a/(\delta^2+\Omega^2)$, so the discrepancy is $O(a)$ in the
squared radius of the resident orbit and vanishes with it, exactly as the
forced example's effect vanishes with $\varepsilon$. What is \emph{not} known
is whether such a bound holds for a general network rather than this one.

\beO[Reaction networks: characterize all structurally invariant invasion thresholds]
\lbl{o:CompleteInvCertificate}
Proposition~\ref{p:maxbuffer} determines the largest OM buffer
disjoint from the coordinates entering a prescribed invasion matrix.
It does not characterize all reaction-rate perturbations preserving
its reproduction number. Indeed, conservation identities may
produce invariances that are not certified by this buffer.
For a class of reaction networks with independently variable
positive rate constants, characterize the parameters $\theta$
for which
\[
\partial_\theta
\rho\bigl(F_\Sigma(x^*(\theta);\theta)
V_\Sigma(x^*(\theta);\theta)^{-1}\bigr)=0
\]
identically at every regular positive resident equilibrium.
Determine whether the OM lattice, together with finitely many
stoichiometric conservation or weighted-flux identities, gives a
necessary and sufficient generic criterion.
If not, construct a minimal epidemic counterexample and identify
the additional algebraic information required.
\eeO

\beO[Multistrain vaccination networks: determine lattice-protected invasion graphs]
\lbl{o:MultiStrainLattice}
Construct a three-strain epidemic network with selective vaccination,
partial cross-immunity and secondary infection, in which distinct
resident faces have non-chain OM buffering lattices and the immune
compartments participate in infection feedback.
For every resident strain set $K$ and absent strain $j$, determine
the maximal buffering set $M_{K,j}$ protecting the corresponding
invasion matrix. Establish conditions under which a family of
reaction-rate perturbations preserves all invasion signs
simultaneously.
For an acyclic invasion graph, determine when this invariance
implies preservation of the permanence or exclusion regime.
For cyclic invasion graphs, determine which additional boundary
invariant measures must be controlled. The objective is a
structural classification of multistrain coexistence regions,
rather than a mere increase in the number of buffering sets.
\eeO

\noindent\textbf{Partly settled in Section~\ref{sec:dynamic}.} The network is
built there, with forty-five species, and Proposition~\ref{p:threeLattice}
computes its core-free resident-face lattices exactly: five elements on a
one-strain face, fifteen on a two-strain face, twenty-seven on the full face,
none of them chains. The maximal core-free certificate $M_{3\mid12}$ and the
eighteen rates it protects are identified, and
Theorem~\ref{t:threeDynamic} upgrades that protection from equilibria to
whole trajectories. The classification of coexistence regions by invasion
graph, acyclic or not, is untouched and remains the content of this problem.

\beO[Ecological reaction networks: continue buffering and invasion certificates through singular faces]
\lbl{o:SingularFaceBuffer}
Theorems~\ref{t:faceRank} and \ref{t:bufInv} use a nonsingular
tangential Jacobian.  Develop a replacement for the localization and
invasion statements when a siphon face carries a nonisolated equilibrium
set, a fold, or another nonhyperbolic resident invariant set.
Specifically, formulate conditions under which selected invasion
quantities are constant on \emph{all} resident equilibria in a
local algebraic component, even though individual resident
coordinates cannot be differentiated uniquely with respect to a
parameter.  Determine whether a quotient-space or determinantal-ideal
version of the buffering index gives a necessary and sufficient
criterion, and how such a criterion behaves when the dimension of
the equilibrium set changes.
Remark~\ref{r:faceScope} supplies the smallest obstruction: on a face
of nonisolated equilibria the restricted index can be negative.
\eeO

\beO[Reinfection and waning: determine when destruction of buffering changes coexistence]
\lbl{o:ReinfLattice}
Section~\ref{ss:prune} shows that introducing reinfection and waning reactions
can reduce a buffering lattice from $83$ elements to $47$, $11$,
and finally $3$. Determine which additions of reactions destroy
a specified maximal invasion-neutral buffer and which actually
change an invasion number.
In particular, characterize minimal sets of reinfection or
waning reactions that simultaneously
\[
\lambda(M)=0\longrightarrow\lambda(M)>0,
\qquad
R_{j\mid K}<1\longleftrightarrow R_{j\mid K}>1
\]
for some resident community $K$ and absent strain $j$.
The first change is structural, whereas the second depends on
the rates and the resident dynamics. Establish conditions under
which the structural change permits, forces, or excludes the
second, and determine its consequences for the permanence region.
\eeO

\beO[Global permanence for all three strains]
\lbl{o:ThreePermanence}
Find verifiable conditions on the invasion exponents of the seven
proper resident communities that ensure uniform permanence of the
$45$-species network of Section~\ref{sec:dynamic}. In particular,
determine whether its invasion graph is acyclic in a useful parameter
region, and whether the rate-independent quotient \eqref{eq:quotient}
allows the permanence criterion to be checked without resolving all
maturation stages. The positive coexistence equilibrium computed
there does not settle this problem.
\eeO

\beO[Converse to dynamic OM protection]
\lbl{o:DynamicConverse}
Characterize those positive reaction networks for which equality of
all periodic or ergodic invasion exponents under a family of
OM-buffered rate perturbations forces a common quotient, or a bounded
cohomological conjugacy of the transversal cocycles. The seasonal
counterexample of Proposition~\ref{p:periodicCounterexample} shows why
a zero buffering index alone is insufficient; what is missing is the
other direction, a characterization of when the hypotheses of
Theorem~\ref{t:dynamicOM} are not merely sufficient but necessary.
\eeO

\beO[Unequal release rates in the three-strain network]
\lbl{o:ThreeUnequal}
Determine the largest parameter families preserving invasion signs
when $h_{k1}\ne h_{k2}$ or $b_{k1}\ne b_{k2}$ in the model of
Section~\ref{sec:dynamic}. Such models retain the equilibrium
buffering certificates but lose the pathwise aggregation
\eqref{eq:quotient}. Establish useful perturbative bounds on the
resulting principal Lyapunov exponents, or exhibit a seasonal
invasion-sign reversal caused by an otherwise equilibrium-neutral
OM-buffered conversion parameter. This is the dynamic counterpart of
Problem~\ref{o:UnequalRelease}, which asks the same question for the
fourteen-species model at equilibrium.
\eeO

\newcommand{\etalchar}[1]{$^{#1}$}
\providecommand{\bysame}{\leavevmode\hbox to3em{\hrulefill}\thinspace}
\providecommand{\MR}{\relax\ifhmode\unskip\space\fi MR }
\providecommand{\MRhref}[2]{%
  \href{http://www.ams.org/mathscinet-getitem?mr=#1}{#2}
}
\providecommand{\href}[2]{#2}


\end{document}